\documentclass{elsart}
\usepackage{epsfig,times,amssymb,amsmath,multirow,rotating}
\usepackage{picinpar}
\usepackage{wrapfig}
\usepackage{floatflt}

\newcommand{\bce}{\begin{center}}
\newcommand{\ece}{\end{center}}

\newcommand{\f}[2]{\frac{#1}{#2}}
\newcommand{\bal}{\begin{eqnarray*}}
\newcommand{\eal}{\end{eqnarray*}}
\newcommand{\bmat}{\begin{math}}
\newcommand{\emat}{\end{math}}
\newcommand{\bec}{\begin{displaymath}}
\newcommand{\eec}{\end{displaymath}}
\newcommand{\becnum}{\begin{equation}}
\newcommand{\eecnum}{\end{equation}}
\newcommand{\hsp}{\hspace*{1cm}}

\begin{document}

\begin{frontmatter}
\title{Propagating leptons through matter with Muon Monte Carlo (MMC)}

\author{Dmitry Chirkin}
\address{University of Wisconsin at Madison, Madison, USA}
\ead{dima@icecube.wisc.edu}
\ead[url]{http://icecube.wisc.edu/\~{ }dima/work/MUONPR/}
\author{Wolfgang Rhode}
\address{Institute of Physics, Dortmund University, Dortmund, Germany}

\begin{abstract}
\label{MMC}
An accurate simulation of the propagation of muons through matter is needed
for the analysis of data produced by muon/neutrino underground experiments. A muon
may sustain hundreds of interactions before it is detected by the experiment.
Since a small systematic uncertainty repeated hundreds of times may lead to sizable
errors, requirements on the precision of the muon propagation code are very
stringent. A new tool for propagating muon and tau charged leptons through matter
that is believed to meet these requirements is presented here. An overview of
the program is given and some results of its application are discussed.
\end{abstract}
\end{frontmatter}

\section{Introduction}

In order to observe atmospheric and cosmic neutrinos with a large underground detector (e.g., AMANDA \cite{AMANDA}), one needs to isolate the neutrino signal from the 3-5 orders of magnitude larger signal from the background of atmospheric muons. Methods that do this have been designed and proven viable \cite{tyce}. In order to prove that these methods work and to derive indirect results such as the spectral index of atmospheric muons, one needs to compare data to the results of the computer simulation. Such a simulation normally contains three parts: propagation of the measured flux of the cosmic particles from the top of the atmosphere down to the surface of the ground (ice, water); propagation of the atmospheric muons from the surface down to and through the detector; and generation of the secondary particles (electrons, Cherenkov photons, etc.) in the vicinity of the detector and their interaction with the detector components. The first part is normally called {\it generator}, since it generates muon flux at the ground surface; the second is {\it propagator}; and the third simulates the detector interaction with the passing muons. To generate atmospheric muon and neutrino fluxes we used CORSIKA \cite{dcors_5}. Results and methods of using CORSIKA as a generator in a neutrino detector (AMANDA-II) were discussed in \cite{mythesis,SLC}. Several muon propagation Monte Carlo programs were used with different degrees of success as propagators. Some are not suited for applications which require the code to propagate muons in a large energy range (e.g., mudedx, a.k.a. LOH \cite{cern85}), and the others seem to work in only some of the interesting energy range ($E > 1$ TeV, propmu, a.k.a. LIP \cite{lip}) \cite{paolo}. Most of the programs use cross section formulae whose precision has been improved since their time of writing. For some applications, one would also like to use the code for the propagation of muons that contain $100 - 1000$ interactions along their track, so the precision of each step should be sufficiently high and the computational errors should accumulate as slowly as possible. Significant discrepancies between the muon propagation codes tested in this work were observed, and are believed to be mostly due to algorithm errors (see Appendix \ref{app_mmc2}). This motivated writing of a new computer program (Muon Monte Carlo: MMC \cite{mmc}), which minimizes calculational errors, leaving only those uncertainties that come from the imperfect knowledge of the cross sections.

\section{Description of the code}

\begin{figure}[!h]\bce \mbox{\epsfig{file=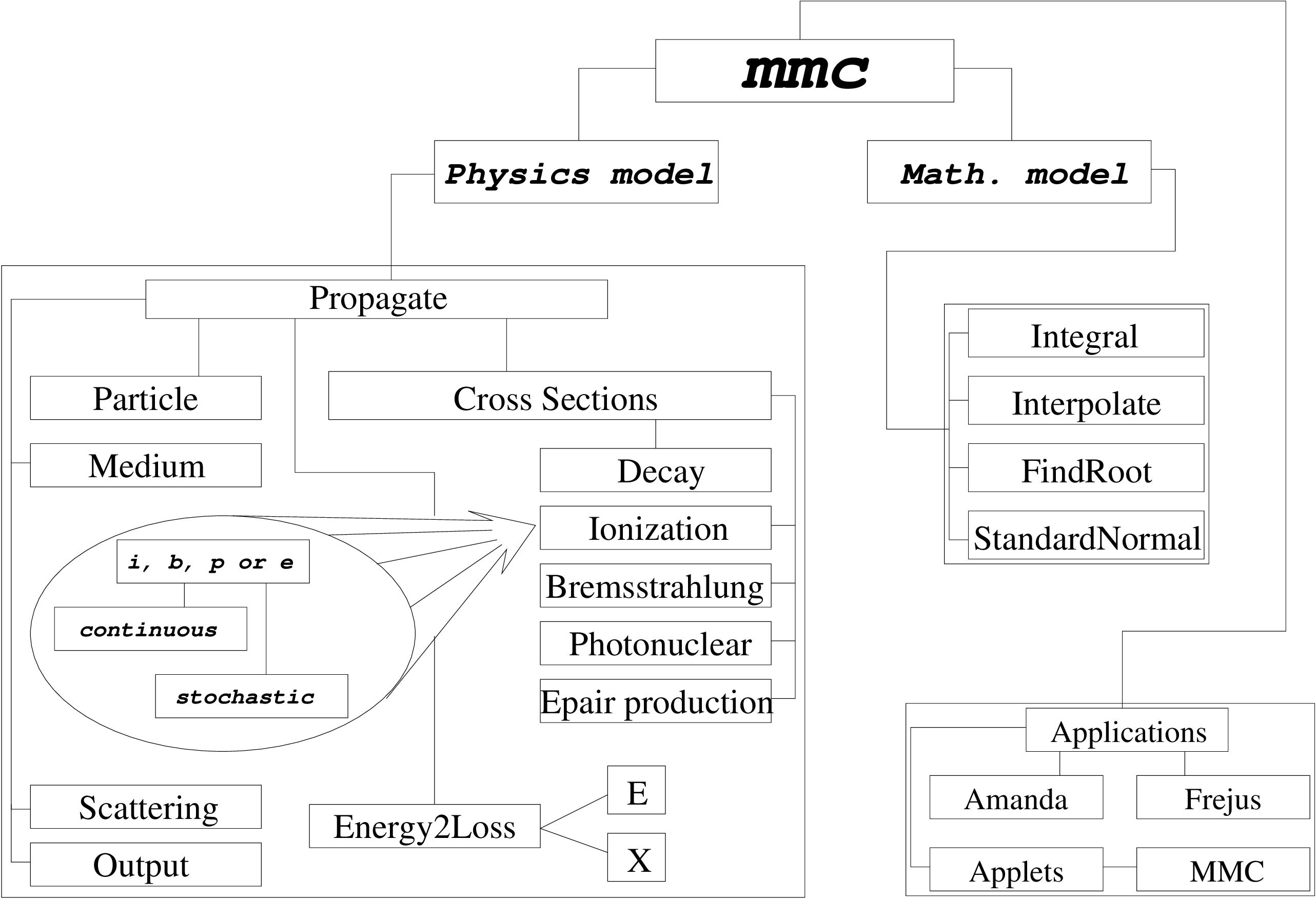,width=.9\textwidth}}\\ {\caption[MMC structure]{\label{mmc_fig_1}MMC structure}} \ece \end{figure}

The primary design goals of MMC were computational precision and code clarity. The program is written in Java, its object-oriented structure being used to improve code readability. MMC consists of pieces of code (classes), each contained in a separate file. These pieces fulfill their separate tasks and are combined in a structured way (Figure \ref{mmc_fig_1}).

The code evaluates many cross-section integrals, as well as several tracking integrals. All integral evaluations are done by the Romberg method of the 5th order (by default) \cite{nure} with a variable substitution (mostly log-exp). If an upper limit of an integral is an unknown (that depends on a random number), an approximation to that limit is found during normalization integral evaluation, and then refined by the Newton-Raphson method combined with bisection \cite{nure}.

Originally, the program was designed to be used in the Massively Parallel Network Computing (SYMPHONY) \cite{symp} framework, and therefore computational speed was considered only a secondary issue. However, parametrization and interpolation routines were implemented for all integrals. These are both polynomial and rational function interpolation routines spanned over a varying number of points (5 by default) \cite{nure}. Inverse interpolation is implemented for root finding (i.e., when $x(f)$ is interpolated to solve $f(x)=y$). Two-dimensional interpolations are implemented as two consecutive one-dimensional ones. It is possible to turn parameterizations on or off for each integral separately at program initialization. The default energy range in which parametrized formulae will work was chosen to be from 105.7 MeV (the muon rest mass; 1777 MeV for taus) to $E_{big}=10^{14}$ MeV, and the program was tested to work with much higher settings of $E_{big}$. With full optimization (parameterizations) this code is at least as fast or even faster than the other muon propagation codes discussed in Appendix \ref{app_mmc2}.

Generally, as a muon travels through matter, it loses energy due to ionization losses, bremsstrah\-lung, photo-nuclear interaction, and pair production. The cross section formulae are summarized in Section \ref{mmc_formulae}. These formulae are claimed to be valid to within about 1\% in the energy range up to $\gtrsim$ 10 TeV. Theoretical uncertainties in the photonuclear cross section above 100 TeV are higher. All of the energy losses have continuous and stochastic components, the division between which is artificial and is chosen in the program by selecting an energy cut ($e_{cut}$, also $E_{cut}$) or a relative energy loss cut ($v_{cut}$). In the following, $v_{cut}$ and $e_{cut}$ are considered to be interchangable and related by $e_{cut}=v_{cut}E$ (even though only one of them is a constant). Ideally, all losses should be treated stochastically. However, that would bring the number of separate energy loss events to a very large value, since the probability of such events to occur diverges as $1/E_{lost}$ for the bremsstrahlung losses, as the lost energy approaches zero, and even faster than that for the other losses. In fact, the reason this number, while being very large, is not infinite, is the existence of kinematic cutoffs (larger than some $e_0$) for all diverging cross sections. A good choice of $v_{cut}$ for the propagation of atmospheric muons should lie in the range $0.05 - 0.1$ (Section \ref{mmc_errors}, also \cite{mum}). For monoenergetic beams of muons, $v_{cut}$ may have to be chosen to be high as $10^{-3}-10^{-4}$.

\subsection{Tracking formulae}
\label{mmc_tracking_section}

Let the continuous part of the energy losses (a sum of all energy losses, integrated from zero to $e_{cut}$) be described by a function $f(E)$: $$-\frac{dE}{dx}=f(E).$$

\begin{figure}[!h]\bce \mbox{\epsfig{file=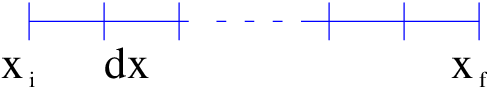,width=.3\textwidth}}\\ {\caption[Derivation of tracking formulae]{\label{mmc_fig_2}Derivation of tracking formulae}} \ece \end{figure}

The stochastic part of the losses is described by the function $\sigma(E)$, which is a probability for any energy loss event (with lost energy $> e_{cut}$) to occur along a path of 1 cm. Consider the particle path from one interaction to the next consisting of small intervals (Figure \ref{mmc_fig_2}). On each of these small intervals the probability of interaction is $dP(E(x_i))=\sigma (E(x_i)) dx$. We now derive an expression for the final energy after this step as a function of the random number $\xi$. The probability to completely avoid stochastic processes on an interval ($x_i$;$x_f$) and then suffer a catastrophic loss on $dx$ at $x_f$ is
$$(1-dP(E(x_i)))\cdot...\cdot(1-dP(E(x_{f-1}))) \cdot dP(E(x_f))$$
$$\approx \exp(-dP(E(x_i)))\cdot...\cdot\exp(-dP(E(x_{f-1}))) \cdot dP(E(x_f))$$
$$\xrightarrow[dx\rightarrow 0]{} \exp\left(-\int^{E_f}_{E_i} dP(E(x))\right) \cdot dP(E(x_f))$$
$$=d_f \left(-\exp(-\int^{E_f}_{E_i} \frac{\sigma(E)}{-f(E)} \cdot dE) \right) = d (-\xi), \quad \xi \in (0;1]$$
To find the final energy after each step the above equation is solved for $E_f$:
$$\int^{E_f}_{E_i} \frac{\sigma(E)}{-f(E)} \cdot dE = - \log(\xi) \quad \mbox{ (energy integral).}$$
This equation has a solution if
$$\xi>\xi_0=\exp\left(-\int^{E_i}_{e_{low}} \frac{\sigma(E)}{f(E)} \cdot dE\right).$$
Here $e_{low}$ is a low energy cutoff, below which the muon is considered to be lost. Note that $f(E)$ is always positive due to ionization losses (unless $e_{cut}\lesssim I(Z)$). The value of $\sigma(E)$ is also always positive because it includes the positive decay probability. If $\xi<\xi_0$, the particle is stopped and its energy is set to $e_{low}$. The corresponding displacement for all $\xi$ can be found from
$$x_f=x_i-\int^{E_f}_{E_i}\frac{dE}{f(E)} \quad \mbox{ (tracking integral),}$$
and time elapsed can be found from
$$t_f=t_i+\int^{x_f}_{x_i}\frac{dx}{v(x)}=t_i-\int^{E_f}_{E_i}\frac{dE}{f(E)v(E)} \quad \mbox{ (time integral).}$$
Evaluation of time integral based on the approximation $v=c$, $t_f=t_i+(x_f-x_i)/c$, is also possible.

\subsection{Continuous randomization}
\label{mmc_cont_section}

\begin{figure}[!h]\begin{center}
\begin{tabular}{ccc}
\mbox{\epsfig{file=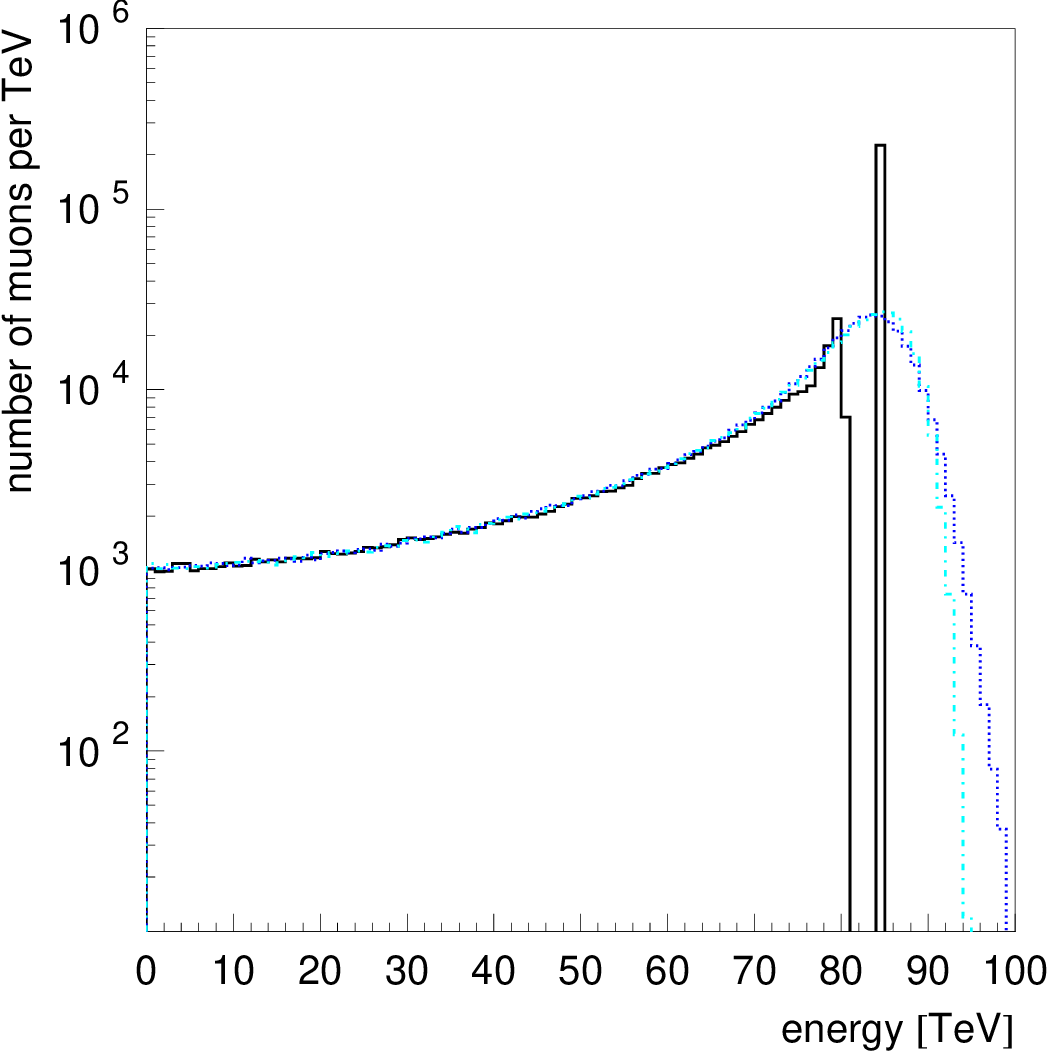,width=.45\textwidth}} & \ & \mbox{\epsfig{file=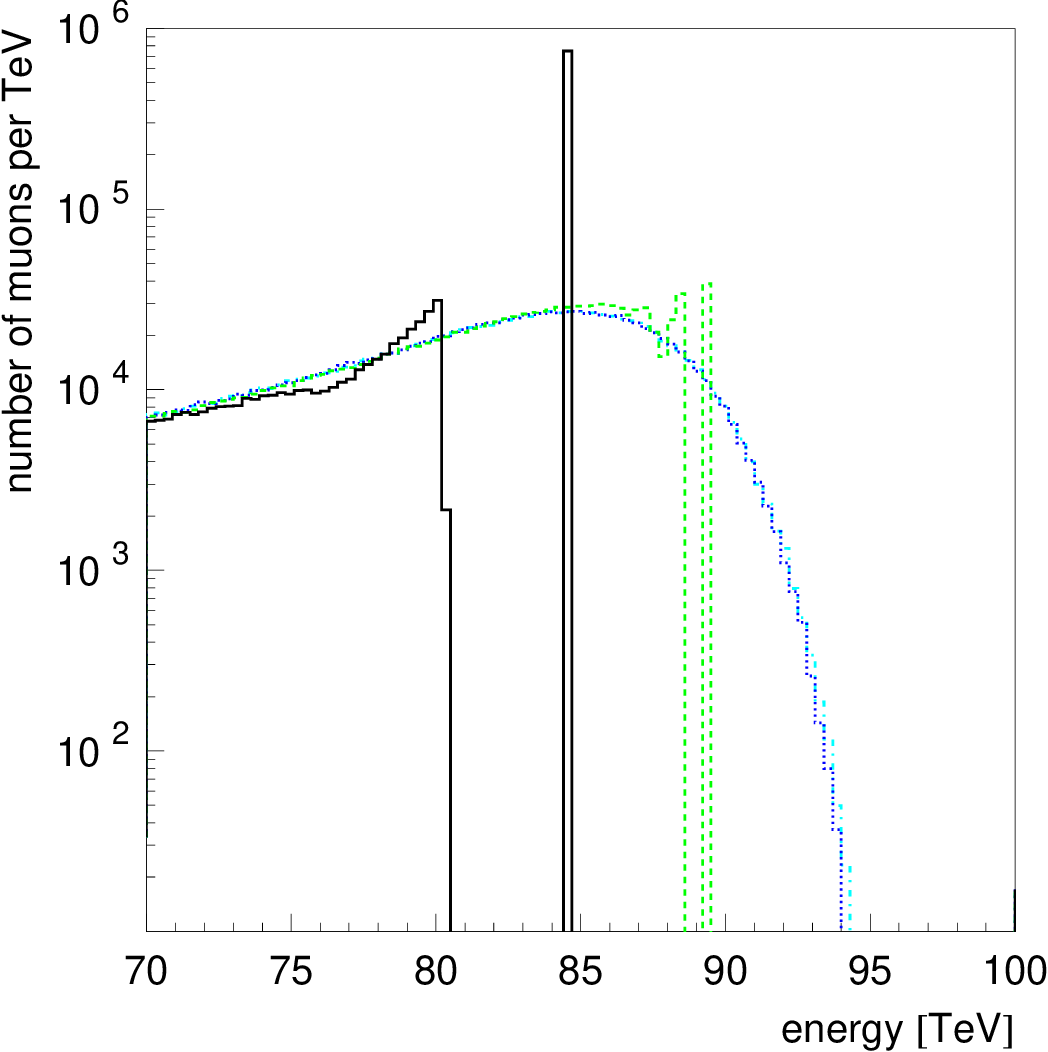,width=.45\textwidth}} \\
\parbox{.45\textwidth}{\caption[Distribution of the final energy of the muons that crossed 300 m of Fr\'{e}jus Rock with initial energy 100 TeV]{\label{mmc_fig_3}Distribution of the final energy of the muons that crossed 300 m of Fr\'{e}jus Rock with initial energy 100 TeV: $v_{cut}=0.05$ (solid), $v_{cut}=10^{-4}$ (dashed-dotted), $v_{cut}=0.05$ and {\it cont} option (dotted)}} & \ & \parbox{.45\textwidth}{\caption[A close-up on the Figure \ref{mmc_fig_3}]{\label{mmc_fig_4}A close-up on the Figure \ref{mmc_fig_3}: $v_{cut}=0.05$ (solid), $v_{cut}=0.01$ (dashed), $v_{cut}=10^{-3}$ (dotted), $v_{cut}=10^{-4}$ (dashed-dotted) \\}} \\
\end{tabular}
\end{center}\end{figure}

It was found that for higher $v_{cut}$ muon spectra are not continuous (Figure \ref{mmc_fig_3}). In fact, there is a large peak (at $E_{peak}$) that collects all particles that did not suffer stochastic losses followed by the main spectrum distribution separated from the peak by at least the value of $v_{cut}E_{peak}$ (the smallest stochastic loss). The appearance of the peak and its prominence are governed by $v_{cut}$, co-relation of initial energy and propagation distance, and the binning of the final energy spectrum histogram. In order to be able to approximate the real spectra with even a large $v_{cut}$ and to study the systematic effect at a large $v_{cut}$, a {\it continuous randomization} feature was introduced.

\begin{figure}[!h]\begin{center}
\begin{tabular}{ccc}
\mbox{\epsfig{file=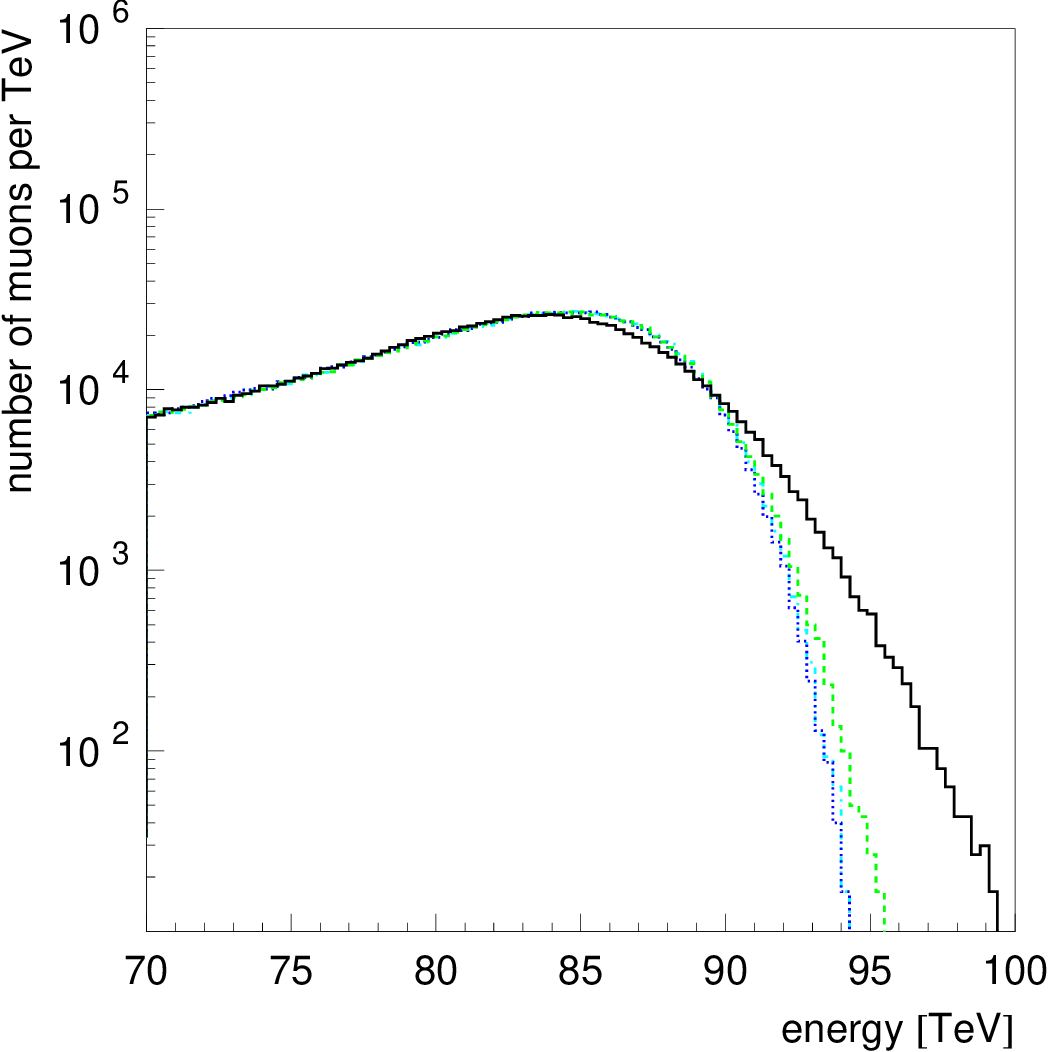,width=.45\textwidth}} & \ & \mbox{\epsfig{file=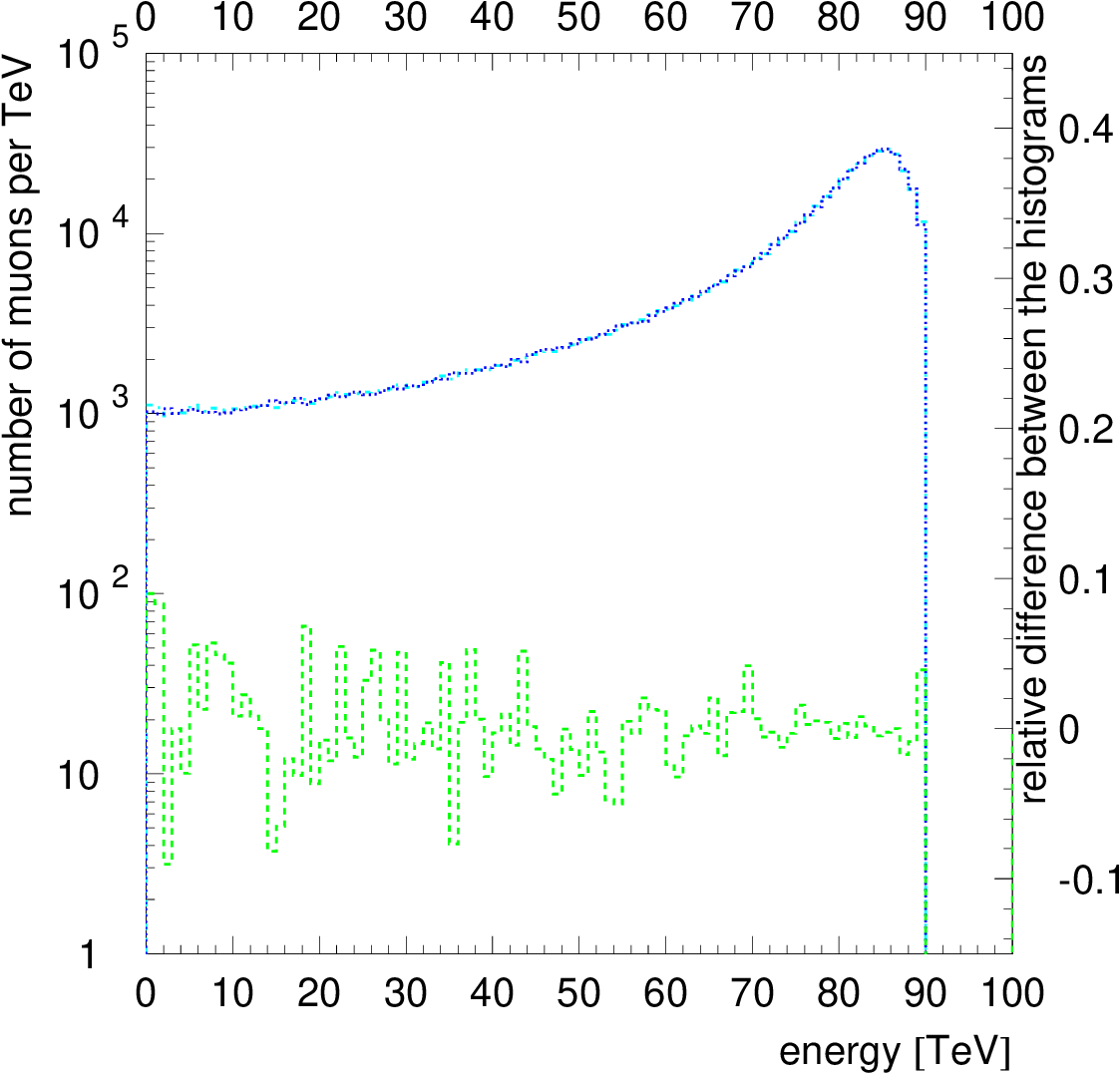,width=.45\textwidth}} \\
\parbox{.45\textwidth}{\caption[Same as in Figure \ref{mmc_fig_4}, but with {\it cont} option enabled ]{\label{mmc_fig_5}Same as in Figure \ref{mmc_fig_4}, but with {\it cont} option enabled \\ \\ \\}} & \ & \parbox{.45\textwidth}{\caption[Comparison of parametrized with exact (non-parametrized) versions for $v_{cut}=0.01$. Also shown is the relative difference of the curves, i.e., ratio of the difference over exact histogram bin values.]{\label{mmc_fig_6}Comparison of parametrized (dashed-dotted) with exact (non-parametrized, dotted) versions for $v_{cut}=0.01$. Also shown is the relative difference of the curves, i.e., ratio of the difference over exact histogram bin values.}} \\
\end{tabular}
\end{center}\end{figure}

For a fixed $v_{cut}$ or $e_{cut}$ a particle is propagated until the algorithm discussed above finds an interaction point, i.e., a point where the particle loses more than the cutoff energy. The average value of the energy decrease due to continuous energy losses is evaluated according to the energy integral formula of the previous section. There will be some fluctuations in this energy loss, which are not described by this formula. Let us assume there is a cutoff for all processes at some small $e_0\ll e_{cut}$. Then the probability $p(e; E)$ for a process with $e_0<e_{lost}<e_{cut}$ on the distance $dx$ is finite. Now choose $dx$ so small that
$$ p_0=\int^{e_{cut}}_{e_0}p(e; E)\ de\cdot dx \ll 1 \mbox{.}$$
Then the probability to not have any losses is $1-p_0$, and the probability to have two or more separate losses is negligible. The standard deviation of the energy loss on $dx$ from the average value
$$<e>=\int^{e_{cut}}_{e_0}e\cdot p(e; E)\ de\cdot dx$$
is then $<(\Delta e)^2>=<e^2>-<e>^2$, where
$$<e^2>=\int^{e_{cut}}_{e_0}e^2\cdot p(e; E)\ de\cdot dx \mbox{.}$$

If the value of $v_{cut}$ or $e_{cut}$ used for the calculation is sufficiently small, the distance $x_f-x_i$ determined by the energy and tracking integrals is so small that the average energy loss $E_i-E_f$ is also small (as compared to the initial energy $E_i$). One may therefore assume $p(e; E)\simeq p(e; E_i)$, i.e., the energy loss distributions on the small intervals $dx_n$ that sum up to the $x_f-x_i$, is the same for all intervals. Since the total energy loss $E_i-E_f=\sum e_n$, the central limit theorem can be applied, and the final energy loss distribution will be Gaussian with the average $\Delta E=E_i-E_f$ and width
$$<(\Delta (\Delta E))^2>=\sum_n \left(<e_n^2>-<e_n>^2\right)$$
$$ =\sum_n \left[\left(\int^{e_{cut}}_{e_0}e_n^2\cdot p(e_n; E_i)\ de_n\right)dx_n - \left(\int^{e_{cut}}_{e_0}e_n\cdot p(e_n; E_i)\ de_n\right)^2 dx_n^2\right] $$
$$\simeq \int_{x_i}^{x_f}dx \cdot \left(\int^{e_{cut}}_{e_0}e^2\cdot p(e; E(x))\ de\right) - \int_{x_i}^{x_f}dx \cdot \left(\int^{e_{cut}}_{e_0}e\cdot p(e; E(x))\ de\right)^2 dx $$
Here $E_i$ was replaced with the average expectation value of energy at $x$, $E(x)$. As $dx \rightarrow 0$, the second term disappears. The lower limit of the integral over $e$ can be replaced with zero, since none of the cross sections diverge faster than or as fast as $1/e^3$. Then,
$$<(\Delta (\Delta E))^2> \simeq \int_{x_i}^{x_f}{dE\over -f(E)} \cdot \left(\int^{e_{cut}}_{0}e^2\cdot p(e; E)\ de\right)  \quad \mbox{ ({\it cont} integral).}$$
This formula is applicable for small $v_{cut}$, as seen from the derivation. Energy spectra calculated with {\it continuous randomization} converge faster than those without as $v_{cut}$ is lowered (see Figures \ref{mmc_fig_4} and \ref{mmc_fig_5}). 

\section{Computational and algorithm errors}
\label{mmc_errors}

\begin{figure}[!h]\begin{center}
\begin{tabular}{ccc}
\mbox{\epsfig{file=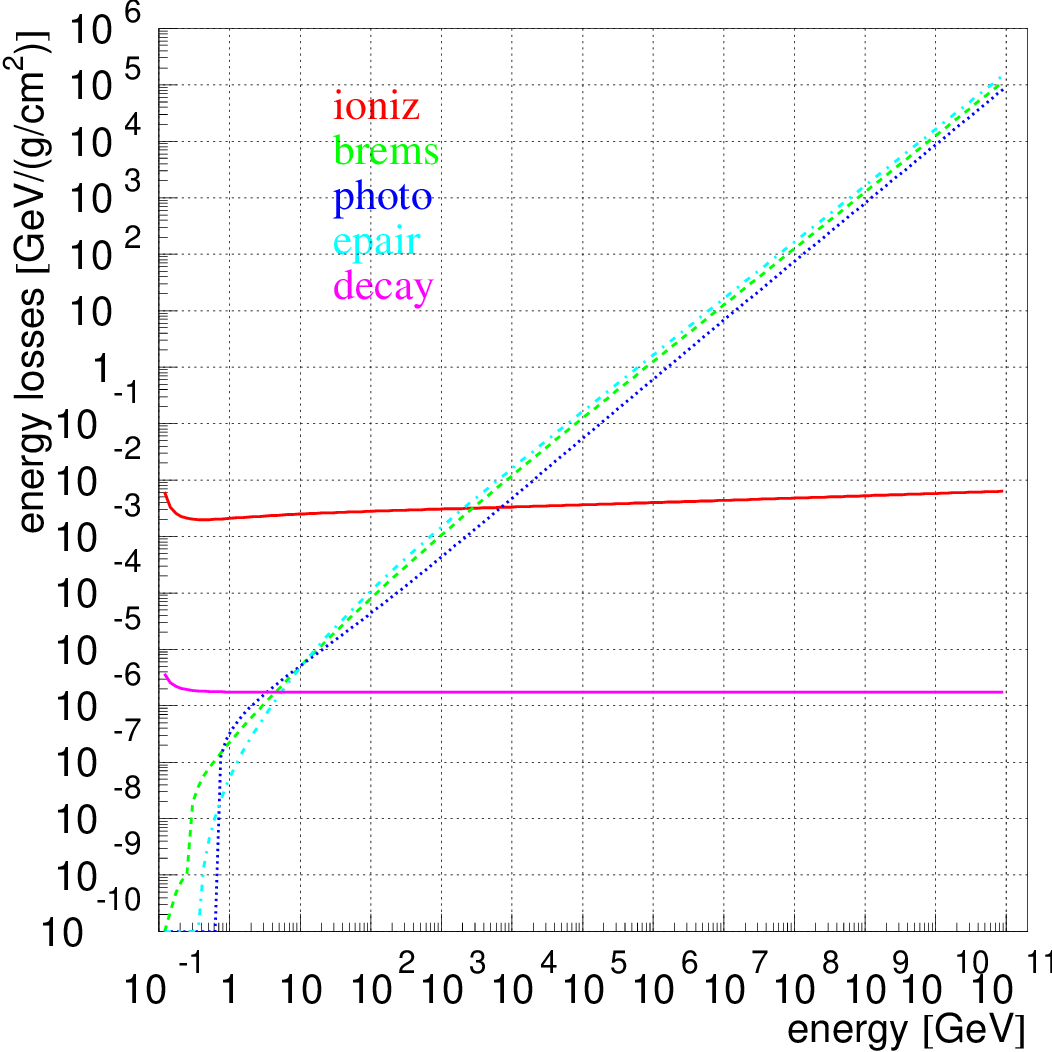,width=.45\textwidth}} & \ & \mbox{\epsfig{file=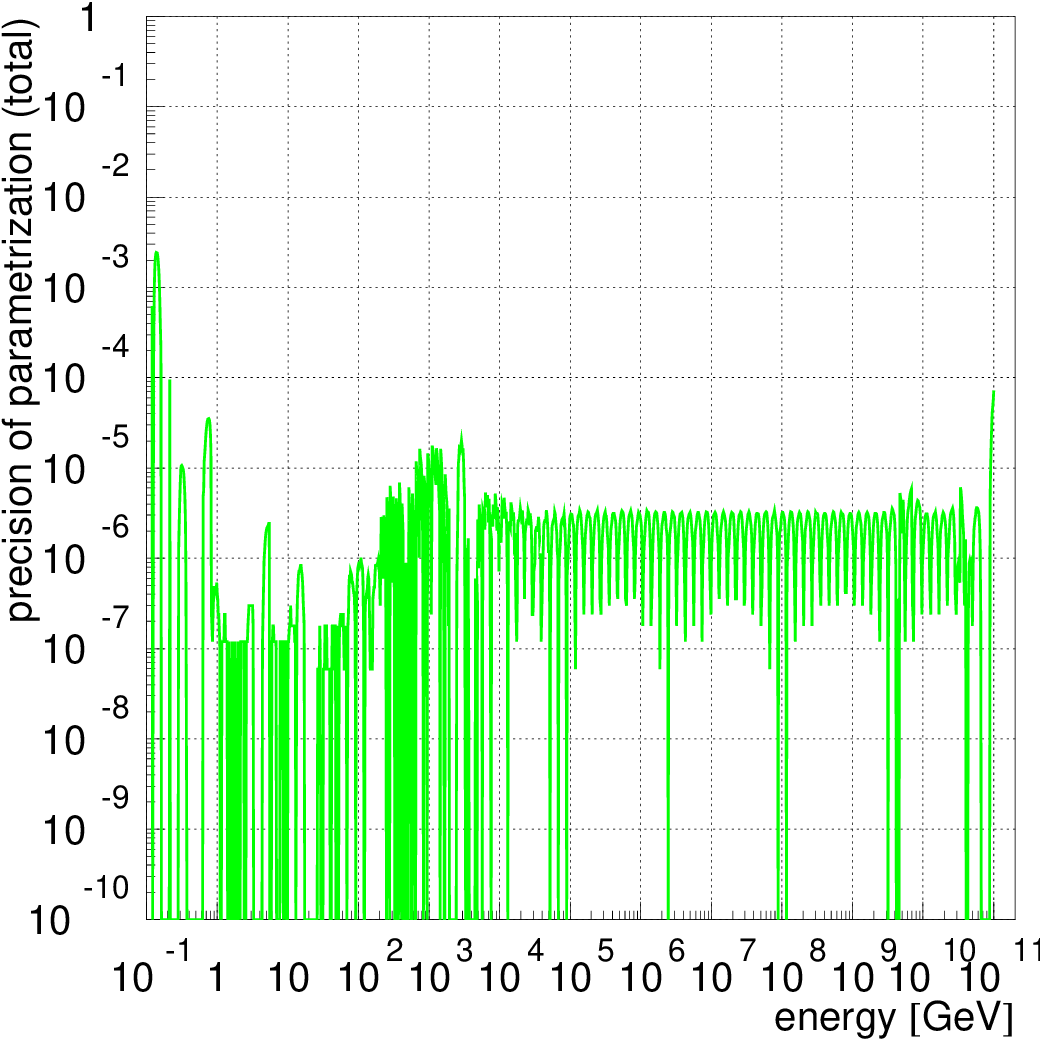,width=.45\textwidth}} \\
\parbox{.45\textwidth}{\caption[Ionization, bremsstrahlung, photonuclear, epair production and decay losses in ice]{\label{mmc_fig_7}Ionization (upper solid curve), bremsstrahlung (dashed), photonuclear (dotted), epair production (dashed-dotted) and decay (lower solid curve) losses in ice}} & \ & \parbox{.45\textwidth}{\caption[Interpolation precision $(e_{pa}-e_{np})/e_{pa}$]{\label{mmc_fig_8}Interpolation precision $(e_{pa}-e_{np})/e_{pa}$\\ \\}} \\
\end{tabular}
\end{center}\end{figure}

\begin{figure}[!h]\begin{center}
\begin{tabular}{ccc}
\mbox{\epsfig{file=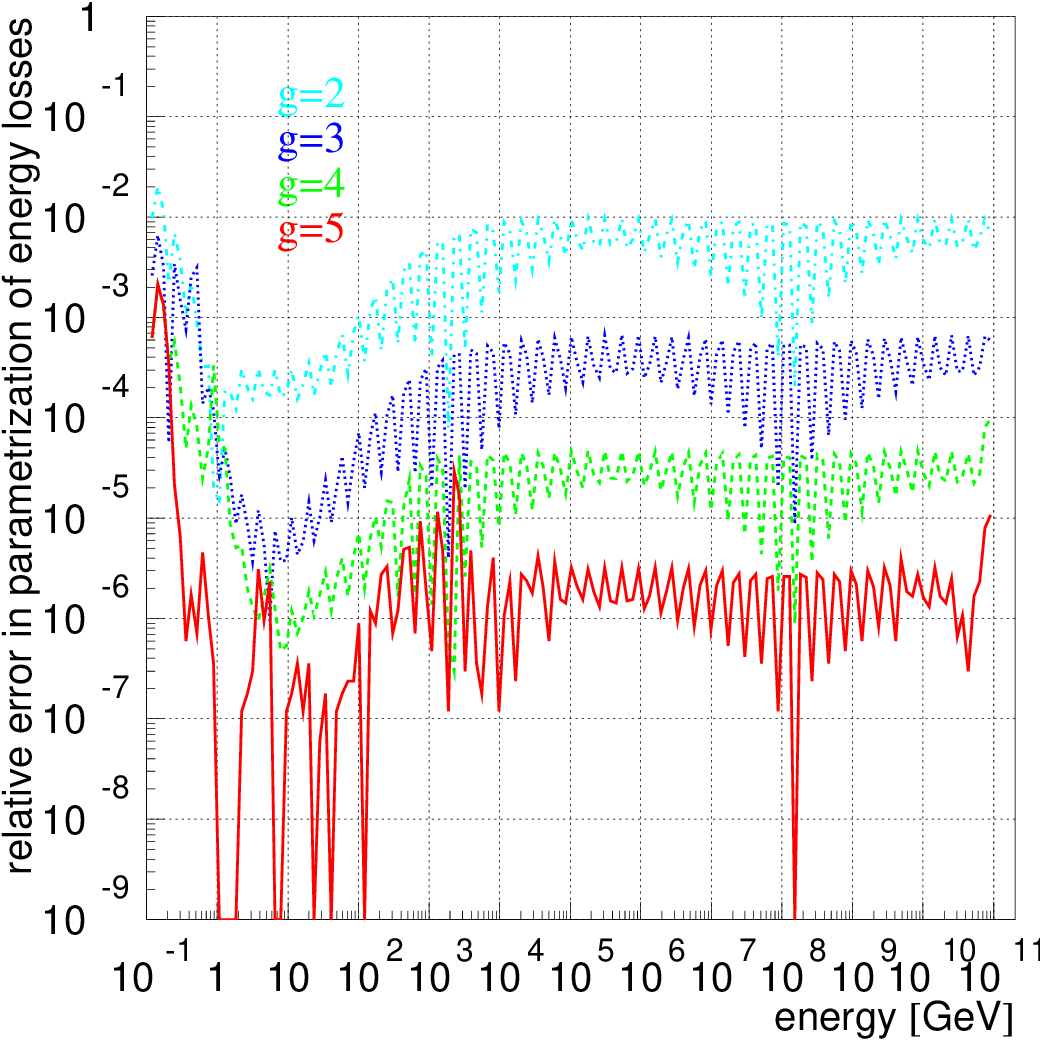,width=.45\textwidth}} & \ & \mbox{\epsfig{file=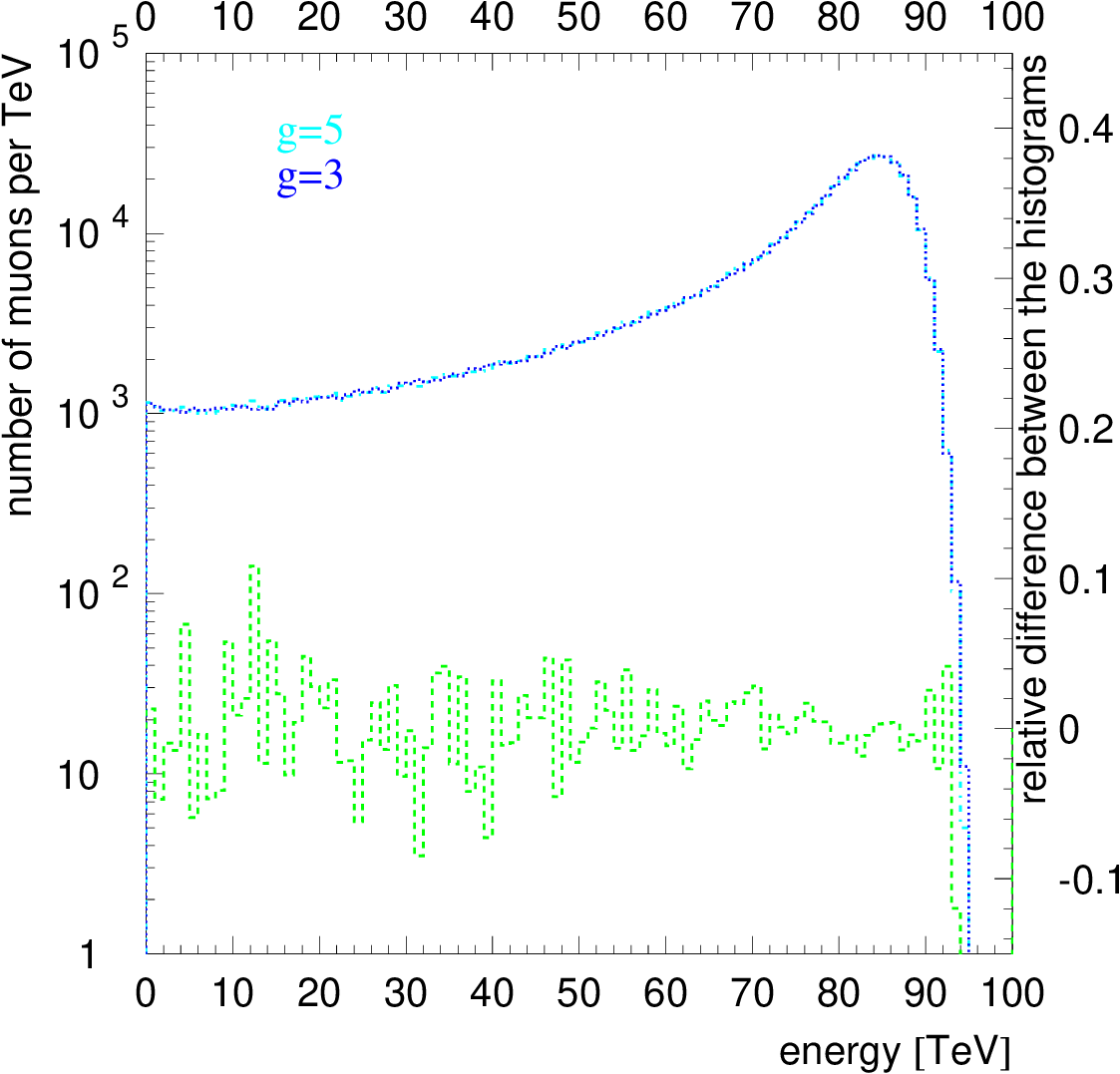,width=.45\textwidth}} \\
\parbox{.45\textwidth}{\caption[Interpolation precision for different orders of the interpolation algorithm]{\label{mmc_fig_a}Interpolation precision for different orders of the interpolation algorithm \\}} & \ & \parbox{.45\textwidth}{\caption[Comparison of the result of the propagation for different orders of the interpolation algorithm]{\label{mmc_fig_b}Comparison of the result of the propagation for different orders of the interpolation algorithm}} \\
\end{tabular}
\end{center}\end{figure}

All cross-section integrals are evaluated to the relative precision of $10^{-6}$; the tracking integrals are functions of these, so their precision was set to a larger value of $10^{-5}$. To check the precision of interpolation routines, results of running with parameterizations enabled were compared to those with parameterizations disabled. Figure \ref{mmc_fig_7} shows relative energy losses for ice due to different mechanisms. Decay energy loss is shown here for comparison and is evaluated by multiplying the probability of decay by the energy of the particle. In the region below 1 GeV, bremsstrahlung energy loss has a double cutoff structure. This is due to a difference in the kinematic restrictions for muon interaction with oxygen and hydrogen atoms. A cutoff (for any process) is a complicated structure to parametrize and with only a few parametrization grid points in the cutoff region, interpolation errors $(e_{pa}-e_{np})/e_{pa}$ may become quite high, reaching 100\% right below the cutoff, where the interpolation routines give non-zero values, whereas the exact values are zero. But since the energy losses due to either bremsstrahlung, photonuclear process, or pair production are very small near the cutoff in comparison to the sum of all losses (mostly ionization energy loss), this large relative error results in a much smaller increase of the relative error of the total energy losses (Figure \ref{mmc_fig_8}). Because of that, parametrization errors never exceed $10^{-4} - 10^{-3}$, for the most part being even much smaller ($10^{-6} - 10^{-5}$), as one can estimate from the plot. These errors are much smaller than the uncertainties in the formulae for the cross sections. Now the question arises whether this precision is sufficient to propagate muons with hundreds of interactions along their way. Figure \ref{mmc_fig_6} is one of the examples that demonstrate that it is sufficient: the final energy distribution did not change after enabling parametrizations. Moreover, different orders of the interpolation algorithm ({\bf g}, corresponding to the number of the grid points over which interpolation is done) were tested (Figure \ref{mmc_fig_a}) and results of propagation with different {\bf g} compared with each other (Figure \ref{mmc_fig_b}). The default value of {\bf g} was chosen to be 5, but can be changed to other acceptable values $3\le${\bf g}$\le6$ at the run time.

\begin{figure}[!h]\begin{center}
\begin{tabular}{ccc}
\mbox{\epsfig{file=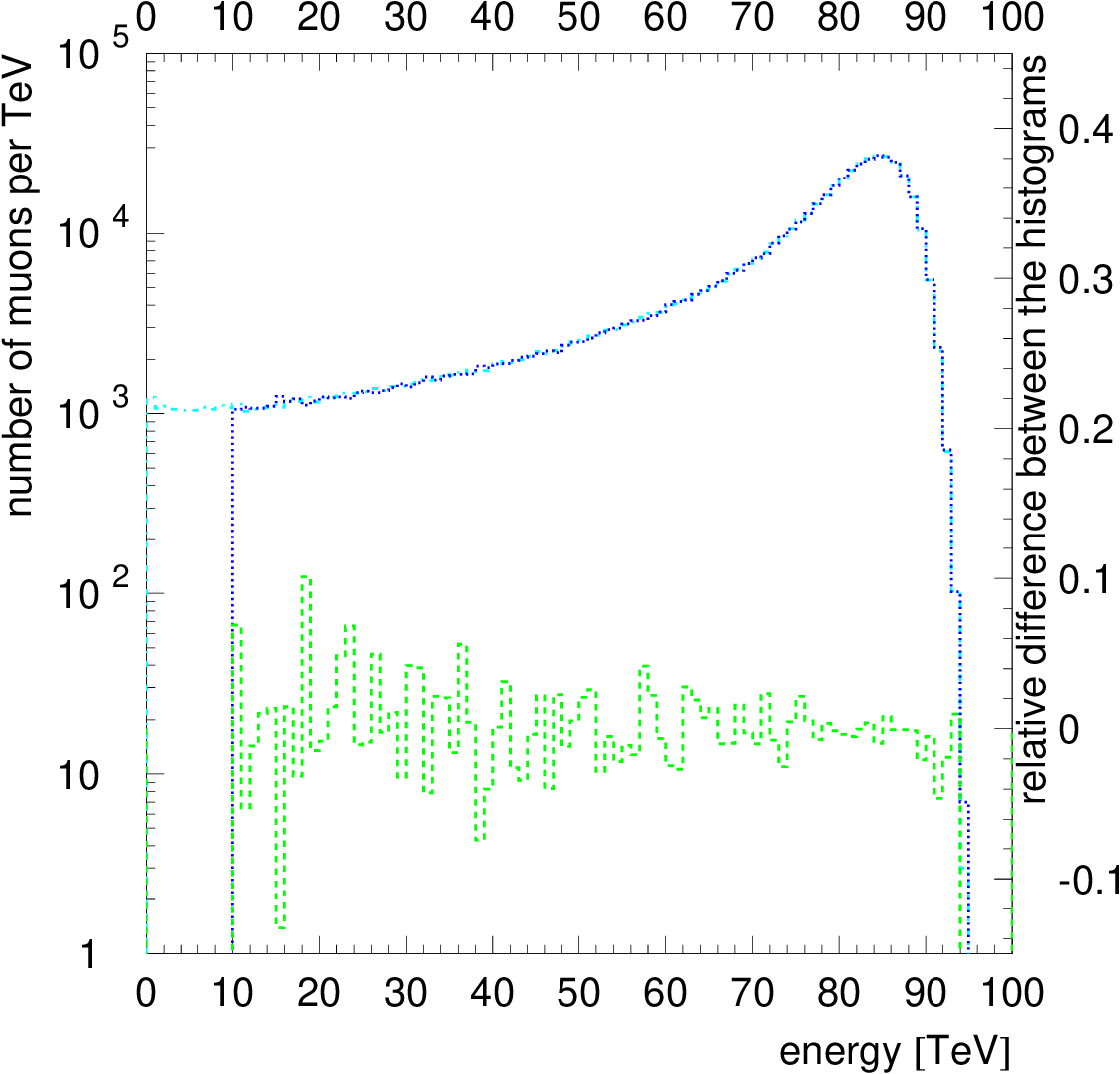,width=.45\textwidth}} & \ & \mbox{\epsfig{file=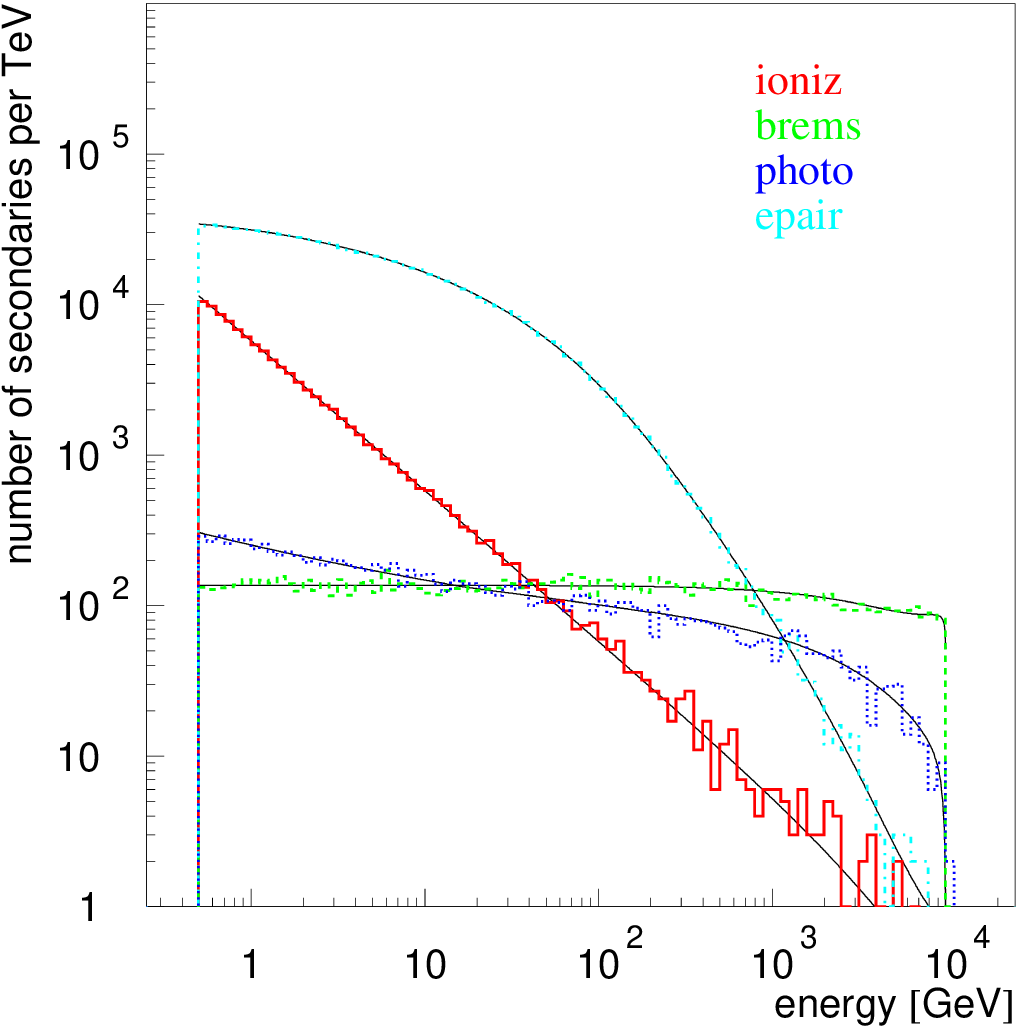,width=.45\textwidth}} \\
\parbox{.45\textwidth}{\caption[Comparison of $e_{low}=m_{\mu}$ with $e_{low}$=10 TeV. Also shown is the relative difference of the curves.]{\label{mmc_fig_9}Comparison of $e_{low}=m_{\mu}$ (dotted-dashed) with $e_{low}$=10 TeV (dotted). Also shown is the relative difference of the curves.}} & \ & \parbox{.45\textwidth}{\caption[Ionization, bremsstrahlung, photonuclear, epair production spectra for $E_{\mu}$=10 TeV in the Fr\'{e}jus rock]{\label{mmc_fig_10}Ionization (upper solid curve), bremsstrahlung (dashed), photonuclear (dotted), epair production (dashed-dotted) spectra for $E_{\mu}$=10 TeV in the Fr\'{e}jus rock}} \\
\end{tabular}
\end{center}\end{figure}

MMC employs a low energy cutoff $e_{low}$ below which the muon is considered to be lost. By default it is equal to the mass of the muon, but can be changed to any higher value. This cutoff enters the calculation in several places, most notably in the initial evaluation of the energy integral. To determine the random number $\xi_0$ below which the particle is considered stopped, the energy integral is first evaluated from $E_i$ to $e_{low}$. It is also used in the parametrization of the energy and tracking integrals, since they are evaluated from this value to $E_i$ and $E_f$, and then the interpolated value for $E_f$ is subtracted from that for $E_i$. Figure \ref{mmc_fig_9} demonstrates the independence of MMC from the value of $e_{low}$. For the curve with $e_{low}=m_{\mu}$ integrals are evaluated in the range 105.7 MeV $-$ 100 TeV, i.e., over six orders of magnitude, and they are as precise as those calculated for the curve with $e_{low}$=10 TeV, with integrals being evaluated over only one order of magnitude.

Figure \ref{mmc_fig_10} demonstrates the spectra of secondaries (delta electrons, bremsstrahlung photons, excited nuclei, and electron pairs) produced by the muon, whose energy is kept constant at 10 TeV. The thin lines superimposed on the histograms are the probability functions (cross sections) used in the calculation. They have been corrected to fit the logarithmically binned histograms (multiplied by the size of the bin which is proportional to the abscissa, i.e., the energy). While the agreement is trivial from the Monte Carlo point of view, it demonstrates that the computational algorithm is correct.

\begin{figure}[!h]\begin{center}
\begin{tabular}{ccc}
\mbox{\epsfig{file=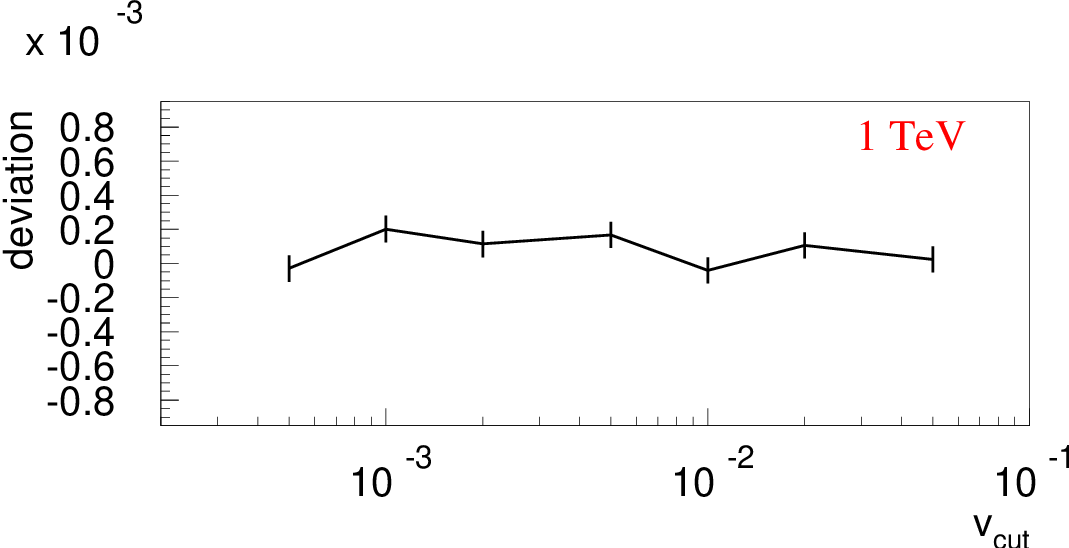,width=.45\textwidth}} & \ & \mbox{\epsfig{file=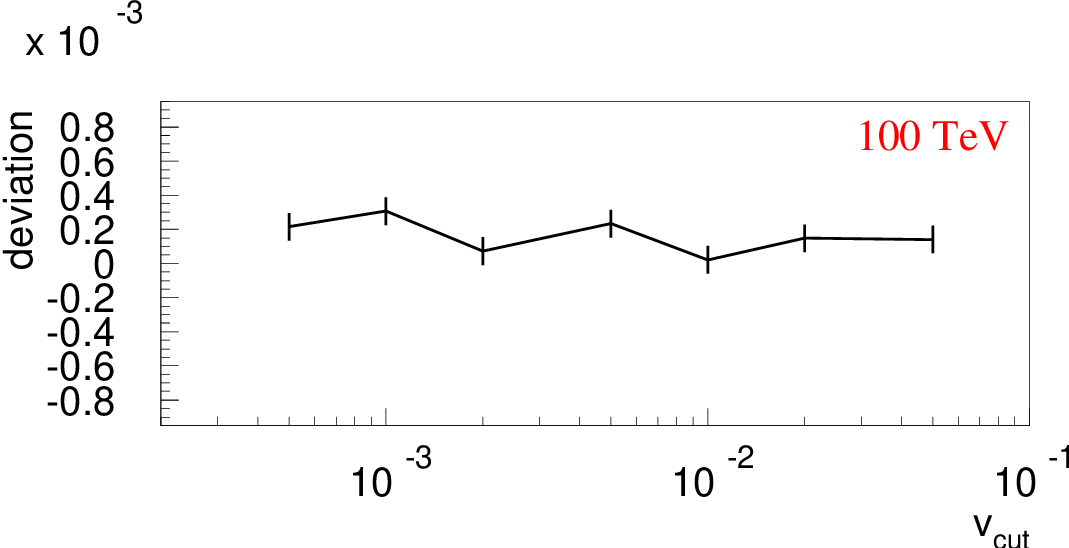,width=.45\textwidth}} \\
\end{tabular}
{\caption[Algorithm errors (average energy loss)]{\label{mmc_fig_11}Algorithm errors (average energy loss)}}
\end{center}\end{figure}

Figure \ref{mmc_fig_11} shows the relative deviation of the average final energy of the $4\cdot 10^6$ 1 TeV and 100 TeV muons propagated through 100 m of Fr\'{e}jus Rock\footnote{A medium with properties similar to that of standard rock (see second table in Appendix \ref{app_mmc}) used for data analysis in the Fr\'{e}jus experiment \cite{frank}.} with the abscissa setting for $v_{cut}$, from the final energy obtained with $v_{cut}=1$. Just like in \cite{mum} the distance was chosen small enough so that only a negligible number of muons stop, while large enough so that the muon suffers a large number of stochastic losses ($>10$ for $v_{cut}\le 10^{-3}$). All points should agree with the result for $v_{cut}=1$, since it should be equal to the integral of all energy losses, and averaging over the energy losses for $v_{cut}<1$ is evaluating such an integral with the Monte Carlo method. There is a visible systematic shift $\lesssim (1-2)\cdot 10^{-4}$ (similar for other muon energies), which can be considered as another measure of the algorithm accuracy \cite{mum}.

\begin{figure}[!h]\begin{center}
\begin{tabular}{ccc}
\mbox{\epsfig{file=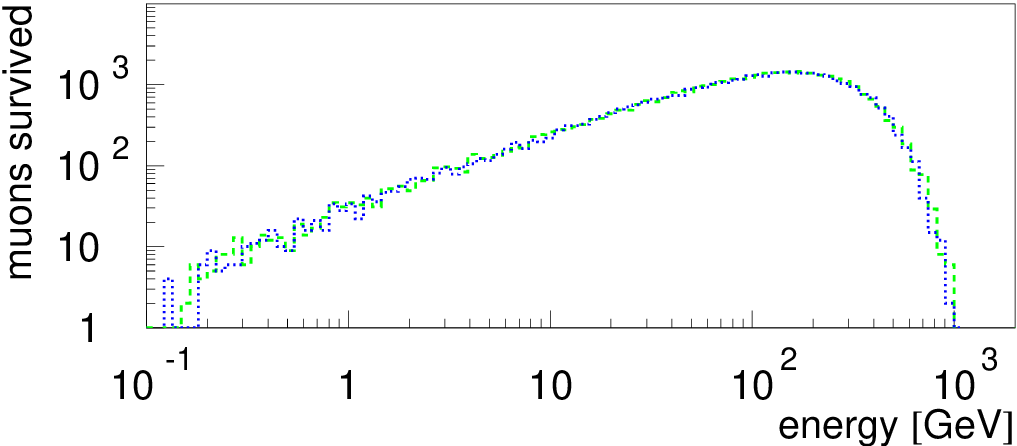,width=.45\textwidth}} & \ & \mbox{\epsfig{file=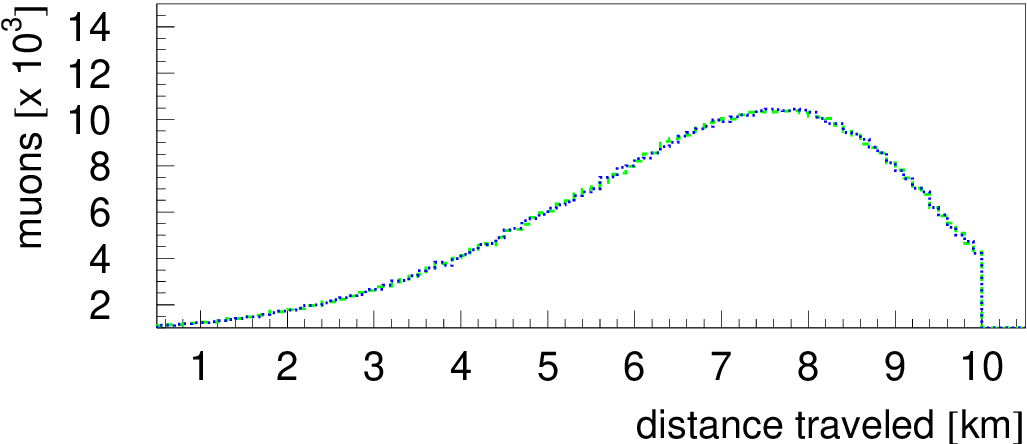,width=.45\textwidth}} \\
\end{tabular}
\parbox{.9\textwidth}{\caption[$10^6$ muons with energy 9 TeV propagated through 10 km of water: regular vs. {\it cont} ]{\label{mmc_fig_12}$10^6$ muons with energy 9 TeV propagated through 10 km of water: regular (dashed) vs. {\it cont} (dotted) }} \\
\end{center}\end{figure}

In the case when almost all muons stop before passing the requested distance (see Figure \ref{mmc_fig_12}), even small algorithm errors may substantially affect survival probabilities.
Table \ref{mmc_tab_1} summarizes the survival probabilities for a monochromatic muon beam of $10^6$ muons with three initial energies (1 TeV, 9 TeV, and $10^6$ TeV) going through three distances (3 km, 10 km, and 40 km) in water. One should note that these numbers are very sensitive to the cross sections used in the calculation; e.g., for $10^9$ GeV muons propagating through 40 km the rate increases 12\% when the BB1981 photonuclear cross section is replaced with the ZEUS parametrization (see Figure \ref{mmc_fig_18}). However, the same set of formulae was used throughout this calculation. The errors of the values in the table are statistical and are $\lesssim\pm 0.001$.
\begin{table}[!h]\begin{center}
{\caption[Survival probabilities]{\label{mmc_tab_1}Survival probabilities}}
\begin{tabular}{|ccccc|}
\hline

$v_{cut}$ & {\it cont} & 1 TeV 3 km & 9 TeV 10 km & $10^6$ TeV 40 km \\

0.2 & no & 0 & 0 & 0.081 \\
0.2 & yes & 0.009 & 0.052 & 0.113 \\

0.05 & no & 0 & 0.028 & 0.076 \\
0.05 & yes & 0.041 & 0.034 & 0.073 \\

0.01 & no & 0.027 & 0.030 & 0.075 \\
0.01 & yes & 0.031 & 0.030 & 0.072 \\

$10^{-3}$ & no & 0.031 & 0.031 & 0.074 \\
$10^{-3}$ & yes & 0.031 & 0.030 & 0.070 \\

\hline
\end{tabular}
\end{center}\end{table}
The survival probabilities converge on the final value for $v_{cut}\lesssim0.01$ in the first two columns. Using the {\it cont} option helped the convergence in the first column. However, the {\it cont} values departed from regular values more in the third column. The relative deviation (5.4\%) can be used as an estimate of the {\it continuous randomization} algorithm precision (not calculational errors) in this case. One should note, however, that with the number of interactions $\gtrsim 10^3$ the {\it continuous randomization} approximation formula was applied $\gtrsim 10^3$ times. It explains why the value of {\it cont} version for $v_{cut}=0.01$ is closer to the converged value of the regular version than for $v_{cut}=10^{-3}$.

\section{Electron, tau, and monopole propagation}

\begin{figure}[!h]\begin{center}
\begin{tabular}{ccc}
\mbox{\epsfig{file=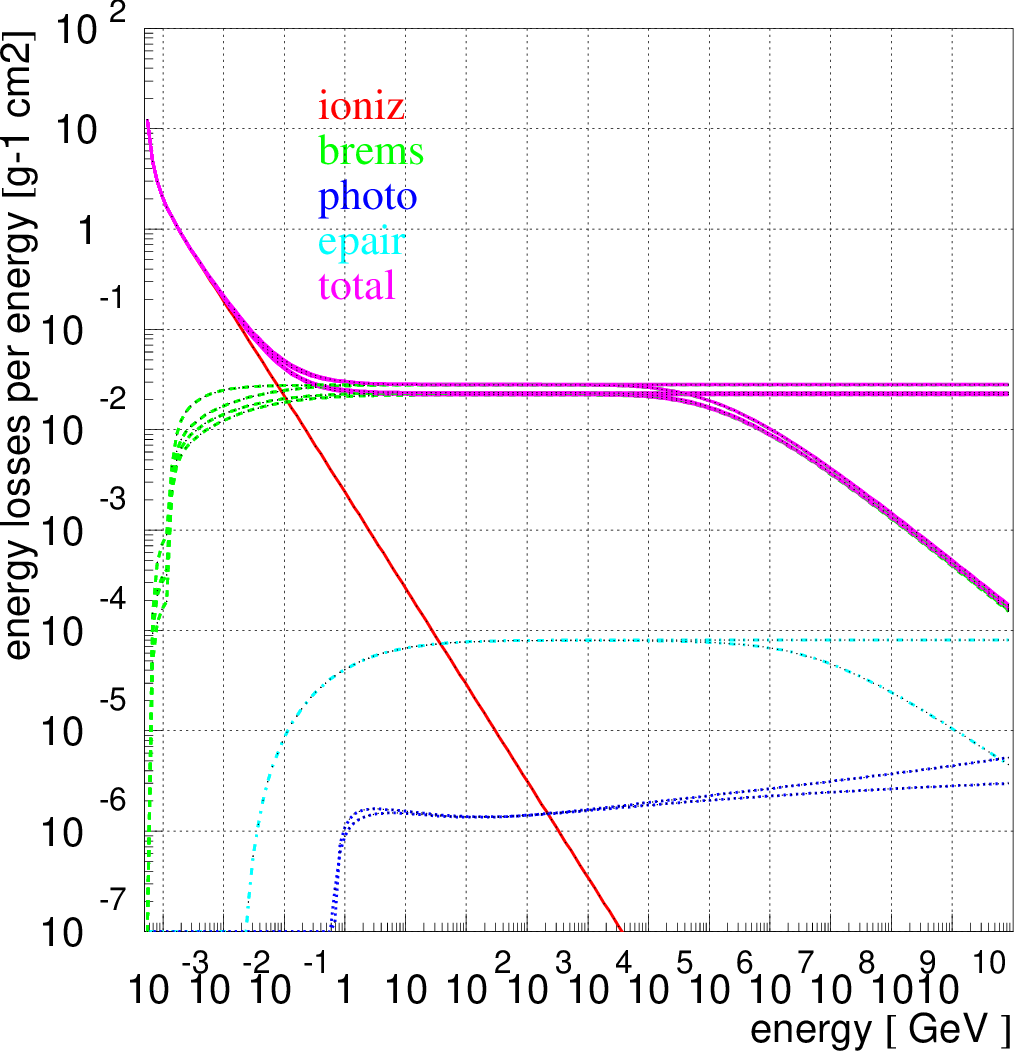,width=.45\textwidth}} & \ & \mbox{\epsfig{file=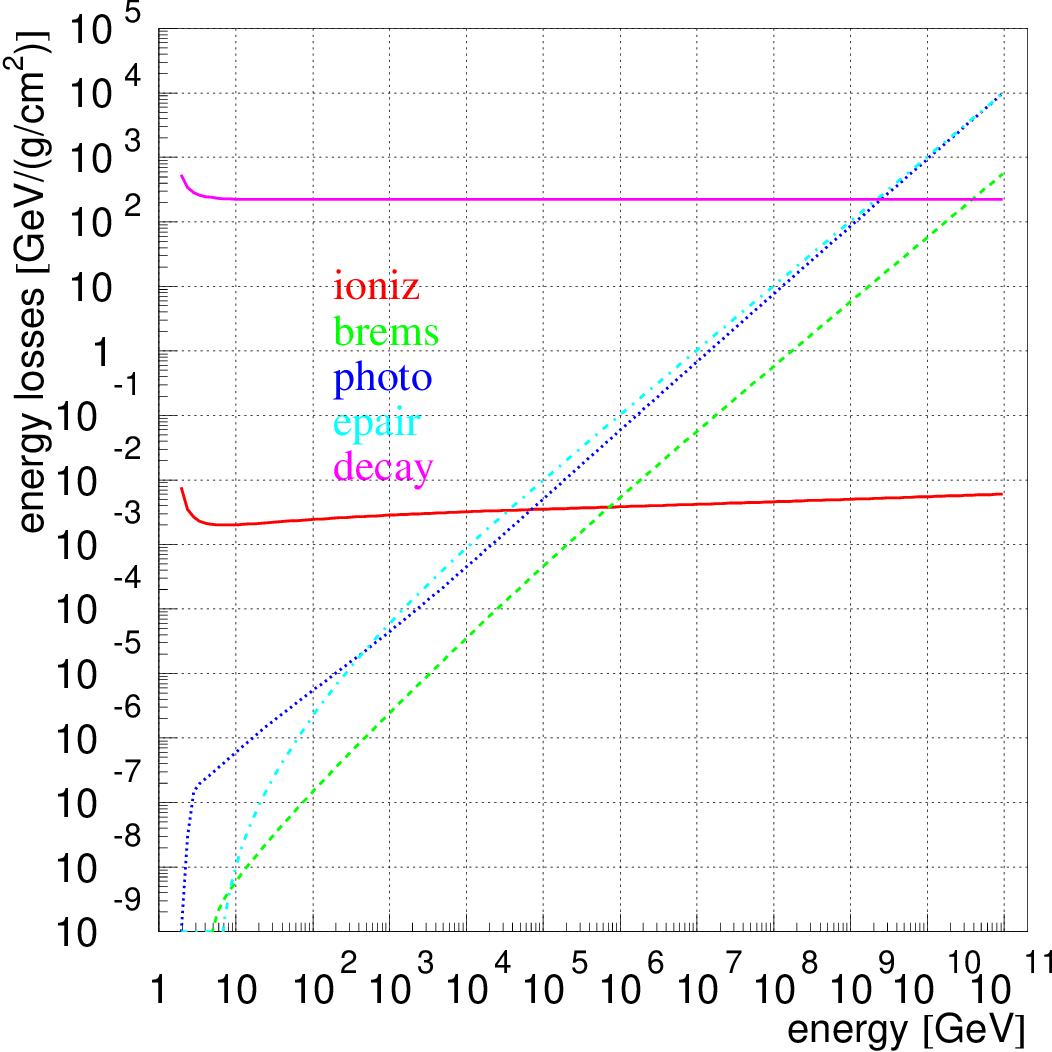,width=.45\textwidth}} \\
\parbox{.45\textwidth}{\caption[Electron energy losses in Ice]{\label{mmc_fig_f}Electron energy losses in Ice}} & \ & \parbox{.45\textwidth}{\caption[Tau energy losses in Fr\'{e}jus Rock]{\label{mmc_fig_c}Tau energy losses in Fr\'{e}jus Rock}} \\
\end{tabular}
\end{center}\end{figure}

\begin{figure}[!h]\begin{center}
\begin{tabular}{ccc}
\mbox{\epsfig{file=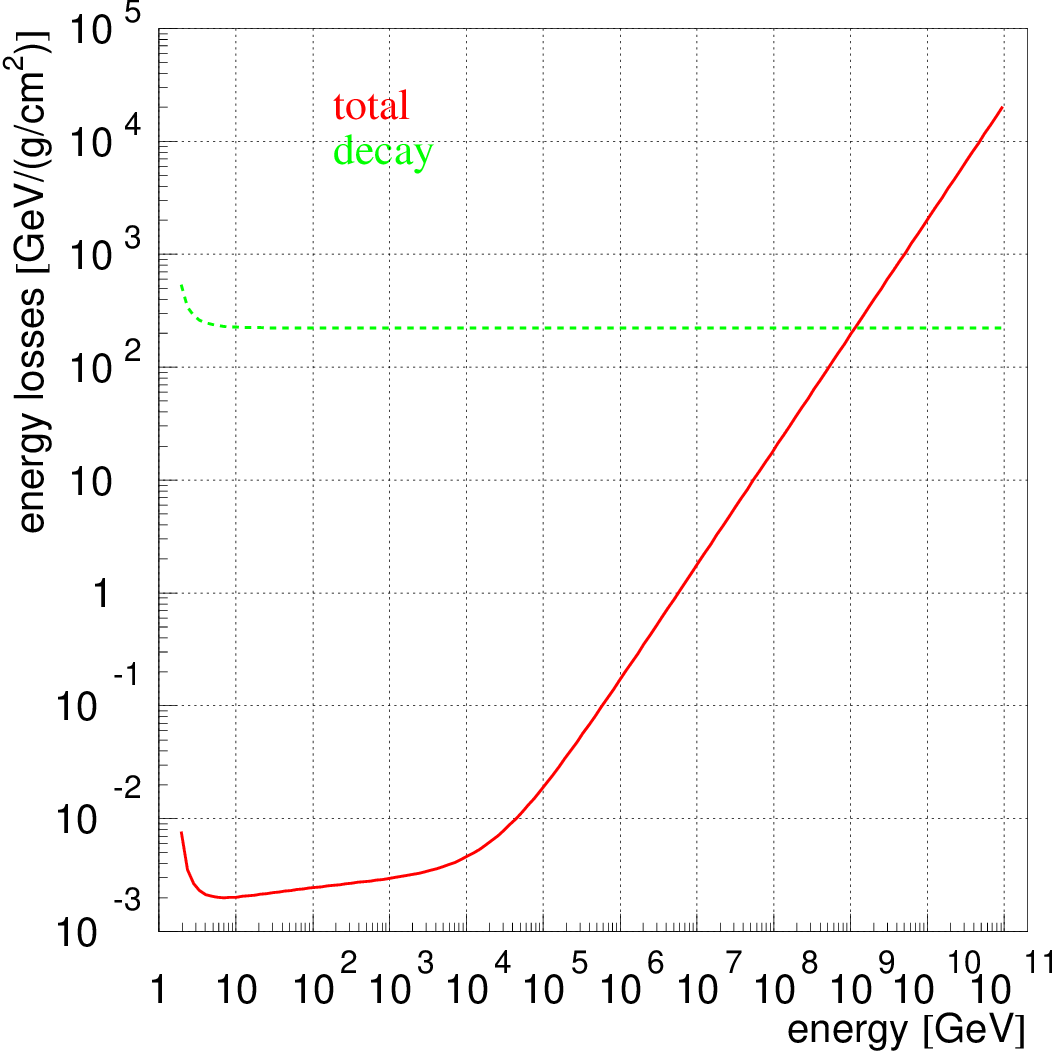,width=.45\textwidth}} & \ & \mbox{\epsfig{file=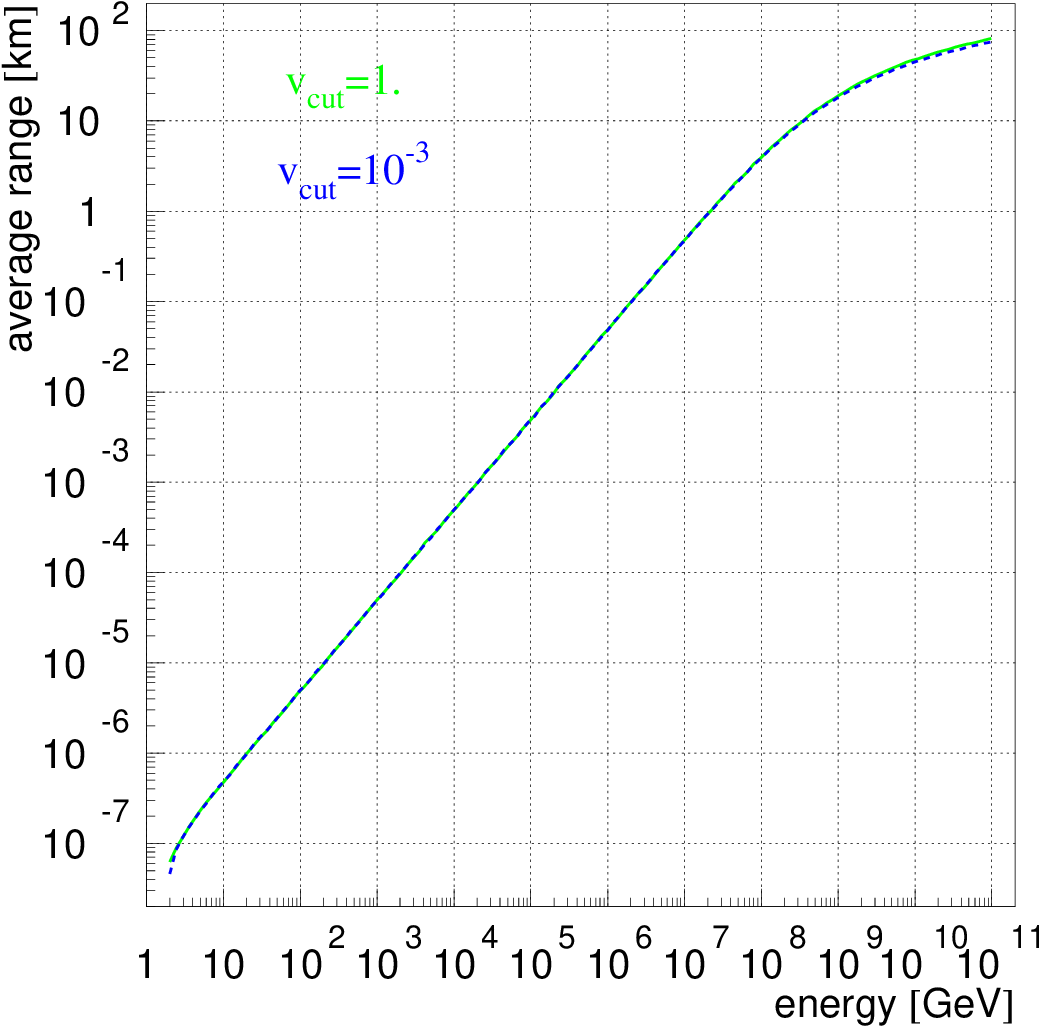,width=.45\textwidth}} \\
\parbox{.45\textwidth}{\caption[Sum of tau energy losses in Fr\'{e}jus Rock]{\label{mmc_fig_d}Sum of tau energy losses in Fr\'{e}jus Rock}} & \ & \parbox{.45\textwidth}{\caption[Average range of taus propagated through Fr\'{e}jus Rock]{\label{mmc_fig_e}Average range of taus propagated through Fr\'{e}jus Rock}} \\
\end{tabular}
\end{center}\end{figure}

Electrons and taus can also be propagated with MMC. Bremsstrahlung is the dominant cross section in case of electron propagation, and the complete screening case cross section should be selected (Section \ref{brems_csc}). Electron energy losses in Ice are shown in Figure \ref{mmc_fig_f} (also showing the LPM suppression of cross sections).

For tau propagation Bezrukov-Bugaev parameterization with the hard component (Section \ref{photo_bb}) or the ALLM parametrization (Section \ref{allm}) should be selected for photonuclear cross section. Tau propagation is quite different from muon propagation because the tau lifetime is 7 orders of magnitude shorter than the muon lifetime. While muon decay can be neglected in most cases of muon propagation, it is the main process to be accounted for in the tau propagation. Figures \ref{mmc_fig_c} and \ref{mmc_fig_d} compare tau energy losses with losses caused by tau decay (given by $E_{\tau}/(\rho v_{\tau}\tau)=m_{\tau}/(\rho v_{\tau}\tau_0)$; this is the energy per mwe deposited by decaying taus in a beam propagating though a medium with density $\rho$). Figure \ref{mmc_fig_e} compares the average range of taus propagated through Fr\'{e}jus Rock with $v_{cut}=1$ (completely continuously) and $v_{cut}=10^{-3}$ (detailed stochastic treatment). Both treatments produce almost identical results. Therefore, tau propagation can be treated continuously for all energies unless one needs to obtain spectra of the secondaries created along the tau track.

Monopoles can also be propagated with MMC. All cross sections except brems\-strahlung (which scales as $z^4$) are scaled up with a factor $z^2$, where $z=1/(2\alpha)$ is the monopole charge, according to \cite{monopole,monopole2}.

\section{Comparison with other propagation codes}

\begin{figure}[!h]\begin{center}
\begin{tabular}{ccc}
\mbox{\epsfig{file=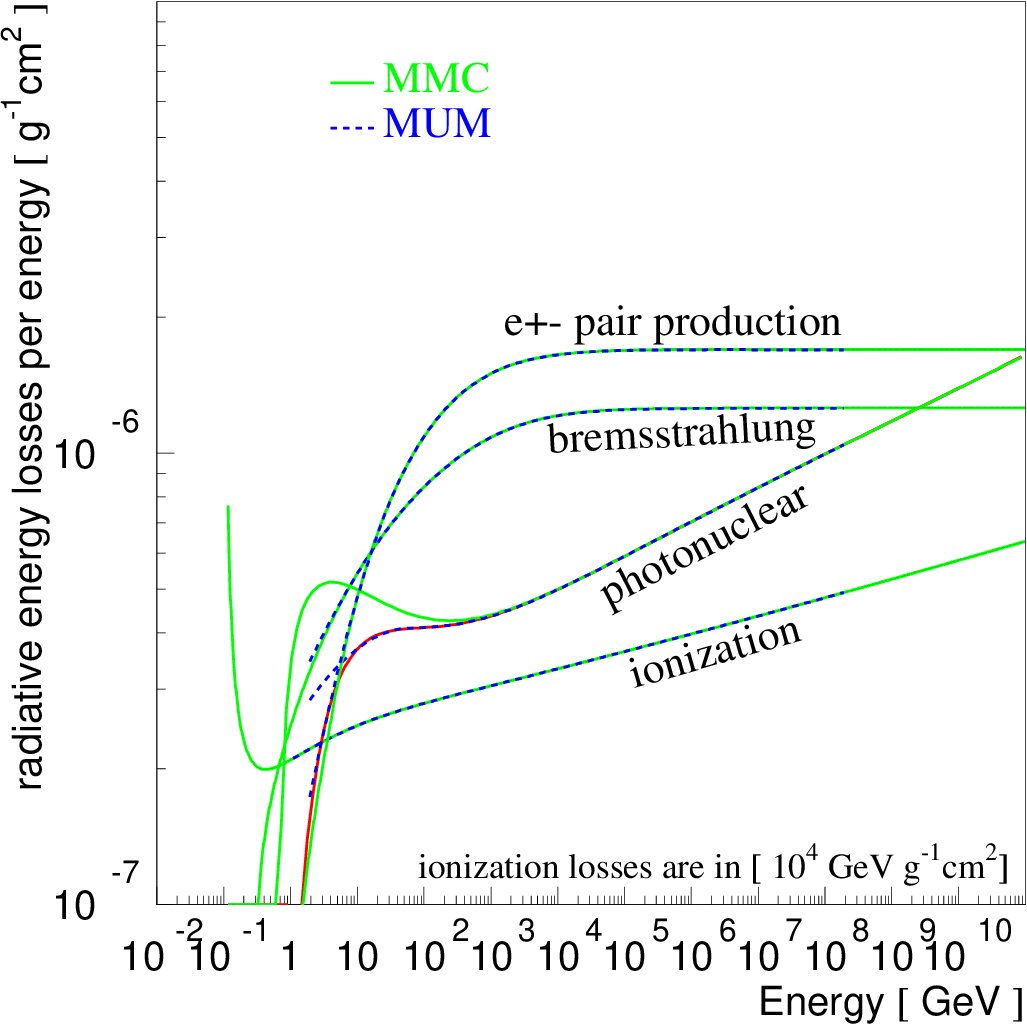,width=.45\textwidth}} & \ & \mbox{\epsfig{file=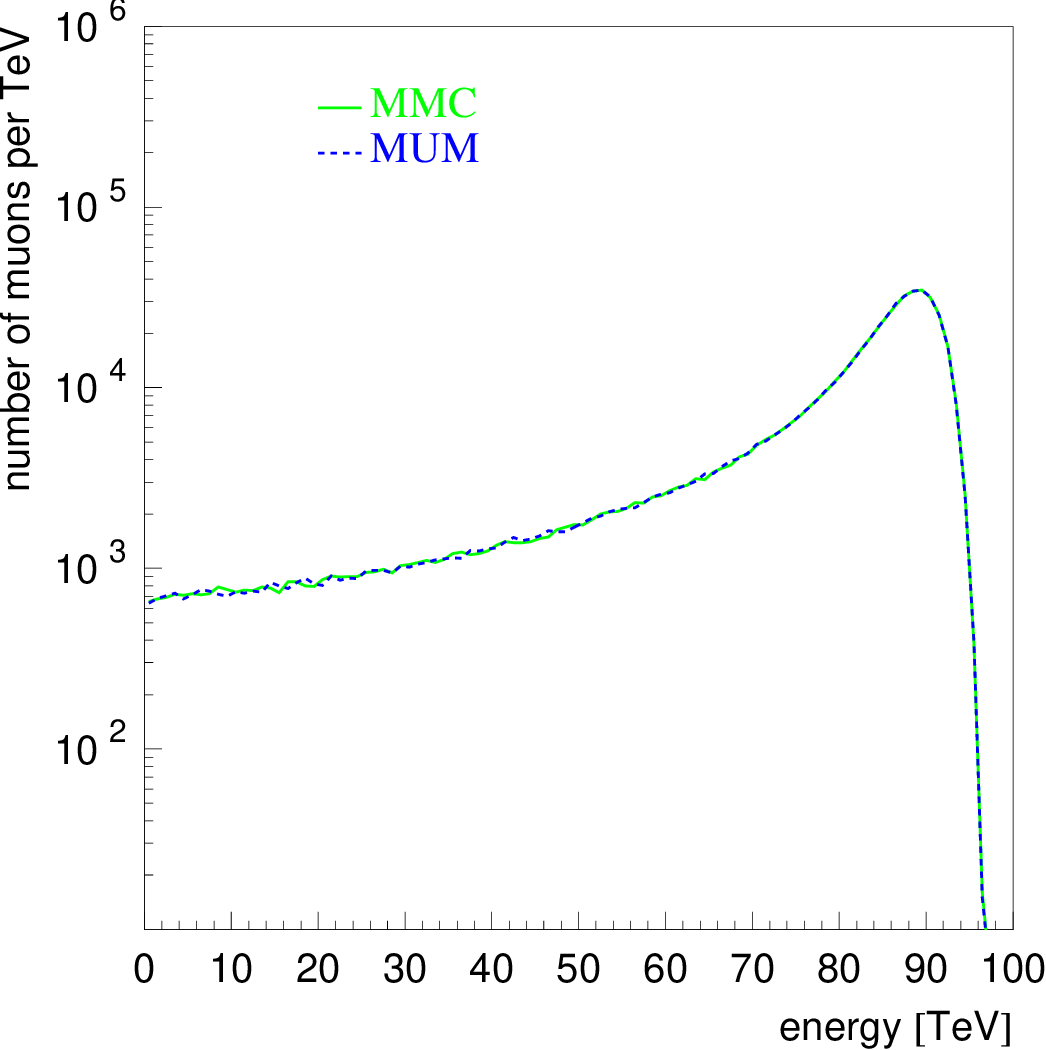,width=.45\textwidth}} \\
\parbox{.45\textwidth}{\caption[Comparison of energy losses in water calculated with MMC and MUM]{\label{mmc_fig_h}Comparison of energy losses in water. Red MMC curve shows photonuclear losses with $\nu$ integration limits as prescribed in \cite{mum}}} & \ & \parbox{.45\textwidth}{\caption[Comparison of muon propagation through 800 m of Ice with MMC and MUM]{\label{mmc_fig_g}Comparison of muon propagation through 800 m of Ice with MMC and MUM}} \\
\end{tabular}
\end{center}\end{figure}

Several propagation codes have been compared with MMC. Where possible MMC settings were changed to match those of the other codes. Figure \ref{mmc_fig_g} compares energy losses calculated with MMC and MUM \cite{mum}, and Figure \ref{mmc_fig_h} compares the results of muon propagation through 800 m of ice with MMC and MUM ($v_{cut}=10^{-3}$, ZEUS parametrization of the photonuclear cross section, Andreev Berzrukov Bugaev parameterization of bremsstrahlung).

Survival probabilities of Table \ref{mmc_tab_1} were compared with results from \cite{mum} in Table \ref{mmc_tab_a}. Survival probabilities are strongly correlated with the distribution of the highest-energy muons in an originally monoenergetic beam. This, in turn, is very sensitive to the algorithm errors and the cross-section implementation used for the calculation.

\begin{table}[!h]\begin{center}
{\caption[Survival probabilities of MMC compared to other codes]{\label{mmc_tab_a}Survival probabilities of MMC compared to other codes}}
\begin{tabular}{|ccccc|}
\hline

$v_{cut}$ & propagation code & 1 TeV 3 km & 9 TeV 10 km & $10^6$ TeV 40 km \\
\hline
$10^{-3}$ & MMC (BB81) & 0.031 & 0.031 & 0.074 \\

$10^{-3}$ & MMC (ZEUS) & 0.031 & 0.030 & 0.083 \\

$10^{-3}$ & MUM \cite{mum} & 0.029 & 0.030 & 0.078 \\

$10^{-3}$ & MUSIC \cite{music} & 0.033 & 0.031 & 0.084 \\

$10^{-3}$ & PROPMU \cite{propmu} & 0.19 & 0.048 & 0.044 \\

\hline
\end{tabular}
\end{center}\end{table}

A detailed comparison between spectra of secondaries produced with MMC, MUM, LOH \cite{cern85}, and LIP \cite{lip,propmu} is given in the Appendix \ref{app_mmc2}. A definite improvement of MMC over the other codes can be seen in the precision of description of spectra of secondaries and the range of energies over which the program works.

\section{Energy losses in ice and rock, some general results}

The code was incorporated into the Monte Carlo chains of three detectors: Fr\'{e}jus \cite{frank,dcors_10}, AMANDA \cite{paolo,mythesis}, and IceCube \cite{icecube}. In this section some general results are presented.

\subsection{Average muon energy loss}
\label{mmc_fit_results}

\begin{figure}[!h]\begin{center}
\begin{tabular}{ccc}
\mbox{\epsfig{file=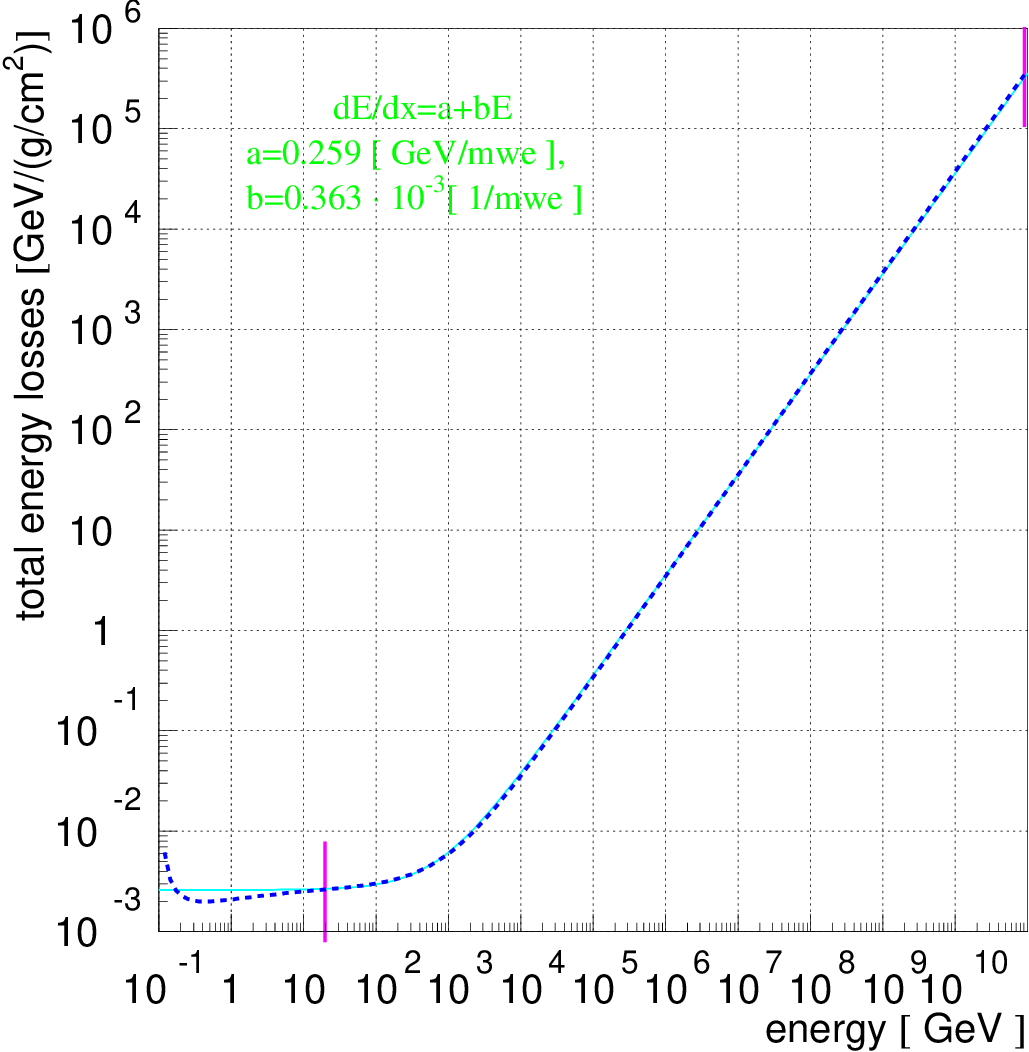,width=.45\textwidth}} & \ & \mbox{\epsfig{file=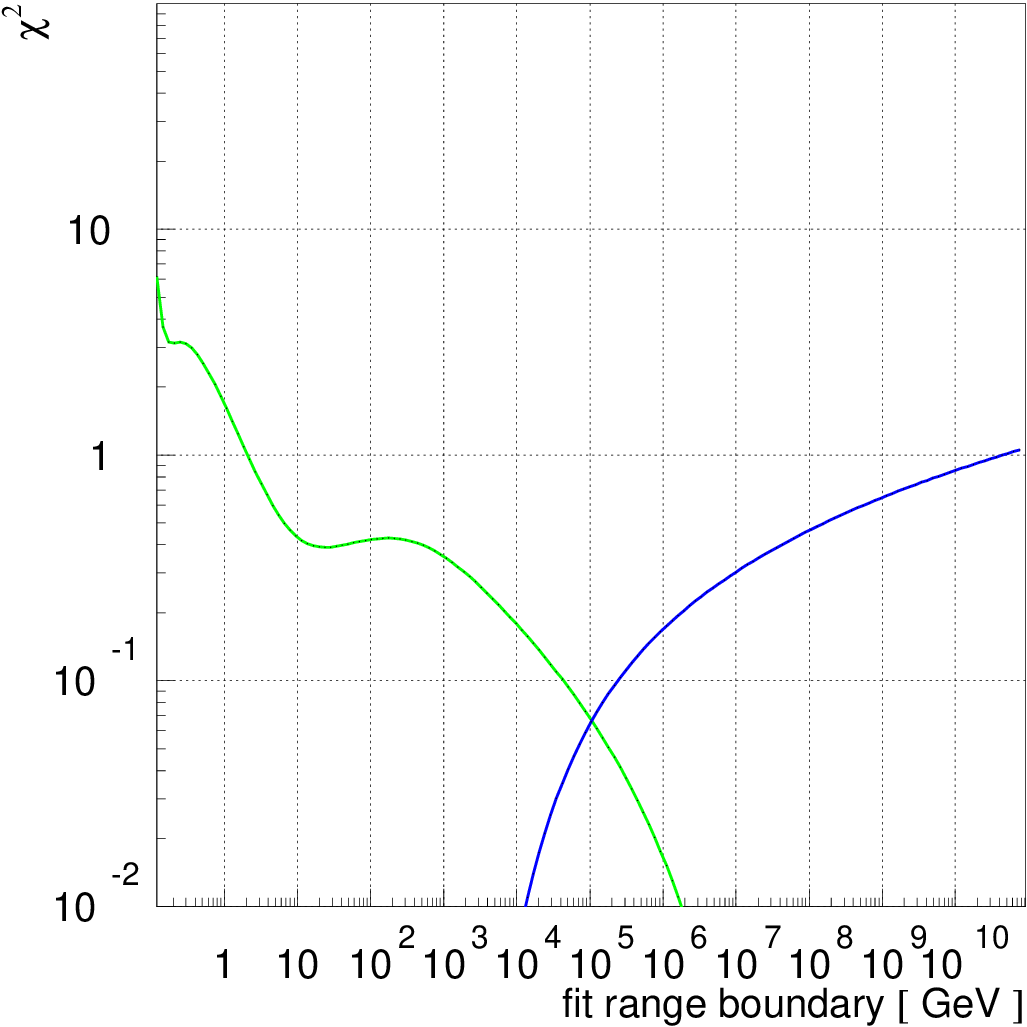,width=.45\textwidth}} \\
\parbox{.45\textwidth}{\caption[Fit to the energy losses in ice ]{\label{mmc_fig_13}Fit to the energy losses in ice }} & \ & \parbox{.45\textwidth}{\caption[$\chi^2$ plot for energy losses in ice ]{\label{mmc_fig_14}$\chi^2$ plot for energy losses in ice }} \\
\end{tabular}
\end{center}\end{figure}

The plot of energy losses was fitted to the function $dE/dx=a+bE$ (Figure \ref{mmc_fig_13}).
The first two formulae for the photonuclear cross section (Section \ref{photo_bb}) can be fitted the best, all others lead to energy losses deviating more at higher energies from this simple linear formula; therefore the numbers given were evaluated using the first photonuclear cross section formula. In order to choose low and high energy limits correctly (to cover the maximum possible range of energies that could be comfortably fitted with a line), a $\chi^2$ plot was generated and analyzed (Figure \ref{mmc_fig_14}). The green curve corresponds to the $\chi^2$ of the fit with a fixed upper bound and a varying lower bound on the fitted energy range. Correspondingly, the blue curve describes the $\chi^2$ of the fit with a fixed lower bound and a varying upper bound. The $\chi^2$ at low energies goes down sharply, then plateaus at around 10 GeV. This corresponds to the point where the linear approximation starts to work. For the high energy boundaries, $\chi^2$ rises monotonically. This means that a linear approximation, though valid, has to describe a growing energy range. An interval of energies from 20 GeV to $10^{11}$ GeV is chosen for the fit. Table \ref{mmc_tab_2} summarizes the found fits to $a$ and $b$;
\begin{table}[!h]\begin{center}
{\caption[Fits to $a$ and $b$ for continuous losses (average energy losses)]{\label{mmc_tab_2}Fits to $a$ and $b$ for continuous losses (average energy losses)}}
\begin{tabular}{|l|c|c|c|c|c|c|c|c|}
\hline
\rule[-1.7mm]{0mm}{6mm}medium & $a$, $\frac{\mbox{GeV}}{\mbox{mwe}}$ & $b$, $\frac{10^{-3}}{\mbox{mwe}}$ & av. dev. & max. dev. & $a$, $\frac{\mbox{GeV}}{\mbox{mwe}}$ & $b$, $\frac{10^{-3}}{\mbox{mwe}}$  & $a$, $\frac{\mbox{GeV}}{\mbox{mwe}}$ & $b$, $\frac{10^{-3}}{\mbox{mwe}}$ \\
\hline
 & \multicolumn{4}{|c|}{$20-10^{11}$ GeV} & \multicolumn{2}{|c|}{$20-10^7$ GeV} & \multicolumn{2}{|c|}{ALLM97} \\
\hline
air & 0.281 & 0.347 & 3.6\% & 6.5\% & 0.284 & 0.335 & 0.282 & 0.344 \\
\hline
ice & 0.259 & 0.363 & 3.7\% & 6.6\% & 0.262 & 0.350 & 0.260 & 0.360 \\
\hline
fr. rock & 0.231 & 0.436 & 3.0\% & 5.1\% & 0.233 & 0.423 & 0.231 & 0.431 \\
\hline
st. rock & 0.223 & 0.463 & 2.9\% & 5.1\% & 0.225 & 0.451 & 0.224 & 0.459 \\
\hline
\end{tabular}
\end{center}\end{table}
the errors in the evaluation of $a$ and $b$ are in the last digit of the given number. However, if the lower energy boundary of the fitted region is raised and/or the upper energy boundary is lowered, each by an order of magnitude, $a$ and $b$ change by about 1\%.

\begin{figure}[!h]\begin{center}
\begin{tabular}{ccc}
\mbox{\epsfig{file=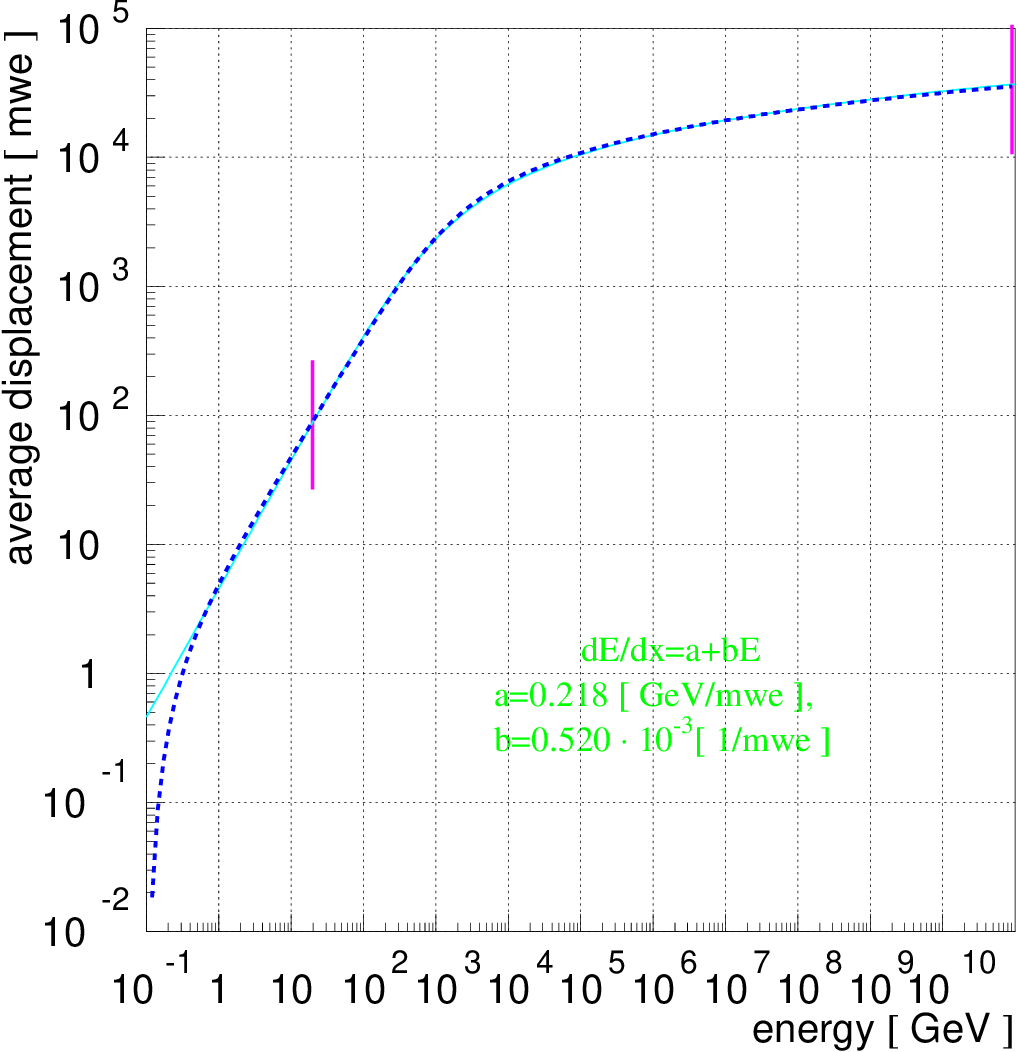,width=.45\textwidth}} & \ & \mbox{\epsfig{file=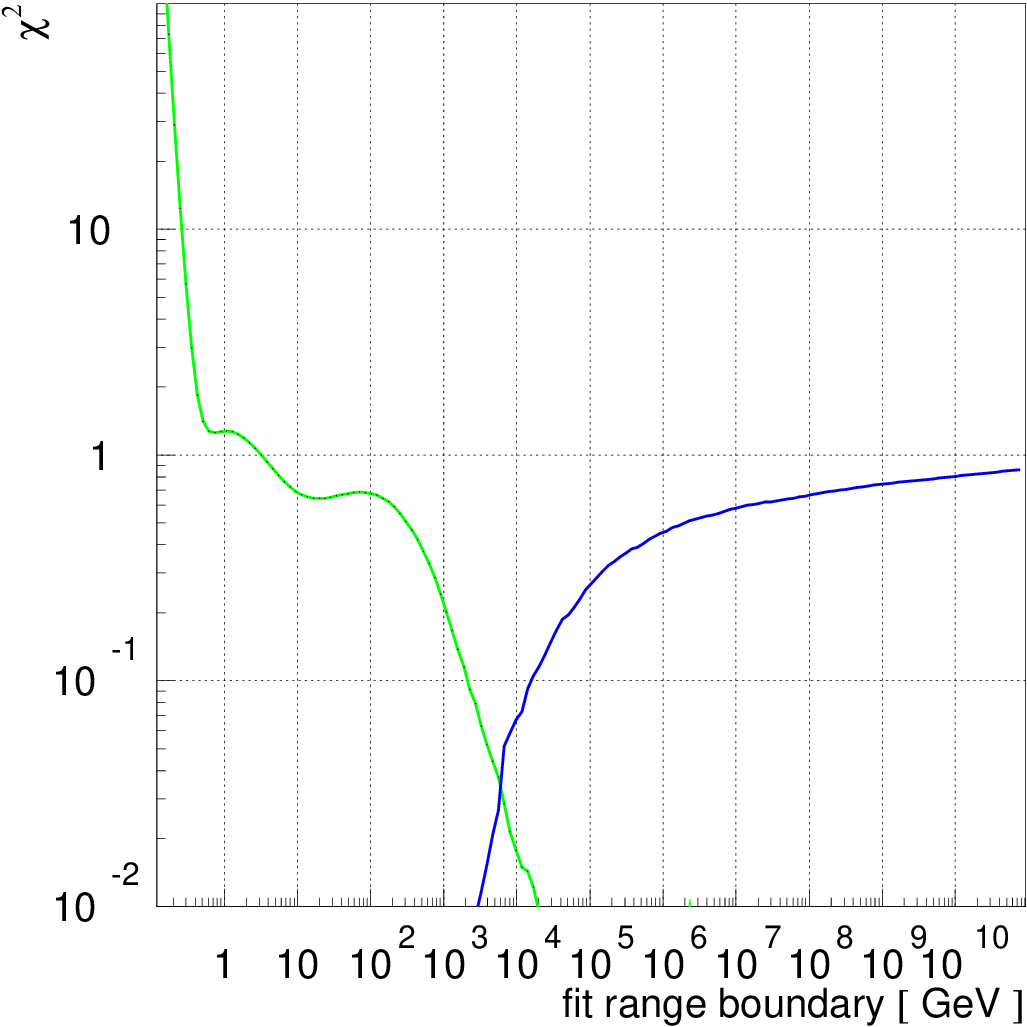,width=.45\textwidth}} \\
\parbox{.45\textwidth}{\caption[Fit to the average range in Fr\'{e}jus rock ]{\label{mmc_fig_15}Fit to the average range in Fr\'{e}jus rock }} & \ & \parbox{.45\textwidth}{\caption[$\chi^2$ plot for average range in Fr\'{e}jus rock ]{\label{mmc_fig_16}$\chi^2$ plot for average range in Fr\'{e}jus rock }} \\
\end{tabular}
\end{center}\end{figure}

To investigate the effect of stochastic processes, muons with energies 105.7 MeV $-$ $10^{11}$ GeV were propagated to the point of their disappearance. The value of $v_{cut}=5\cdot 10^{-3}$ was used in this calculation; using the the {\it continuous randomization} option did not change the final numbers. The average final distance (range) for each energy was fitted to the solution of the energy loss equation $dE/dx=a+bE$: $$x_f=\log(1+E_i \cdot b/a)/b$$ (Figure \ref{mmc_fig_15}). The same analysis of the $\chi^2$ plot as above was done in this case (Figure \ref{mmc_fig_16}). A region of initial energies from 20 GeV to $10^{11}$ GeV was chosen for the fit. Table \ref{mmc_tab_3} summarizes the results of these fits.
\begin{table}[!h]\begin{center}
{\caption[Fits to $a$ and $b$ for stochastic losses (average range estimation)]{\label{mmc_tab_3}Fits to $a$ and $b$ for stochastic losses (average range estimation)}}
\begin{tabular}{|l|c|c|c|}
\hline
\rule[-1.7mm]{0mm}{6mm}medium & $a$, $\frac{\mbox{GeV}}{\mbox{mwe}}$ & $b$, $\frac{10^{-3}}{\mbox{mwe}}$ & av. dev. \\
\hline
ice & 0.268 & 0.470 & 3.0\% \\
\hline
fr\'{e}jus rock & 0.218 & 0.520 & 2.8\% \\
\hline
\end{tabular}
\end{center}\end{table}

\begin{figure}[!h]\begin{center}
\begin{tabular}{ccc}
\mbox{\epsfig{file=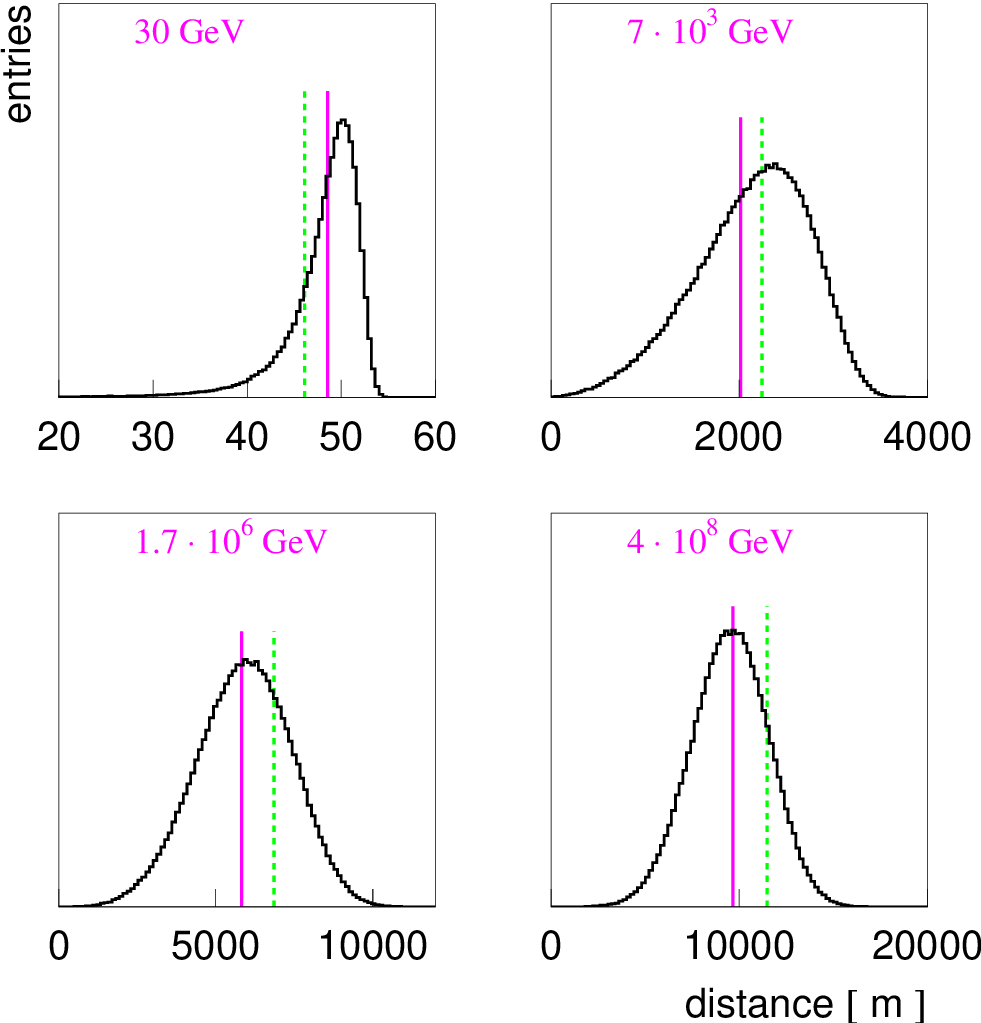,width=.45\textwidth}} & \ & \mbox{\epsfig{file=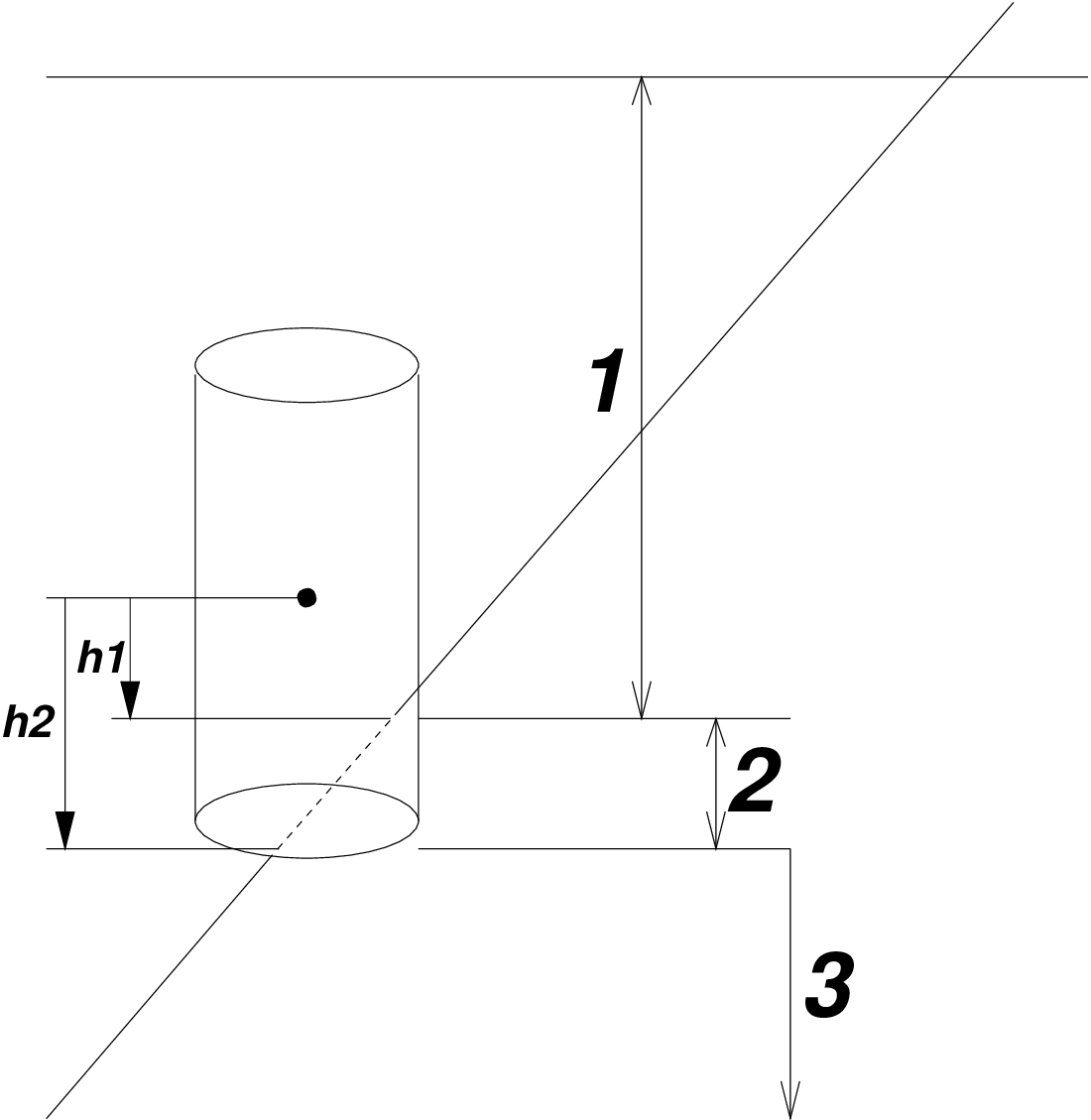,width=.45\textwidth}} \\
\parbox{.45\textwidth}{\caption[Range distributions in Fr\'{e}jus rock ]{\label{mmc_fig_17}Range distributions in Fr\'{e}jus rock: solid line designates the value of the range evaluated with the second table (continuous and stochastic losses) and the broken line shows the range evaluated with the first table (continuous losses only). }} & \ & \parbox{.45\textwidth}{\caption[ 3 regions of propagation defined for AMANDA-II simulation ]{\label{mmc_fig_4a}3 regions of propagation defined for AMANDA-II simulation }}
\end{tabular}
\end{center}\end{figure}

As the energy of the muon increases, it suffers more stochastic losses before it is lost\footnote{As considered by the algorithm, here: stopped.} and the range distribution becomes more Gaussian-like (Figure \ref{mmc_fig_17}). It is also shown in the figure (vertical lines) that the inclusion of stochastic processes makes the muons on average travel a shorter distance.

\subsection{Muon range}

\label{dcors_sec_opt}
In certain cases it is necessary to find the maximum range $x$ of (the majority of) muons of certain energy $E$, or find what is the minimum energy $E_{cut}$ muons must have in order to cross distance $x$.

\begin{figure}[!h]\begin{center}
\begin{tabular}{ccc}
\mbox{\epsfig{file=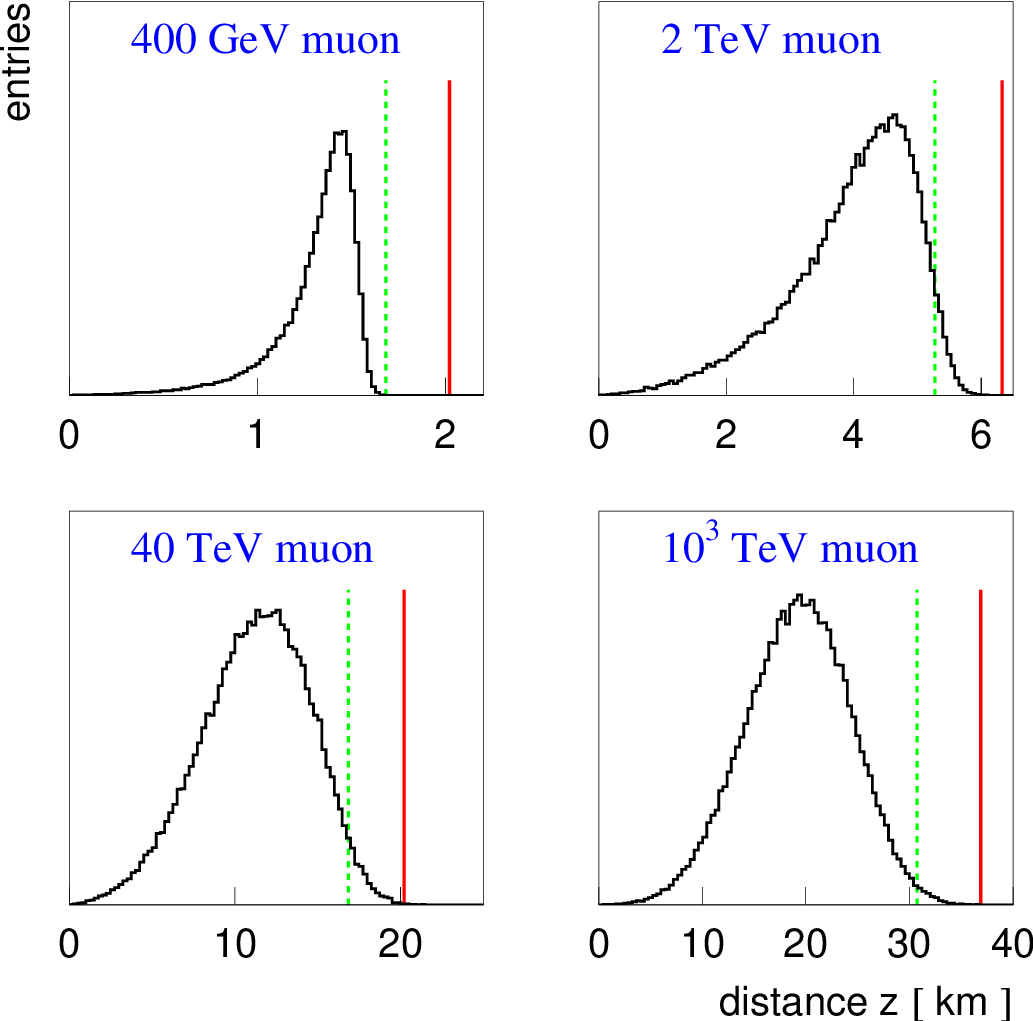,width=.45\textwidth}} & \ & \mbox{\epsfig{file=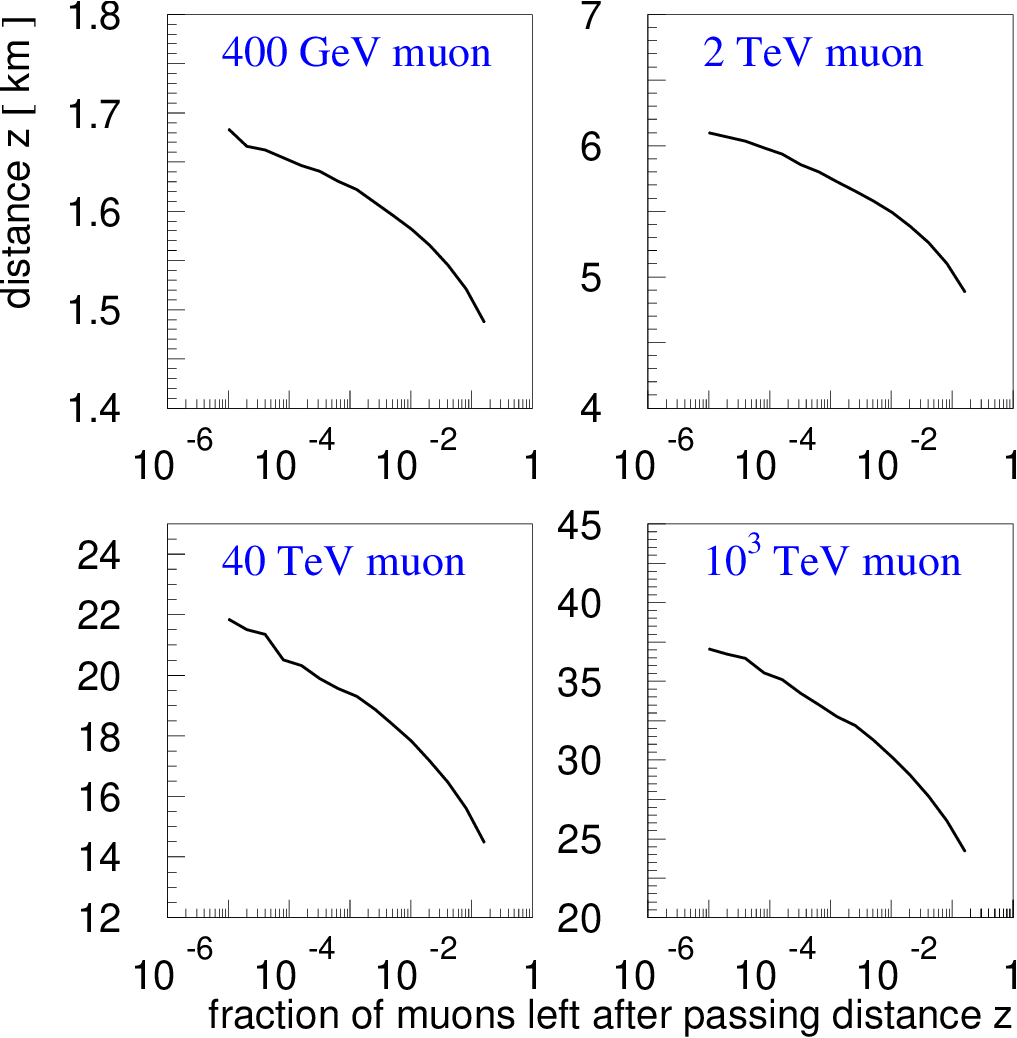,width=.45\textwidth}} \\
\parbox{.45\textwidth}{\caption[Muon range distributions in ice]{\label{dcors_fig_17}Muon range distributions in ice }} & \ & \parbox{.45\textwidth}{\caption[Distance in ice vs.\ fraction of survived muons ]{\label{dcors_fig_18}Distance in ice vs.\ fraction of survived muons }} \\
\end{tabular}
\end{center}\end{figure}

\begin{figure}[!h]\begin{center}
\begin{tabular}{ccc}
\mbox{\epsfig{file=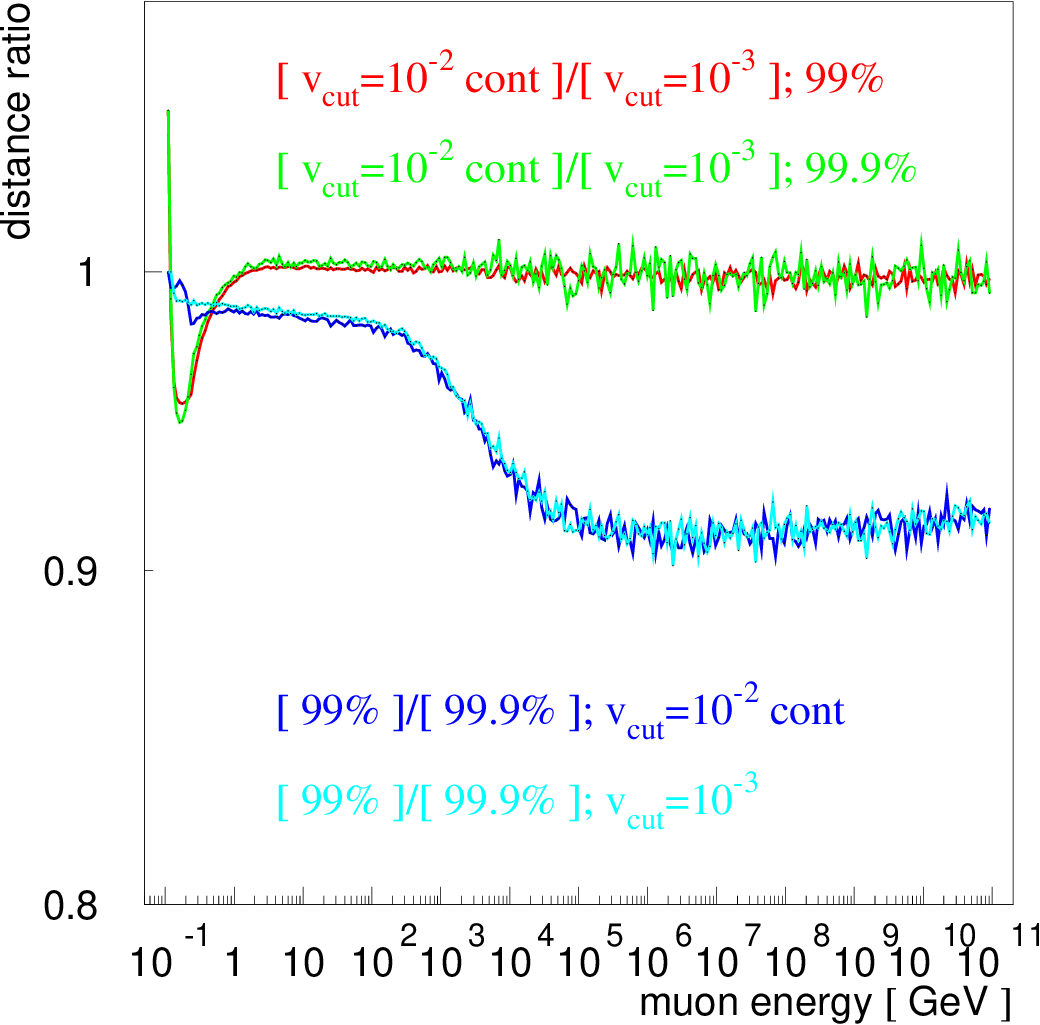,width=.45\textwidth}} & \ & \mbox{\epsfig{file=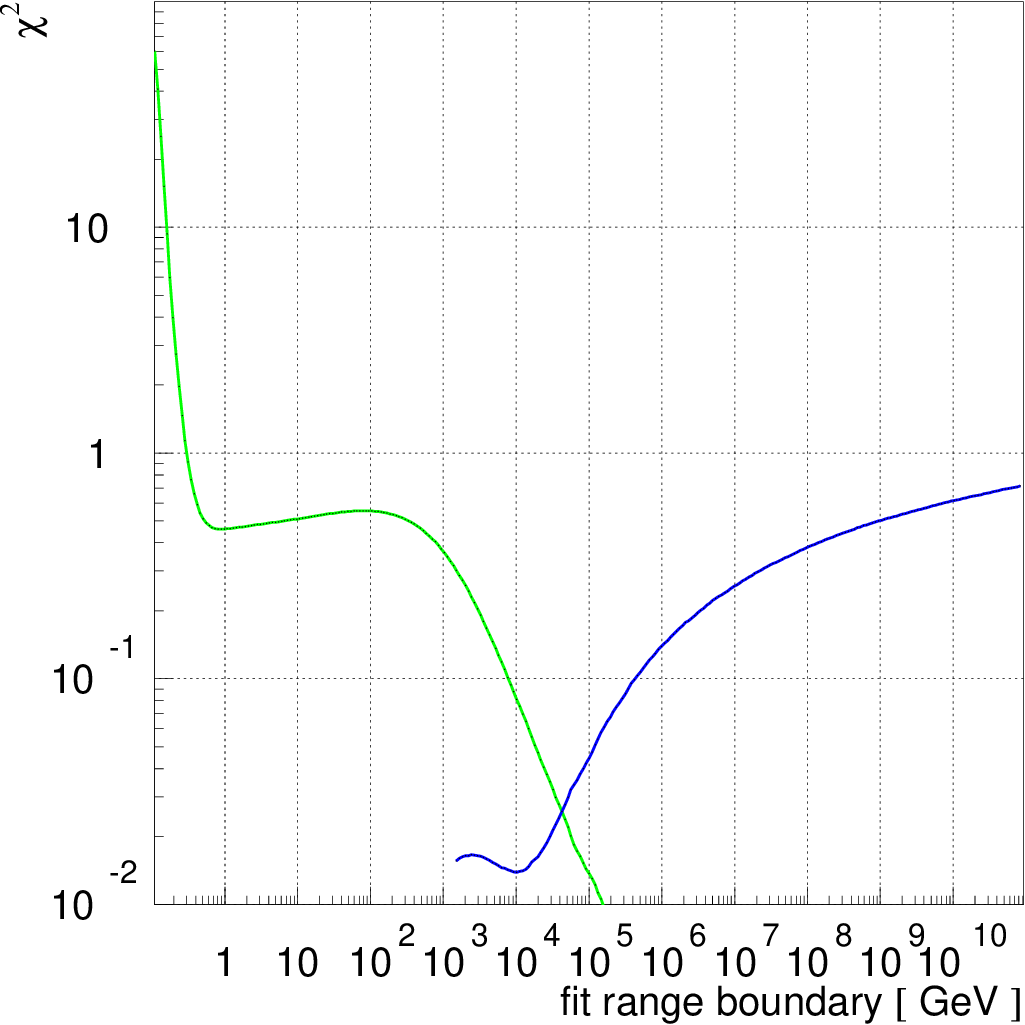,width=.45\textwidth}} \\
\parbox{.45\textwidth}{\caption[Comparison between different surviving fraction and MMC configuration settings ]{\label{dcors_fig_19}Comparison between different surviving fraction and MMC configuration settings }} & \ & \parbox{.45\textwidth}{\caption[$\chi^2$ of the fit as a function of fit boundaries ]{\label{dcors_fig_20}$\chi^2$ of the fit as a function of fit boundaries }} \\
\end{tabular}
\end{center}\end{figure}

To determine such function $E_{cut}(x)$, MMC was run for ice as propagation medium, with muon energies from 105 MeV to $10^{20}$ eV. For each energy $10^5$ muons were propagated to the point of their disappearance and the distance traveled was histogrammed (Figure \ref{dcors_fig_17}). This is similar to the analysis done in Section \ref{mmc_fit_results}. However, instead of the average distance traveled, the distance at which only a fraction of muons survives was determined for each muon energy (Figure \ref{dcors_fig_18}). Two fixed fractions were used: 99\% and 99.9\%. MMC was run with 2 different settings: $v_{cut}=10^{-2}$ with the {\it cont} ({\it continuous randomization} feature described in Section \ref{mmc_cont_section}) option and $v_{cut}=10^{-3}$ without {\it cont}. In Figure \ref{dcors_fig_19} the ratio of distances determined with both settings is displayed for 99\% of surviving muons (red line) and for 99.9\% (green line). Both lines are very close to 1.0 in most of the energy range except the very low energy part (below 2 GeV) where the muon does not suffer enough interactions with the $v_{cut}=10^{-2}$ setting before stopping (which means $v_{cut}$ has to be lowered for a reliable estimation of the shape of the travelled distance histogram). The ratio of 99\% distance to 99.9\% distance is also plotted (dark and light blue lines). This ratio is within 10\% of 1, i.e., 0.1\% of muons travel less than 10\% farther than 1\% of muons.

\begin{figure}[!h]\begin{center}
\begin{tabular}{ccc}
\mbox{\epsfig{file=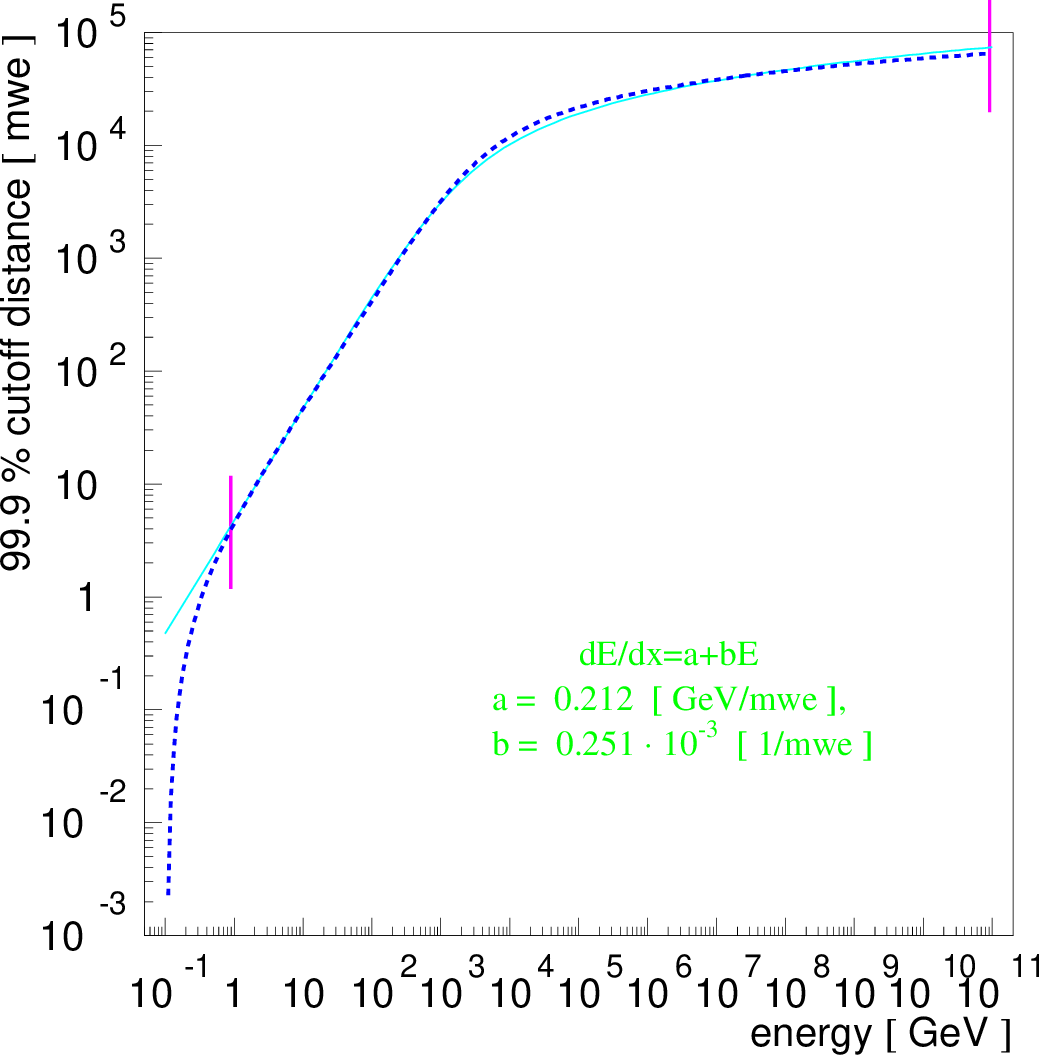,width=.45\textwidth}} & \ & \mbox{\epsfig{file=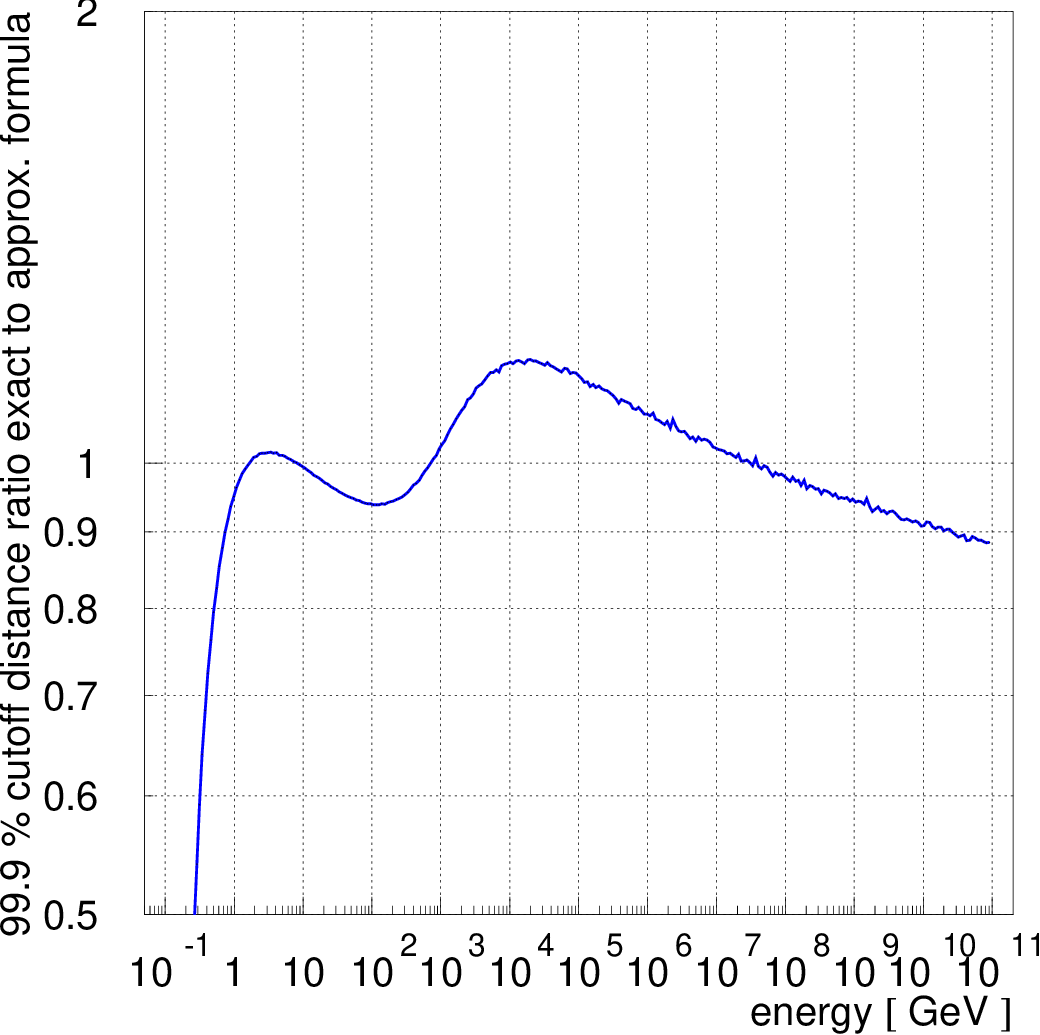,width=.45\textwidth}} \\
\parbox{.45\textwidth}{\caption[Fit to the $E_{cut}(x)$ ]{\label{dcors_fig_21}Fit to the $E_{cut}(x)$ }} & \ & \parbox{.45\textwidth}{\caption[Deviation of the $E_{cut}(x)$ from the fit ]{\label{dcors_fig_22}Deviation of the $E_{cut}(x)$ from the fit }} \\
\end{tabular}
\end{center}\end{figure}

The value $v_{cut}=10^{-3}$ with no {\it cont} setting, used to determine the maximum range of the 99.9\% of the muons, was chosen for the estimate of the function $E_{cut}(x)$. The function $$x_f=\log(1+E_i \cdot b/a)/b,$$ which is a solution to the equation represented by the usual approximation to the energy losses: $dE/dx=a+bE$, was fitted to $E_{cut}(x)$. Figure \ref{dcors_fig_20} shows the $\chi^2$ of the fit as function of the lower (green) and upper (blue) boundaries of the fitted energy range. Using the same argument as in Section \ref{mmc_fit_results} the lower limit is chosen at just below 1 GeV while the upper limit was left at $10^{11}$ GeV. As seen from the plot, raising the lower boundary to as high as 400 GeV would not lower the $\chi^2$ of the fit (and the root mean square of the deviation from it), so the lower boundary was left at 1 GeV for generality of the result. The fit is displayed in Figure \ref{dcors_fig_21} and the deviation of the actual $x_f$ from the fit is shown on Figure \ref{dcors_fig_22}. The maximum deviation is less than 20\%, which can be accounted for by lowering $a$ and $b$ by 20\%. Therefore, the final values quoted here for the function $$E_{cut}(x)=(e^{bx}-1)a/b$$ $$\mbox{are} \quad a=0.212/1.2 \mbox{ } \frac{\mbox{GeV}}{\mbox{mwe}} \quad \mbox{and} \quad b=0.251 \cdot 10^{-3}/1.2 \mbox{ } \frac{1}{\mbox{mwe}} \mbox{.}$$ The distances obtained with these values for four different muon energies are shown by red solid lines in Figure \ref{dcors_fig_17}. The distances obtained with values of $a$ and $b$ not containing the 20\% correction are shown with green dashed lines.

\section{Phenomenological lepton generation and neutrino propagation}

MMC allows one to generate fluxes of atmospheric leptons according to parameterizations given in \cite{fluxes}. Earth surface (important for detectors at depth) and atmospheric curvature are accounted for, and so are muon energy losses and probability of decay. Although the reference \cite{fluxes} provides flux parameterization, which is accurate in the region of energies from 600 GeV to 60 TeV, it is possible to introduce a correction to spectral index and normalization of each leptonic component and extrapolate the results to the desired energy range. One can also add an ad-hoc prompt component, specify $E^{-\gamma}$-like fluxes of neutrinos of all flavors, or inject leptons with specified location and momenta into the simulation.

\begin{figure}[!h]\begin{center}
\begin{tabular}{ccc}
\mbox{\epsfig{file=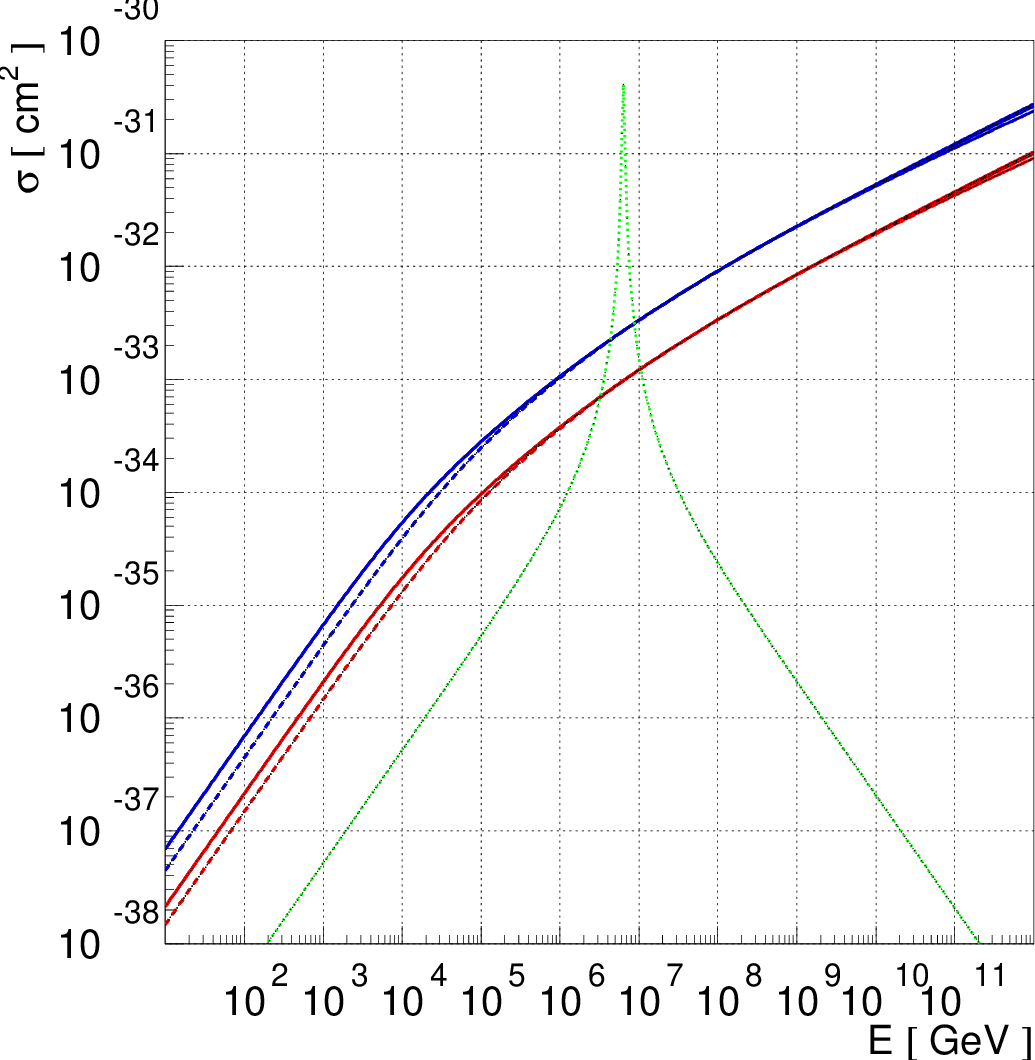,width=.45\textwidth}} & \ & \mbox{\epsfig{file=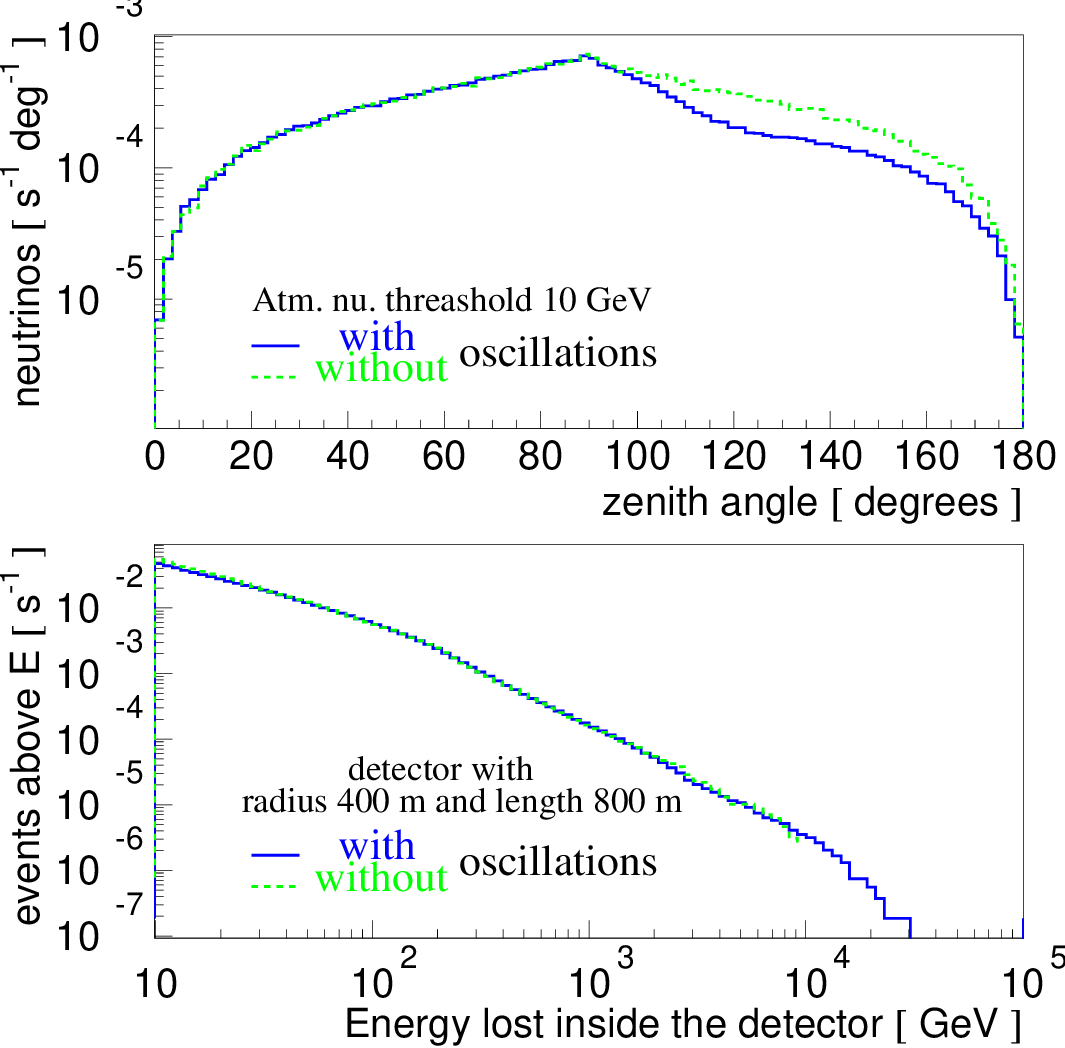,width=.45\textwidth}} \\
\parbox{.45\textwidth}{\caption[Simultated neutrino cross sections]{\label{mmc_fig_n1}Simulated neutrino cross sections: higher blue curves are CC, lower red curves are NC; solid are $\nu$, dashed are $\bar{\nu}$; green dotted is $\bar{\nu}_ee^-\rightarrow W^-$ }} & \ & \parbox{.45\textwidth}{\caption[ Neutrino flavor oscillations: muon events ]{\label{mmc_fig_n2}Neutrino flavor oscillations: muon events. Since latitude-dependent geomagnetic cutoff is not calculated, a fixed 10 GeV cutoff is applied (cf. \cite{anis}). }}
\end{tabular}
\end{center}\end{figure}

Neutrino cross sections are evaluated according to \cite{gandhi,gandhi96,ghill} with CTEQ6 parton distribution functions \cite{cteq6} (Figure \ref{mmc_fig_n1}). Neutrino and anti-neutrino neutral and charged current interaction, as well as Glashow resonance $\bar{\nu}_ee^-$ cross sections are taken into account. Power-law extrapolation of the CTEQ PDFs to small x is implemented to extend the cross section applicability range to high energies. Earth density is calculated according to \cite{earth,gandhi96}, with a possibility of adding layers of different media. All secondary leptons are propagated, therefore it is possible to simulate particle oscillations, e.g., $\tau \leftrightarrow \nu_{\tau}$. Additionally, atmospheric neutrino $\nu_{\tau} \leftrightarrow \nu_{\mu}$ oscillations are simulated (Figure \ref{mmc_fig_n2}).

Alternatively, MMC allows integration with other neutrino generators/propagators, such as NUSIM \cite{ghill-thesis}, ANIS \cite{anis}, or Juliet \cite{juliet}, or lepton generators, e.g., CORSIKA \cite{dcors_5}.

\section{MMC implementation for AMANDA-II}

Most light observed by AMANDA-II is produced by muons passing through a cylinder with radius 400 and length 800 meters around the detector \cite{mythesis}. Inside this cylinder, the Cherenkov radiation from the muon and all secondary showers along its track with energies below 500 MeV (a somewhat loose convention) are estimated together. In addition to light produced by such a ``dressed'' muon, all secondary showers with energies above 500 MeV produced in the cylinder create their own Cherenkov radiation, which is considered separately for each secondary. So in the active region of the detector muons are propagated with $E_{cut}=500$ MeV, creating secondaries on the way. This is shown as region 2 in the Figure \ref{mmc_fig_4a}.

In region 1, which is where the muon is propagated from the Earth's surface (or from under the detector) to the point of intersection of its track with the detector cylinder, muons should be propagated as fast as possible with the best accuracy. For downgoing muons, values of $v_{cut}=0.05$ with the {\it continuous randomization} option enabled were found to work best. These values should also work for muons propagated from points which are sufficiently far from the detector. For muons created in the vicinity of the detector, values of $v_{cut}=0.01$ with {\it cont} or even $v_{cut}=0.001$ without {\it cont} should be used.

In region 3, which is where the muon exits the detector cylinder, it is propagated in one step ($v_{cut}=1.0$, no {\it cont}) to the point of its disappearance, thus only resulting in an estimate of its average range.

It is possible to define multiple concentric media to describe both ice and rock below the ice, which is important for the study of the muons, which might be created in either medium in or around the detector and then propagated toward it. Definition of spherical, cylindrical, and cuboid detector and media geometries is possible. This can be easily extended to describe other shapes.

Although the ALLM97 with nuclear structure function as described in Section \ref{ckmt} parametrization of the photonuclear cross section was chosen to be the default for the simulation of AMANDA-II, other cross sections were also tested. No significant changes in the overall simulated data rate or the number of channels ($N_{ch}$) distribution (important for the background muon analysis of \cite{mythesis,my03icrc}) were found between the parameterizations described in Section \ref{photo}. This is to be expected since for the background muons (most of which have energies of 0.5-10 TeV on the surface) all photonuclear cross section parameterizations are very close to each other (see Figure \ref{mmc_fig_18}). Also the effects of the Moli\`{e}re scattering and LPM-related effects (Section \ref{lpm}) can be completely ignored (although they have been left on for the default settings of the simulation).

\section{Formulae}
\label{mmc_formulae}

This section summarizes cross-section formulae used in MMC. In the formulae below, $E$ is the energy of the incident muon, while $\nu=vE$ is the energy of the secondary particle: knock-on electron for ionization, photon for bremsstrahlung, virtual photon for photonuclear process, and electron pair for the pair production. As usual, $\beta=v/c$ and $\gamma=\left(1-\beta^2\right)^{-1/2}$; also $\mu$ is muon mass (or tau mass, except in the expression for $q_c$ of Section \ref{brems_abb}, where $\mu$ is just a mass-dimension scale factor equal to the muon mass \cite{bugaev_montaruli}), $m=m_e$ is electron mass, and $M$ is proton mass. Values of constants used below are summarized in Appendix \ref{app_mmc}.

\subsection{Ionization}
\label{ioniz}

A standard Bethe-Bloch equation \cite{pdb} was modified for muon and tau charged leptons (massive particles with spin 1/2 different from electron) following the procedure outlined in \cite{rossi}. The result is given below (and is consistent with \cite{tables}):
$$- {dE  \over dx}=K z^2 {Z \over A\beta^2}\left[{1 \over 2}
\ln\left({2m_{e}\beta^2 \gamma^2 \nu_{upper}\over
I(Z)^2}\right)
-{\beta^2 \over 2} \left(1+{\nu_{upper}\over \nu_{max}}\right) \right.$$
$$\left. +{1 \over 2}\left({\nu_{upper} \over 2E(1+1/\gamma)}\right)^2
-{\delta \over 2}\right],$$
$$\mbox{where}\quad \nu_{max}={2m_{e}(\gamma^2-1) \over
1+2\gamma {m_e \over \mu}+\left({m_e \over \mu}\right)^2}\quad \mbox{and} \quad
\nu_{upper}=\min(\nu_{cut},\nu_{max}) \mbox{.}$$
The density correction $\delta$ is computed as follows:
\bal
\delta&=& \delta_0 10^{2(X-X_0)},\quad\mbox{if}\  X<X_0\nonumber\\
\delta&=&2(\ln 10) X +C + a(X_1-X)^m,\quad\mbox{if}\ X_0\le X<X_1\nonumber\\
\delta&=&2(\ln 10) X+C,\quad\mbox{if}\ X\ge X_1,\nonumber \quad\mbox{where}\quad X=\log_{10}(\beta\gamma)\eal
$${{d^2N}\over{d\nu dx}} = {{1 \over 2} K z^2 {Z \over A} {1\over \beta^2}
{1\over \nu^2 } {\left[ 1 \ - \beta^2 {\nu \over \nu_{max}} + {1
\over 2 } \left( \nu \over E(1+1/\gamma) \right)^2 \right]}} \mbox{.}$$
This formula, integrated from $\nu_{min}={1 \over 2m_{e} } \cdot \left({I(Z) \over  \beta \gamma}\right)^2$ to $\nu_{upper}$, gives the expression for energy loss above, less the density correction and $\beta^2$ terms (plus two more terms which vanish if $\nu_{min} \ll \nu_{upper}$).

\subsection{Bremsstrahlung}
\label{brems}

According to \cite{kelner95}, the bremsstrahlung cross section may be represented by the sum of an elastic component ($\sigma_{el}$, discussed in \cite{heitler,bethe}) and two inelastic components ($\Delta\sigma_{a,n}^{in}$),
$$\sigma=\sigma_{el} +
\Delta\sigma_{a}^{in}+\Delta\sigma_{n}^{in}
\label{eq:sigma} \mbox{.}$$

\subsubsection[Elastic Bremsstrahlung: Kelner Kokoulin Petrukhin parameterization]{Elastic Bremsstrahlung (Kelner Kokoulin Petrukhin parameterization):}
\bce \mbox{\epsfig{file=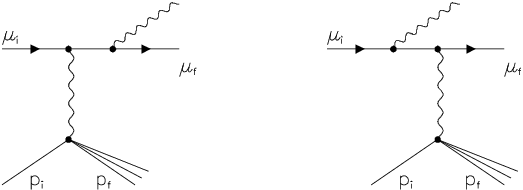,width=.5\textwidth}} \ece
$$\sigma_{el}(E,v)=\f{\alpha}{v}\left(2 Z
\f{m}{\mu}r_{e}\right)^{2} \left(\f{4}{3}-\f{4}{3}v+v^{2}\right)
\left[\ln\left[\f{\mu}{\delta}\right] -\f{1}{2}
-\Delta_{a}^{el}-\Delta_{n}^{el}\right],$$
$$\mbox{where}\quad \delta\approx\f{\mu^{2}\omega}{2E(E-\omega)},$$
is the minimum momentum transfer. The formfactors (atomic $\Delta_{a}^{el}$ and nuclear $\Delta_{n}^{el}$) are
\bal \Delta_{a}^{el}(\delta)
&=&\ln\left[1+\f{1}{\delta\sqrt{e}BZ^{-1/3}/m}\right]
\nonumber\\
\Delta_{n}^{el}(\delta)&=&\ln\left[\f{D_{n}}{1+\delta(D_{n}\sqrt{e}-2)/\mu}
\right];\hsp D_{n}=1.54 A^{0.27}  \mbox{.} \nonumber \eal

Integration limits for this cross section are
$$ v_{min}=0\le v\le v_{max}=1-\f{3\sqrt{e}}{4}\f{\mu}{E}Z^{1/3} $$

\subsubsection[Petrukhin Shestakov form factor parameterization]{Petrukhin Shestakov form factor parameterization:}
\label{brems_ps}
Somewhat older parameterization of the form factors in the Bethe-Heitler formula \cite{heitler} is given in \cite{ps-brems}:
$$\sigma_{el}(E,v)=\f{\alpha}{v}\left(2 Z
\f{m}{\mu}r_{e}\right)^{2} \left(\f{4}{3}-\f{4}{3}v+v^{2}\right)
\phi(\delta),$$
\bal \phi(\delta)
&=&\ln\left[\f{{189 \mu \over m } Z^{-1/3}}{1+{189 \sqrt{e} \over m } \delta Z^{-1/3}}\right], \quad Z \le 10
\nonumber\\
\phi(\delta)
&=&\ln\left[\f{{2\over 3} {189 \mu \over m } Z^{-2/3}}{1+{189 \sqrt{e} \over m } \delta Z^{-1/3}}\right], \quad Z>10
\mbox{.} \nonumber \eal

\subsubsection[Andreev Berzrukov Bugaev parameterization]{Andreev Berzrukov Bugaev parameterization:}
\label{brems_abb}
Another parameterization of the bremsstrahlung cross section, both elastic and inelastic $\mu$-diagram contributions (not the $e$-diagram, which is included with the ionization cross section) is implemented according to \cite{abb,abb2,mum}.

$$\sigma(E,v)=\alpha \left( 2r_e Z {m_e \over \mu} \right)^2 {1 \over v} \left[ (2-2v+v^2)\Psi_1(q_{min},Z)-{2\over 3}(1-v)\Psi_2(q_{min},Z) \right],$$
$$\Psi_{1,2}(q_{min},Z)=\Psi_{1,2}^0(q_{min},Z)-\Delta_{1,2}(q_{min},Z)$$
$$\Psi_{1}^0(q_{min},Z)={1\over 2}\left( 1+\ln{\mu^2a_1^2 \over 1+x_1^2}\right) -x_1 \arctan{1\over x_1}+{1\over Z} \left[ {1\over 2}\left( 1+\ln {\mu^2 a_2^2\over 1+x_2^2}\right) \right.$$ $$\left. -x_2 \arctan{1\over x_2}\right],$$
\bal
\Psi_{2}^0(q_{min},&Z&)={1\over 2}\left( {2\over 3}+\ln{\mu^2a_1^2 \over 1+x_1^2}\right)+2x_1^2\left(1-x_1 \arctan{1\over x_1} +{3\over 4}\ln {x_1^2\over 1+x_1^2} \right)\nonumber\\
+&{1\over Z}&\left[ {1\over 2}\left( {2\over 3}+\ln {\mu^2 a_2^2\over 1+x_2^2}\right)+2x_2^2\left( 1-x_2 \arctan{1\over x_2} +{3\over 4}\ln {x_2^2\over 1+x_2^2}\right)\right],
\eal
\bal
\Delta_1(q_{min}, Z\ne 1)&=&\ln{\mu \over q_c}+{\zeta \over 2}\ln{\zeta +1 \over \zeta -1},\nonumber\\
\Delta_2(q_{min}, Z\ne 1)&=&\ln{\mu \over q_c}+{\zeta \over 4}(3-\zeta^2)\ln{\zeta +1 \over \zeta -1}+{2\mu^2 \over q_c^2},\nonumber\\
\Delta_{1,2}(q_{min}, Z=1)&=&0,
\eal
$$q_{min}={\mu^2 v \over 2E(1-v)}, \quad x_i=a_i q_{min},$$
$$a_1={111.7 \over Z^{1/3} m_e}, \quad a_2={724.2 \over Z^{2/3} m_e}, \quad \zeta=\sqrt{1+{4\mu^2\over q_c^2}}, \quad q_c={1.9\mu \over Z^{1/3}}.$$

\subsubsection[Complete screening case]{Complete screening case:}
\label{brems_csc}
This parameterization is given in \cite{pdb} (based on \cite{tsai,davies}) and is most suitable for electrons:
$$\sigma_{el}(E,v)=\f{\alpha}{v}\left(2 Z
\f{m}{\mu}r_{e}\right)^{2} \left\{\left(\f{4}{3}-\f{4}{3}v+v^{2}\right)\left[ Z^2 (L_{rad}-f(Z))+ZL^{\prime}_{rad} \right]\right.$$ $$\left.+{1\over 9}(1-v)(Z^2+Z)\right\},$$
$$f(Z)=a^2\left[ {1 \over 1+a^2} + 0.20206-0.0369a^2+0.0083a^4-0.002a^6\right], \mbox{ with } a=\alpha Z.$$

\begin{figure}[!h]\begin{center}
\begin{tabular}{ccc}
\mbox{\epsfig{file=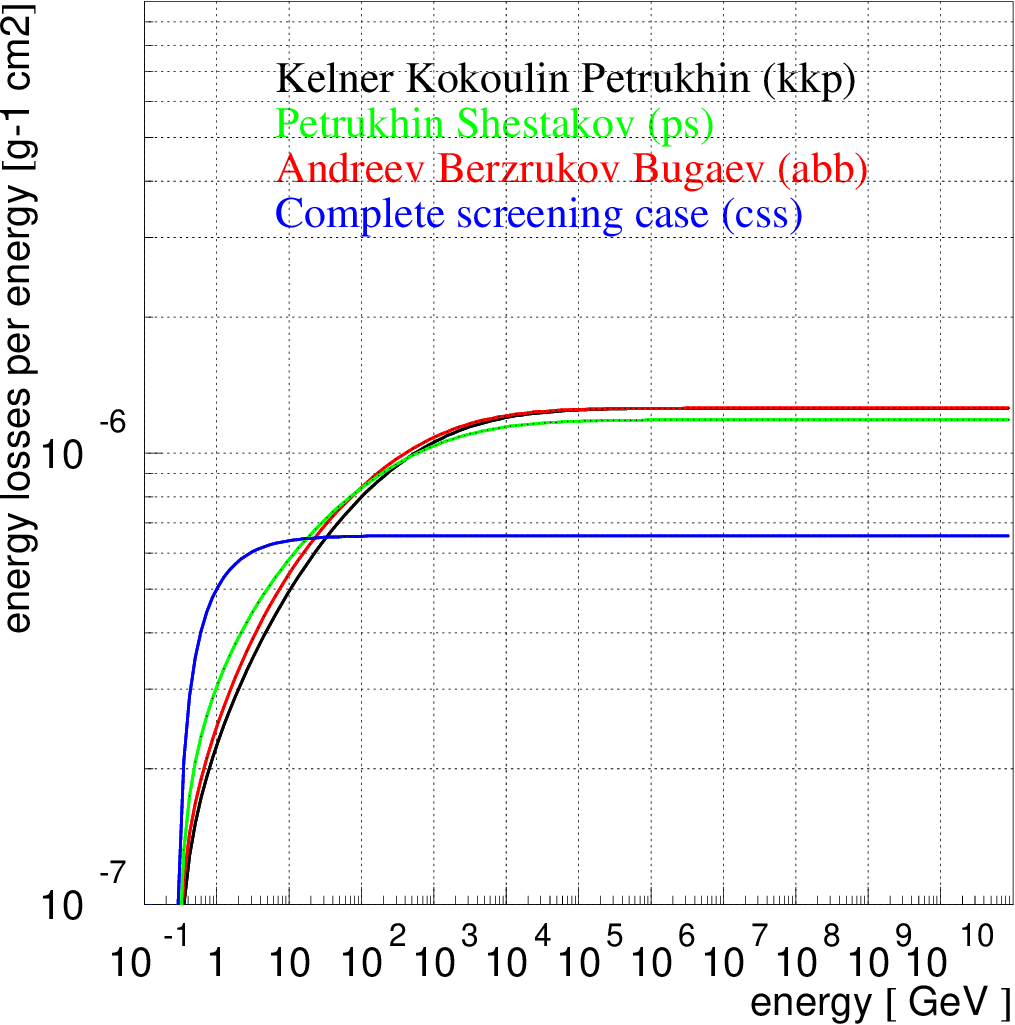,width=.45\textwidth}} & \ & \mbox{\epsfig{file=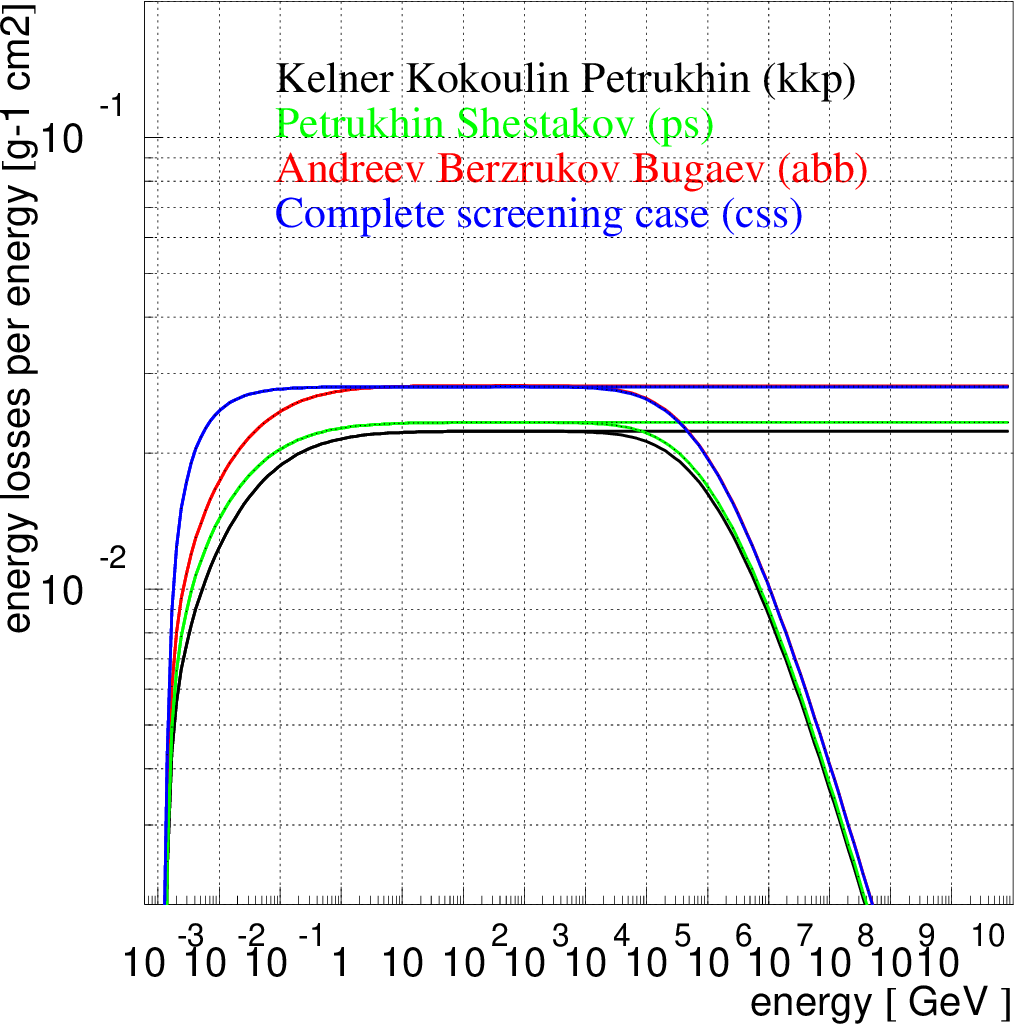,width=.45\textwidth}} \\
\parbox{.45\textwidth}{\caption[Bremsstrahlung cross section parameterizations for muons ]{\label{mmc_fig_n3}Bremsstrahlung cross section parameterizations for muons }} & \ & \parbox{.45\textwidth}{\caption[Bremsstrahlung cross section parameterizations for electrons ]{\label{mmc_fig_n4}Bremsstrahlung cross section parameterizations for electrons }} \\
\end{tabular}
\end{center}\end{figure}

All bremsstrahlung parameterizations are compared in Figures \ref{mmc_fig_n3} and \ref{mmc_fig_n4}. Parameterization of Section \ref{brems_abb} (abb) agrees best with the complete screening case of electrons and with the other two cross sections for muons, thereby providing the most comprehensive description of bremsstrahlung cross section.

\subsubsection[Inelastic Bremsstrahlung]{Inelastic Bremsstrahlung:}
The effect of nucleus excitation can be evaluated as
\bec
\Delta_{n}^{in}=\f{1}{Z}\Delta^{el}_{n};\ (Z\ne 1) \mbox{.} \eec

Bremsstrahlung on the atomic electrons can be described by the diagrams below;
e-diagram is included with ionization losses (because of its sharp $1/v^2$ energy loss spectrum), as described in \cite{kelner96}:
$$\Delta {{d^2N}\over{d\nu dx}} = \left( {{d^2N}\over{d\nu dx}} \right)_{I_0} \cdot {\alpha \over {2 \pi}} (a(2b+c)-b^2)$$
$$a=\log(1+2\nu/m_e), \quad b=\log((1-\nu/\nu_{max})/(1-\nu/E)),$$ $$c=\log((2 \gamma (1-\nu/E) m_e)/(m_\mu \nu/E)) \mbox{.}$$
The maximum energy lost by a muon is the same as in the pure ionization (knock-on) energy losses. The minimum energy is taken as
$\nu_{min}=I(Z)$. In the above formula $\nu$ is the energy lost by the muon, i.e., the sum of energies transferred to both electron and photon. On the output all of this energy is assigned to the electron.
\bce \mbox{\epsfig{file=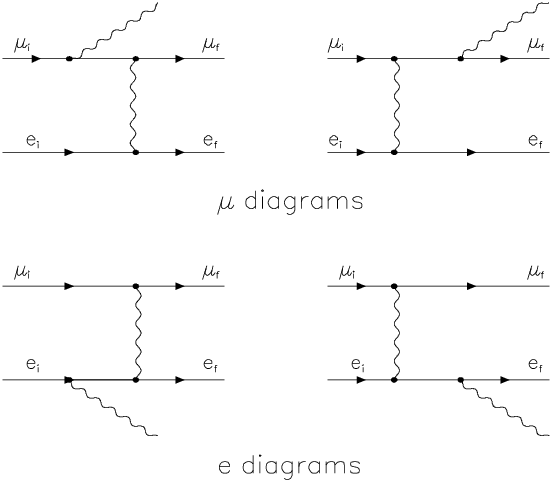,width=.5\textwidth}} \ece

The contribution of the $\mu$-diagram (included with bremsstrahlung) is discussed in \cite{kelner95}:
$$\Delta\sigma_{a}^{in}(E,v)\approx\f{\alpha}{v}
\left(2Z\f{m}{\mu}r_{e}\right)^{2}
\left(\f{4}{3}-\f{4}{3}v+v^{2}\right)\Delta_{a}^{in}\nonumber$$
$$\Delta_{a}^{in}\approx\f{1}{Z}\tilde\Phi_{a}^{in}(\delta)\nonumber \quad\mbox{with}\quad
\tilde\Phi_{a}^{in}(\delta)=\ln\left[\f{\mu/\delta}{\delta\mu/m^{2}+\sqrt{e}}
\right]- \ln\left[1+\f{m}{\delta\sqrt{e}B^{\prime}Z^{-2/3}}\right]$$
\bce $B^\prime$=1429 for $Z\ge 2$ and $B^\prime$=446 for Z=1.\ece
The maximum energy transferred to the photon is
$$v_{max}=\f{m(E-\mu)}{E(E-p+m)} \mbox{.}$$
On the output all of the energy lost by a muon is assigned to the bremsstrahlung photon.

\subsection{Photonuclear interaction}
\label{photo}
\bce \mbox{\epsfig{file=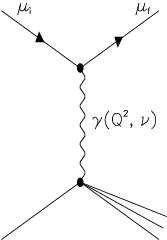,width=.15\textwidth}} \ece
\subsubsection{Bezrukov Bugaev parameterization of the photonuclear interaction}

\begin{figure}[!h]\begin{center}
\begin{tabular}{ccc}
\mbox{\epsfig{file=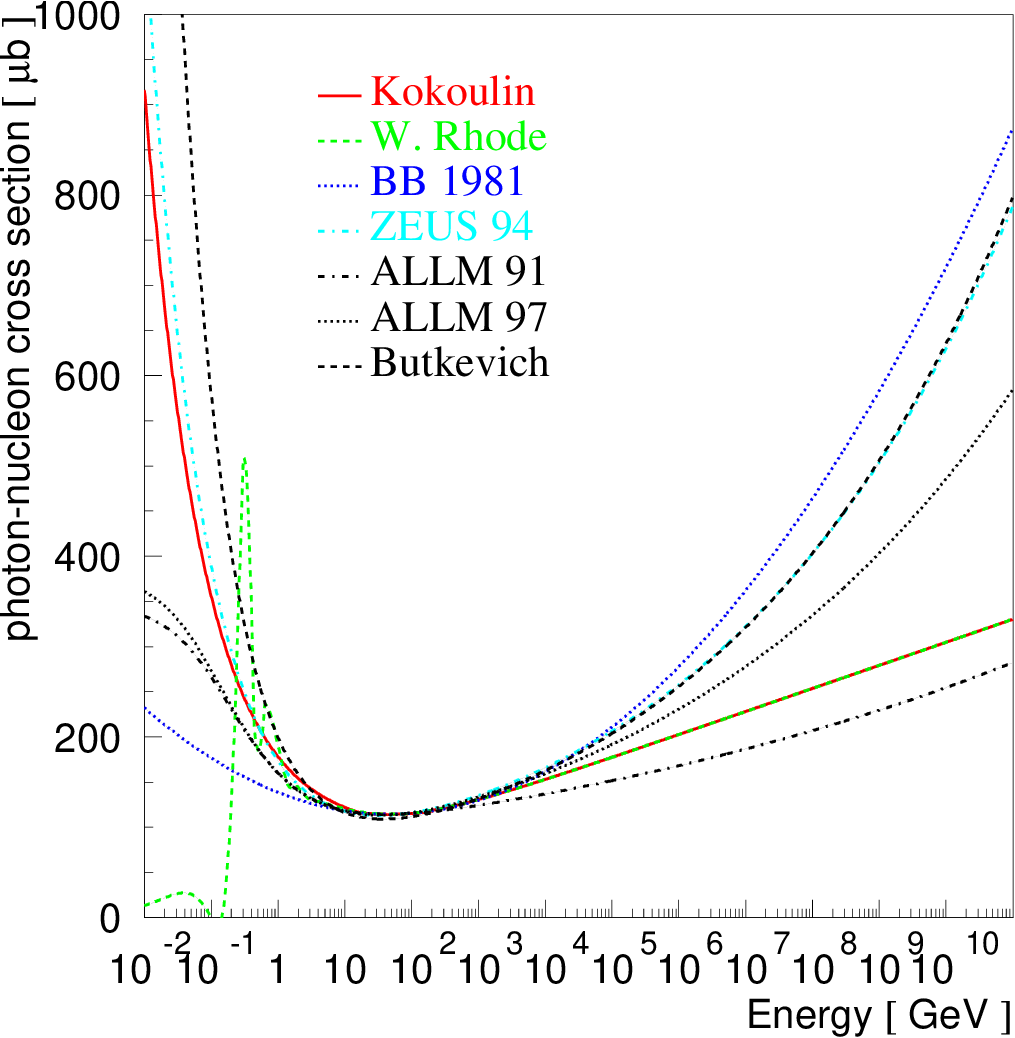,width=.45\textwidth}} & \ & \mbox{\epsfig{file=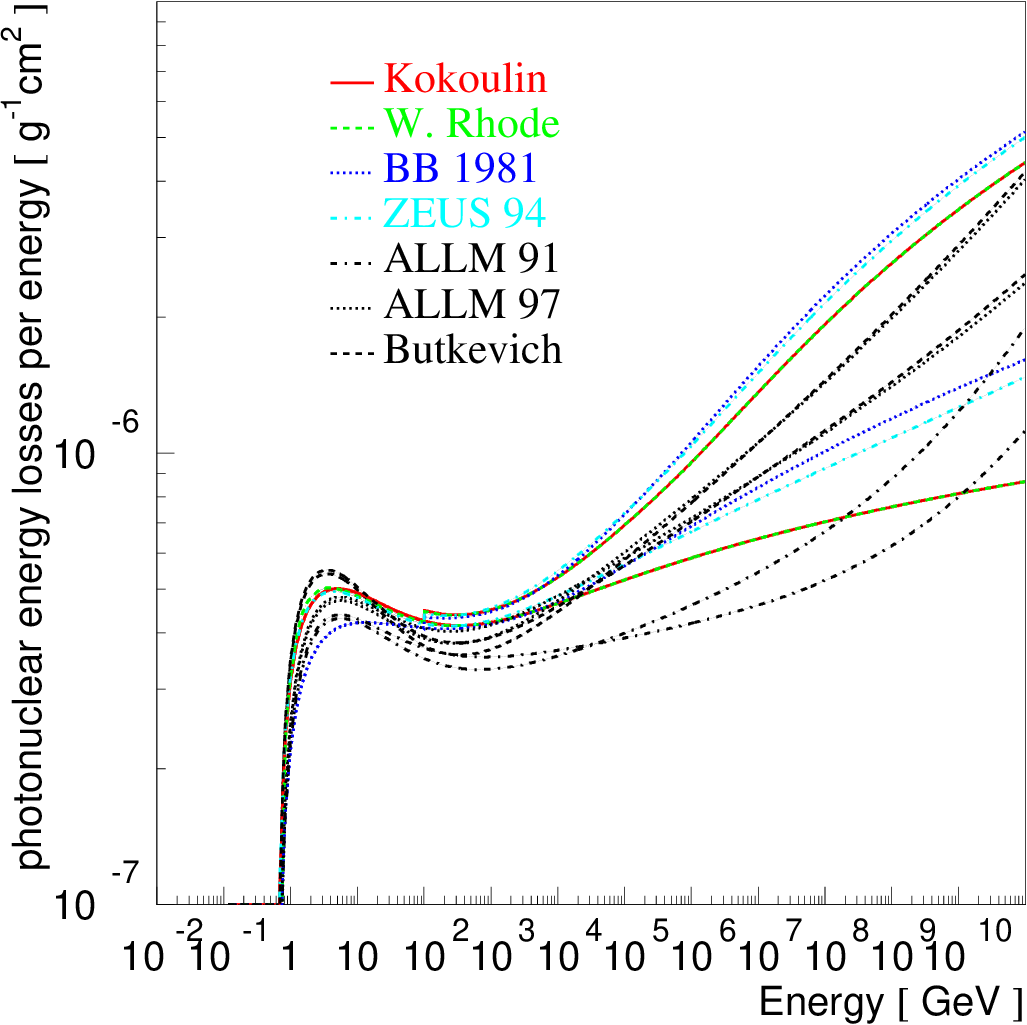,width=.45\textwidth}} \\
\parbox{.45\textwidth}{\caption[Photon-nucleon cross sections, as described in the text ]{\label{mmc_fig_18}Photon-nucleon cross sections, as described in the text: Kokoulin \cite{kokgs}, W. Rhode \cite{rhode}, BB 1981 \cite{bugaev}, ZEUS 94 \cite{zeus}, ALLM 91 and 97 \cite{allm2}, Butkevich \cite{butmikh}. Curves 5-7 are calculated according to $\sigma_{\gamma N}=\lim \limits_{Q^2\rightarrow 0}{4\pi^2\alpha F_2^N\over Q^2}$ }} & \ & \parbox{.45\textwidth}{\caption[Photonuclear energy losses (divided by energy). ]{\label{mmc_fig_19}Photonuclear energy losses (divided by energy), according to formulae from Section \ref{photo}. Higher lines for the parameterizations 1-4 include the hard component \cite{bugaev_shlepin}, higher lines for 5-7 calculate shadowing effects as in Section \ref{ckmt}, lower as in Section \ref{allm}}} \\
\end{tabular}
\end{center}\end{figure}
\label{photo_bb}
The soft part of the photonuclear cross section is used as parametrized in \cite{bugaev} (underlined terms taken from \cite{bugaev_shlepin,bugaev_montaruli} are important for tau propagation):
\bal \f{d\sigma}{d
v}&=&\f{\alpha}{2\pi} A\sigma_{\gamma N}v\left\{ 0.75 G(x)\left[\kappa
\ln\left(1+\f{m_1^2}{t}\right) \right. \right.\\
 &-&\left. \left.\f{\kappa
m_1^2}{m_1^2+t}-\f{2\mu^2}{t}+\underline{{4\mu^2\over m_1^2} \ln\left(1+{m_1^2 \over t}\right)}\right]+
\right.\nonumber\\
&+&\left.  0.25 \left[\left(\kappa+\underline{2\mu^2 \over m_2^2}\right) \ln\left(1+\f{m_2^2}{t}\right)-
\f{2\mu^2}{t}\right] \right. \\
&+&\left. \f{\mu^2}{2t}\left[0.75 G(x)\f{m_1^2-\underline{4t}}{m_1^2+t}+
0.25\f{m_2^2}{t}\ln\left(1+\f{t}{m_2^2}\right)\right]
\right\},\eal
$$\mbox{where}\quad t=Q^{2}_{max}=\f{\mu^2 v^2}{1-v} \ \ ,\quad
 \kappa=1-\f{2}{v}+\f{2}{v^2}\ \ ,$$ $$ m_1^2=0.54\ \mbox{GeV}^2\ \ ,\ \ \mbox{and}\ \ \ \ \ m_2^2=1.8\
\mbox{GeV}^2 \mbox{.} \nonumber$$
Nucleon shadowing is calculated according to
\bal \sigma_{\gamma A}(\nu)&=&A\sigma_{\gamma
N}(\nu)\{0.75 G(x)+0.25\} \nonumber\\
\mbox{with}\ \  G(x)&=&\f{3}{x^3}\left(\f{x^2}{2}-1+e^{-x}(1+x)\right),\quad
\mbox{for}\ Z \neq 1 \mbox{, and }G(x)=1\mbox{ for Z=1}
\nonumber\\
x&=&R n \sigma_{\rho N}\simeq 0.00282 A^{\f{1}{3}}\sigma_{\gamma N}(\nu) \mbox{.}
\nonumber \eal
Several parametrization schemes for the photon-nucleon cross section are implemented. The first is
\bal \sigma_{\gamma N}(\nu)&=&96.1+\f{82}{\sqrt{\nu}},\quad\mbox{for}\ \nu\le 17 \mbox{ GeV}\nonumber\\
\sigma_{\gamma N}(\nu)&=&114.3 +1.647 \ln^2[0.0213\nu]\ \mu \mbox{b},\quad \mbox{for } \nu\in[17,200 \mbox{ GeV}] \mbox{ \cite{bugaev}}\nonumber\\
\sigma_{\gamma N}(\nu)&=&49.2 +11.1\ln[\nu]+151.8/\sqrt\nu\ \mu \mbox{b}, \quad \mbox{above 200 GeV \cite{kokgs}} \mbox{.} \eal
The second is based on the table parametrization of \cite{rhode} below 17 GeV. Since the second formula from above is valid for energies up to $10^6$ GeV, it is taken to describe the whole energy range alone as the third case. Formula \cite{zeus}
$$ \sigma_{\gamma N}(\nu)=63.5s^{0.097}+145s^{-0.5} \ \mu \mbox{b}\quad\mbox{with}\quad s=2M \nu$$
can also be used in the whole energy range, representing the fourth case (see Figure \ref{mmc_fig_18}). Finally, the ALLM parametrization (discussed in Section \ref{allm}) or Butkevich-Mikhailov parameterization (discussed in Section \ref{ckmt}) can be enabled. It does not rely on ``nearly-real'' exchange photon assumption and involves integration over the square of the photon 4-momentum ($Q^2$). Also, treatment of the hard component within the Bezrukov-Bugaev parameterization can optionally be enabled. The hard component of photonuclear cross section was calculated in \cite{bugaev_shlepin} and parametrized in \cite{bugaev_montaruli} as
$${d\sigma_{hard} \over dv}=A \cdot {1 \over v} \sum_{k=0}^7 a_k \log_{10}^k v, \mbox{ used for } 10^{-7} \leqslant v \leqslant 1, \mbox{ }  10^2 \mbox{ GeV} \leqslant E \leqslant 10^9 \mbox{ GeV}.$$

Integration limits used for the photonuclear cross section are (kinematic limits for $Q^2$ are used for the ALLM and Butkevich-Mikhailov cross section formulae)
\bal m_\pi+{m_\pi^2 \over 2M} &<& \nu\ < E-{M \over 2} \cdot
\left( 1+{m_\mu^2 \over
M^2}\right)\nonumber \\
{m_\mu^2 \nu^2 \over EE'}-{m_\mu^4 \over 2EE'} &<& Q^2 <
2M(\nu-m_\pi)-m_\pi^2 \ \ ,\quad E'=E-\nu \mbox{.} \eal

\subsubsection[Abramowicz Levin Levy Maor parametrization of the photonuclear cross section]{Abramowicz Levin Levy Maor (ALLM) parametrization of the photonuclear cross section}
\label{allm}
The ALLM formula is based on the parametrization \cite{allm,allm2,dutta}
$$ {d\sigma(v, Q^2)\over dvdQ^2}={4\pi \alpha^2\over Q^4} {F_2\over v} \left[1-v-{Mxv\over 2E}+\left(1-{2\mu^2\over Q^2}\right)
{v^2(1+4M^2x^2/Q^2)\over 2(1+R))} \right] $$
$$ x={Q^2\over 2MEv} \mbox{.} $$
The limits of integration over $Q^2$ are given in the section for photonuclear cross section.
$$ F_2=a(Z+(A-Z)P)F_2^p \quad \mbox{Here,} \quad a(A, x, Q^2)\simeq a(A,x) $$
\bal
a(A,x)&=&A^{-0.1} \quad \mbox{for} \quad x<0.0014\\
a(A,x)&=&A^{0.069\log_{10}x+0.097} \quad \mbox{for} \quad 0.0014 \le x<0.04\\
a(A,x)&=&1 \quad \mbox{for} \quad x \ge 0.04
\eal
$$P(x)=1-1.85x+2.45x^2-2.35x^3+x^4$$
$$ F_2^p(x,Q^2)={Q^2\over Q^2+m_0^2} (F_2^P+F_2^R) $$
$$ F_2^i(x, Q^2)=c_i x_i^{a_i}(1-x)^{b_i} \quad \mbox{for} \quad  i=P,R $$
\bal
\mbox{For}\quad f=c_R, a_R, b_R, b_P & \quad & f(t)=f_1+f_2 t^{f_3}\\
\mbox{For}\quad g=c_P, a_P & \quad & g(t)=g_1+(g_1-g_2)\left[{1\over 1+t^{g_3}}-1\right]
\eal
$$t=\ln{\ln{Q^2+Q_0^2\over \Lambda^2} \over \ln {Q_0^2\over \Lambda^2}} $$
$$x_i={Q^2+m_i^2\over Q^2+m_i^2+W^2-M^2} \quad \mbox{for} \quad  i=P,R, $$
where $W$ is the invariant mass of the nucleus plus virtual photon \cite{badelek}: $W^2=M^2+2MEv-Q^2$. Figure \ref{mmc_fig_19} compares ALLM-parametrized cross section with formulae of Bezrukov and Bugaev from Section \ref{photo_bb}.

\begin{figure}[!h]\begin{center}
\begin{tabular}{ccc}
\mbox{\epsfig{file=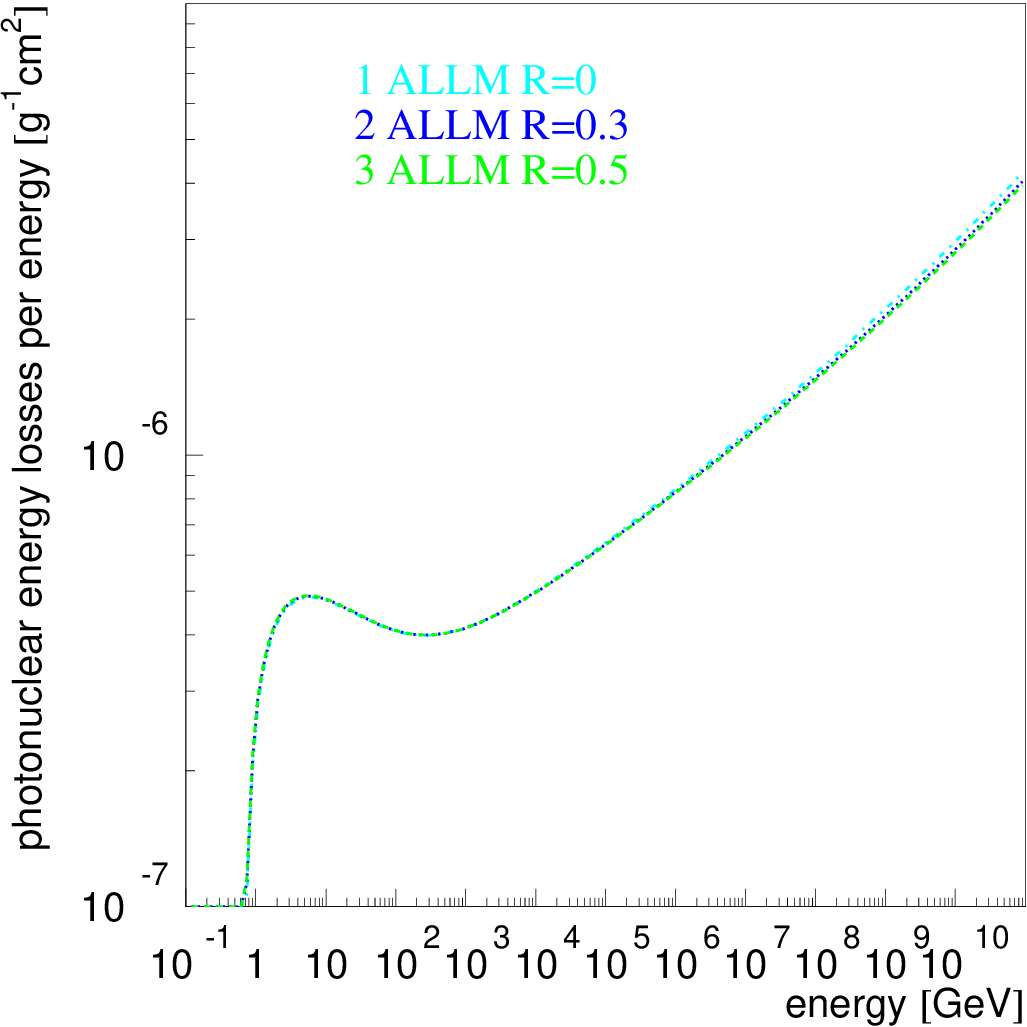,width=.45\textwidth}} & \ & \mbox{\epsfig{file=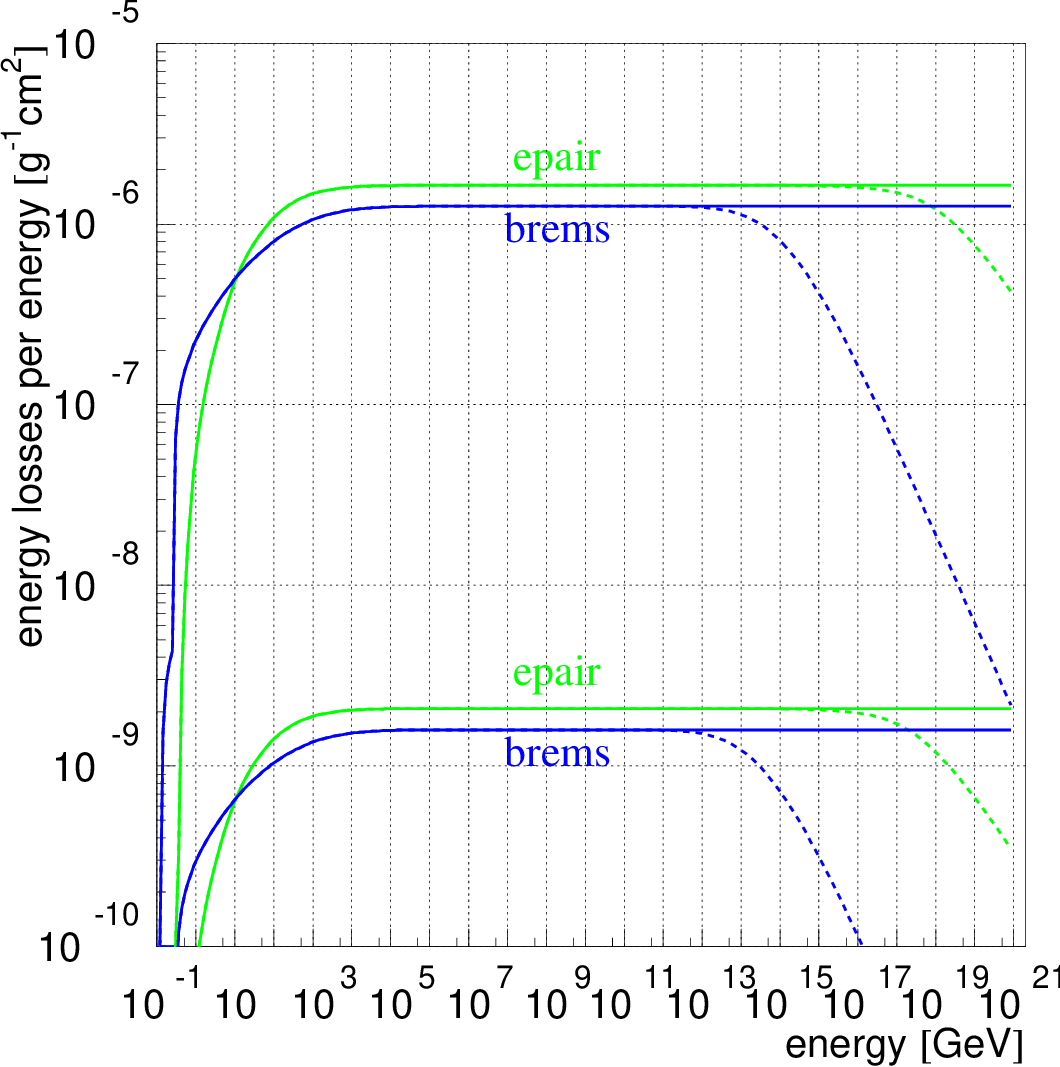,width=.45\textwidth}} \\
\parbox{.45\textwidth}{\caption[Comparison of ALLM energy loss (divided by energy) for $R$=0, $R$=0.3, $R$=0.5 ]{\label{mmc_fig_20}Comparison of ALLM energy loss (divided by energy) for $R$=0 (dashed-dotted), $R$=0.3 (dotted), $R$=0.5 (dashed) }} & \ & \parbox{.45\textwidth}{\caption[LPM effect in ice and Fr\'{e}jus rock ]{\label{mmc_fig_21}LPM effect in ice (higher plots) and Fr\'{e}jus rock (lower plots, multiplied by $10^{-3}$) }} \\
\end{tabular}
\end{center}\end{figure}

The quantity $R(x,Q^2)$ is not very well known, although it has been measured for high $x$ ($x>0.1$) \cite{whitlow} and modeled for small $x$ ($10^{-7}<x<0.1$, $0.01 \mbox{ GeV}^2 <Q^2<50 \mbox{ GeV}^2$) \cite{badelek2}. It is of the order $\sim 0.1-0.3$ and even smaller for small $Q^2$ (behaves as $O(Q^2)$). In Figure \ref{mmc_fig_20} three photonuclear energy loss curves for $R$=0, 0.3, and 0.5 are shown. The difference between the curves never exceeds 7\%. In the absence of a convenient parametrization for $R$ at the moment, it is set to zero in MMC.

The values of cross sections in Figures \ref{mmc_fig_18}$-$\ref{mmc_fig_20} should not be trusted at energies below 10 GeV. However, their exact values at these energies are not important for the muon propagation since the contribution of the photonuclear cross section to the muon energy losses in this energy range is negligible.

\subsubsection{Butkevich-Mikhailov parametrization of the photonuclear cross section}
\label{ckmt}
Following the parameterization of the proton ($p$) and neutron ($n$) structure functions according to the CKMT model \cite{ckmt,butmikh},
$$F_2^{p,n}(x,Q^2)=F_S^{p,n}(x,Q^2)+F_{NS}^{p,n}(x,Q^2)$$

$$F_S^p(x,Q^2)=A_S x^{-\Delta (Q^2)} (1-x)^{n(Q^2)+4}\left( {Q^2 \over Q^2+a} \right)^{1+\Delta(Q^2)}$$
$$F_S^n(x,Q^2)=A_S x^{-\Delta (Q^2)} (1-x)^{n(Q^2)+\tau}\left( {Q^2 \over Q^2+a} \right)^{1+\Delta(Q^2)}$$

$$F_{NS}^p(x,Q^2)=xU_V(x,Q^2)+xD_V(x,Q^2)$$
$$F_{NS}^n(x,Q^2)={1\over 4} xU_V(x,Q^2)+4xD_V(x,Q^2)$$

$$xU_V(x,Q^2)=B_u x^{(1-\alpha_R)} (1-x)^{n(Q^2)} \left( {Q^2 \over Q^2+b} \right)^{\alpha_R}$$
$$xD_V(x,Q^2)=B_d x^{(1-\alpha_R)} (1-x)^{n(Q^2)+1} \left( {Q^2 \over Q^2+b} \right)^{\alpha_R},$$

$$\mbox{where} \quad \Delta(Q^2)=\Delta_0 \left( 1+{2Q^2 \over Q^2+d}\right), \quad \mbox{and} \quad n(Q^2)={3 \over 2} \left( 1+{Q^2 \over Q^2+c}\right)$$

$$F_2^A(x,Q^2)=r^{A/d}[ZF_2^p(x,Q^2)+(A-Z)F_2^n(x,Q^2)]$$

The nuclear structure function $r^{A/d}$ can be evaluated as the shadowing function $a$ from the previous section, or can optionally be calculated as follows \cite{smirnov_1,smirnov_2,butmikh}.
At $x>0.3$, $r^{A/d}=1-m_b(A)a_{osc}(x)$, with $m_b(A)=M_b[1-N_s(A)/A]$ and $M_b=0.437$. $N_s(A)$ is the Wood-Saxon potential
$$N_s(A)=4\pi \rho_0 \int\limits_{r_0(A)}^{\infty} {r^2 dr \over 1+\exp\{[r-r_0(A)]/a\}},$$
where $\rho_0=0.17 \mbox{ fm}^{-3}$, $a=0.54 \mbox{ fm}$, and $r_0(A)=1.12 A^{1/3}-0.86 A^{-1/3}$.
$$a_{osc}(x)=(1-\lambda x)\left\{ \left( {1\over u}-{1\over c}\right)-\mu \left({1\over u^2}-{1\over c^2} \right) \right\},$$
where $u=1-x$, $c=1-x_2$, $x_2=0.278$, $\lambda=0.5$, and $\mu=m_{\pi}/M$.

At $10^{-3}\gtrsim x_0 \leqslant x \leqslant 0.3$, $r^{A/d}(x)=x^{m_1}(1+m_2)(1-m_3 x)$ with $m_i=M_i[1-N_s(A)/A]$, where $M_1=0.129$, $M_2=0.456$, and $M_3=0.553$. Here
$$x_0=\left[ {1 \over 1+m_2} (0.75G(\nu)+0.25) \right]^{1/m_1},$$
where $G(\nu)$ is given by expression of Section \ref{photo_bb} with $\sigma_{\gamma N}=112.2(0.609 \nu^{0.0988}+1.037 \nu^{-0.5944})$. At $x<x_0$ function $r^{A/d}(x)=r^{A/d}(x_0)$.

\subsection{Electron pair production}
\label{epair}
Two out of four diagrams describing pair production are shown below. These describe the dominant ``electron'' term. The two diagrams not shown here describe the muon interacting with the atom and represent the ``muon'' term. The cross section formulae used here were first derived in \cite{pair1,pair2,pair}.
\\
\bce \mbox{\epsfig{file=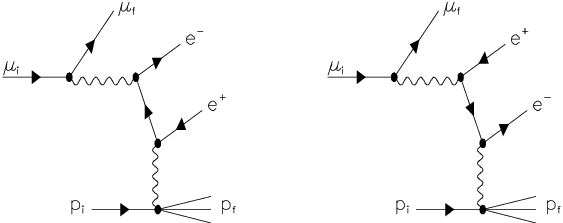,width=.5\textwidth}} \ece
\bec
\f{d\sigma(E,v,\rho)}{dv d\rho}=\f{2}{3\pi}Z(Z+\zeta)(\alpha r_{e})^2
\f{1-v}{v}\left(\Phi_{e}+\f{m^2}{\mu^2}\Phi_{\mu}\right)
\eec
$$v=(\epsilon_{+}+\epsilon_{-})/E, \quad \rho=(\epsilon_{+}-\epsilon_{-})/(\epsilon_{+}+\epsilon_{-})$$
\bal \Phi_{e}&=&\left\{
[(2+\rho^2)(1+\beta)+\xi(3+\rho^2)]\ln\left(1+\f{1}{\xi}\right)
+\f{1-\rho^2-\beta}{1+\xi}-(3+\rho^2)\right\} L_{e}\nonumber\\
\Phi_{\mu}&=&\left\{\left[ (1+\rho^2) \left(1+\f{3}{2}\beta\right) -
\f{1}{\xi}(1+2\beta)(1-\rho^2)\right]\ln(1+\xi)+ \right .\nonumber\\
&+&\left.\f{\xi(1-\rho^2-\beta)}{1+\xi}+(1+2\beta)(1-\rho^2)
\right\}L_\mu\nonumber\\
L_{e}&=&\ln\left(\f{B Z^{-1/3}\sqrt{(1+\xi)(1+Y_e)}} {1+\f{2m\sqrt{e} B
Z^{-1/3}(1+\xi)(1+Y_e)}{E v(1-\rho^2)}}\right)-
\f{1}{2}\ln\left[1+\left(\f{3m}{2\mu}Z^{1/3}\right)^2(1+\xi)(1+Y_e)\right]
\nonumber\\
L_{\mu}&=&\ln\left(\f{\f{2}{3}\f{\mu}{m}BZ^{-2/3}} {1+\f{2m\sqrt{e} B
Z^{-1/3}(1+\xi)(1+Y_\mu)}{E v(1-\rho^2)}}\right)
\nonumber\\
Y_{e}&=&\f{5-\rho^2+4\beta(1+\rho^2)}
{2(1+3\beta)\ln(3+1/\xi)-\rho^2-2\beta(2-\rho^2)}\nonumber\\
Y_{\mu}&=&\f{4+\rho^2+3\beta(1+\rho^2)}
{(1+\rho^2)(3/2+2\beta)\ln(3+\xi)+1-\f{3}{2}\rho^2}\nonumber
\eal
$$\beta=\f{v^2}{2(1-v)}, \quad
\xi=\left(\f{\mu v}{2 m}\right)^2 \f{1-\rho^2}{1-v}$$
\bec \zeta_{loss}^{pair}(E, Z)\sim
\f{0.073\ln\left(\f{E/\mu}{1+\gamma_1Z^{2/3} E/\mu}\right)-0.26} {0.058
\ln\left(\f{E/\mu}{1+\gamma_2Z^{1/3} E/\mu}\right)-0.14} \eec
\bal
\gamma_1=1.95\ 10^{-5} \quad &\mbox{and}& \quad \gamma_2=5.3\ 10^{-5} \quad \mbox{for} \quad Z\ne1\\
\gamma_1=4.4\ 10^{-5} \quad &\mbox{and}& \quad \gamma_2=4.8\ 10^{-5} \quad \mbox{for} \quad Z=1 \mbox{.}
\eal
Integration limits for this cross section are
\bal \f{4m}{E}&=& v_{min}\le v\le
v_{max}=1-\f{3\sqrt{e}}{4}\f{\mu}{E}Z^{1/3}
\nonumber\\
0&\le&|\rho|\le\rho_{max}=\sqrt{1-\f{4m}{Ev}}\left[ 1-\f{6\mu^2}{E^2(1-v)}
\right]\nonumber \eal
Muon pair production is discussed in detail in \cite{mupair} and is not considered by MMC. Its cross section is estimated to be $\sim 2 \cdot 10^4$ times smaller than the direct electron pair production cross section discussed above.

\subsection{Landau-Pomeranchuk-Migdal and Ter-Mikaelian effects}
\label{lpm}
These affect brems\-strahlung and pair production. See Figure \ref{mmc_fig_21} for the combined effect in ice and Fr\'{e}jus rock.
\subsubsection[LPM suppression of the bremsstrahlung cross section]{LPM suppression of the bremsstrahlung cross section:}
The bremsstrahlung cross section is modified as follows \cite{klein,migdal,polityko,polityko2}:
$$ {4 \over 3}(1-v)+v^2 \rightarrow {\xi(s) \over 3} \left( v^2 G(s)+2[1+(1-v)^2]\phi(s)\right) \mbox{.}$$
The regions of the following expressions for $\phi(s)$ and $G(s)$ were chosen to represent the best continuous approximation to the actual functions:
\bal
\phi(s)&=&1-\exp \left(-6s\left[1+(3-\pi)s\right]+{s^3 \over 0.623+0.796s+0.658s^2}\right) \quad \mbox{for} \quad s<1.54954\\
\phi(s)&=&1-0.012/s^4 \quad \mbox{for} \quad s\ge 1.54954\\
\psi(s)&=&1-\exp \left(-4s-{8s^2 \over 1+3.936s+4.97s^2-0.05s^3+7.50s^4}\right)\\
G(s)&=&3\psi(s)-2\phi(s) \quad \mbox{for} \quad s<0.710390\\
G(s)&=&36s^2/(36s^2+1) \quad \mbox{for} \quad 0.710390 \le s <0.904912\\
G(s)&=&1-0.022/s^4 \quad \mbox{for} \quad s \ge 0.904912 \mbox{.}
\eal
Here the SEB scheme \cite{stanev} is employed for evaluation of $\phi(s)$, $\psi(s)$, and $\xi(s)$ below:
\bal
\xi(s^{\prime})&=&2 \quad \mbox{for} \quad s^{\prime}<s_1 \\
\xi(s^{\prime})&=&1+h-{0.08(1-h)[1-(1-h)^2] \over \ln s_1} \quad \mbox{for} \quad s_1\le s^{\prime}<1\\
\xi(s^{\prime})&=&1 \quad \mbox{for} \quad s^{\prime}\ge1
\eal
$$ E_{LPM}={\alpha (\mu c^2)^2 X_0 \over 4 \pi \hbar c} \mbox{.} $$
$X_0$ is the same as in Section \ref{scat}. Here are the rest of the definitions:
$$s={s^{\prime} \over \sqrt{\xi}} \quad
s_1=\sqrt{2} {Z^{1/3} D_n \over B}{m_e \over \mu} \quad
s^{\prime}=\sqrt{{E_{LPM} v \over 8 E (1-v)}} \quad
h={\ln s^{\prime} \over \ln s_1} \mbox{.} $$

\subsubsection[Dielectric (Longitudinal) suppression effect]{Dielectric (Longitudinal) suppression effect:}
In addition to the above change of the brems\-strahlung cross section, s is replaced by $\Gamma \cdot s$ and functions $\xi(s)$, $\phi(s)$, and $G(s)$ are scaled as \cite{polityko}
$$ \xi(s) \rightarrow \xi(\Gamma s) \quad \phi(s) \rightarrow \phi(\Gamma s)/\Gamma \quad G(s) \rightarrow G(\Gamma s)/\Gamma^2 \mbox{.} $$
Therefore the first formula in the previous section is modified as
$$ {4 \over 3}(1-v)+v^2 \rightarrow {\xi(\Gamma s) \over 3} \left( v^2 {G(\Gamma s) \over \Gamma^2} +2[1+(1-v)^2]{\phi(\Gamma s) \over \Gamma} \right) \mbox{.} $$
$$\Gamma \quad \mbox{is defined as} \quad \Gamma=1+\gamma^2 \left( {\hbar \omega_p \over v E} \right)^2 , $$
where $\omega_p=\sqrt{4\pi NZe^2/m}$ is the plasma frequency of the medium and $vE$ is the photon energy.
The dielectric suppression affects only processes with small photon transfer energy, therefore it is not directly applicable to the direct pair production suppression.

\subsubsection[LPM suppression of the direct pair production cross section]{LPM suppression of the direct pair production cross section:}
$\Phi_e$ from the pair production cross section is modified as follows \cite{polityko,ternovskii}:
$$\Phi_e \rightarrow \left( (1+\beta)(A+[1+\rho^2]B)+\beta(C+[1+\rho^2]D)+(1-\rho^2)E \right) \cdot L_e $$
$$ s={1\over 4} \sqrt{{E_{LPM}\over E_{\mu}} {1\over v(1-\rho^2)}} \mbox{.} $$
The $E_{LPM}$ energy definition is different than in the bremsstrahlung case:
$$ E_{LPM}={\mu^4\over 2\pi n \alpha^2 \sum Z^2 L}, \quad \mbox{where} \quad L=\ln(3.25BZ^{-1/3}) \mbox{.} $$
Functions $A(s, \xi)$, $B(s, \xi)$, $C(s, \xi)$, and $D(s, \xi)$ are based on the approximation formulae
$$\Phi(s)={6s\over 6s+1}  \quad \mbox{and} \quad G(s)={(6s)^2\over (6s)^2+1} $$
and are given below:
$$A(s, x)={G\over 2}(1+2Gx)\ln{36s^2(1+x)^2+1\over 36s^2x^2}-G$$ $$+6Gs\left(1+{36s^2-1\over 36s^2+1}x\right)\left(\arctan(6s[x+1])-{\pi \over 2}\right)$$
$$B(s, x)=\Phi(1+\Phi x)\ln{6s(1+x)+1\over 6sx}-\Phi$$
$$C(s, x)=-G^2x\ln{36s^2(1+x)^2+1\over 36s^2x^2}+G-{G^2(36s^2-1)\over 6s}x\left(\arctan(6s[x+1])-{\pi \over 2}\right)$$
$$D(s, x)=\Phi-\Phi^2x\ln{6s(1+x)+1\over 6sx}$$
$$E(s, x)=-6s\left(\arctan(6s[x+1])-{\pi \over 2}\right) \mbox{.} $$

\subsection{Muon and tau decay}
\label{decay}
Muon decay probability is calculated according to 
$${dN \over dx}={1 \over {\gamma \beta c \tau}} \mbox{.}$$
The energy of the outgoing electron is evaluated as
$$\nu_e=\gamma \left(\nu_{rest}+\beta\sqrt{\nu_{rest}^2-m_e^2} \cos(\theta) \right) \mbox{.}$$
The value of $\cos(\theta)$ is distributed uniformly on $(-1,1)$ and
$\nu_{rest}$ is determined at random from the distribution
$$ {dN \over dx}={G^2 \mu^5 \over 192 \pi^3} (3-2x)x^2
\mbox{,}\quad x={\nu \over \nu_{max}}
\quad\mbox{with} \quad \nu_{min}=m_e \quad \mbox{and} \quad
\nu_{max}={\mu^2+m_e^2 \over 2 \mu} \mbox{.} $$

Tau leptonic decays, into a muon (17.37\%) and electron (17.83\%), are treated similarily. Hardronic decays are approximated by two-body decays into a neutrino and a hardonic part, which is assumed to be one of the particles or resonances: $\pi$ (11.09\%), $\rho$-770 (25.40\%, $M=769.3$ MeV), $a_1$-1260 (18.26\%, $M=1230$ MeV), and the rest into $\rho$-1465 (10.05\%, $M=1465$ MeV). The energy of the hardronic part in the tau rest frame is evaluated as $\nu_{rest}=(m_{\tau}^2+M^2)/(2m_{\tau})$.

\begin{figure}[!h]\begin{center}
\begin{tabular}{ccc}
\mbox{\epsfig{file=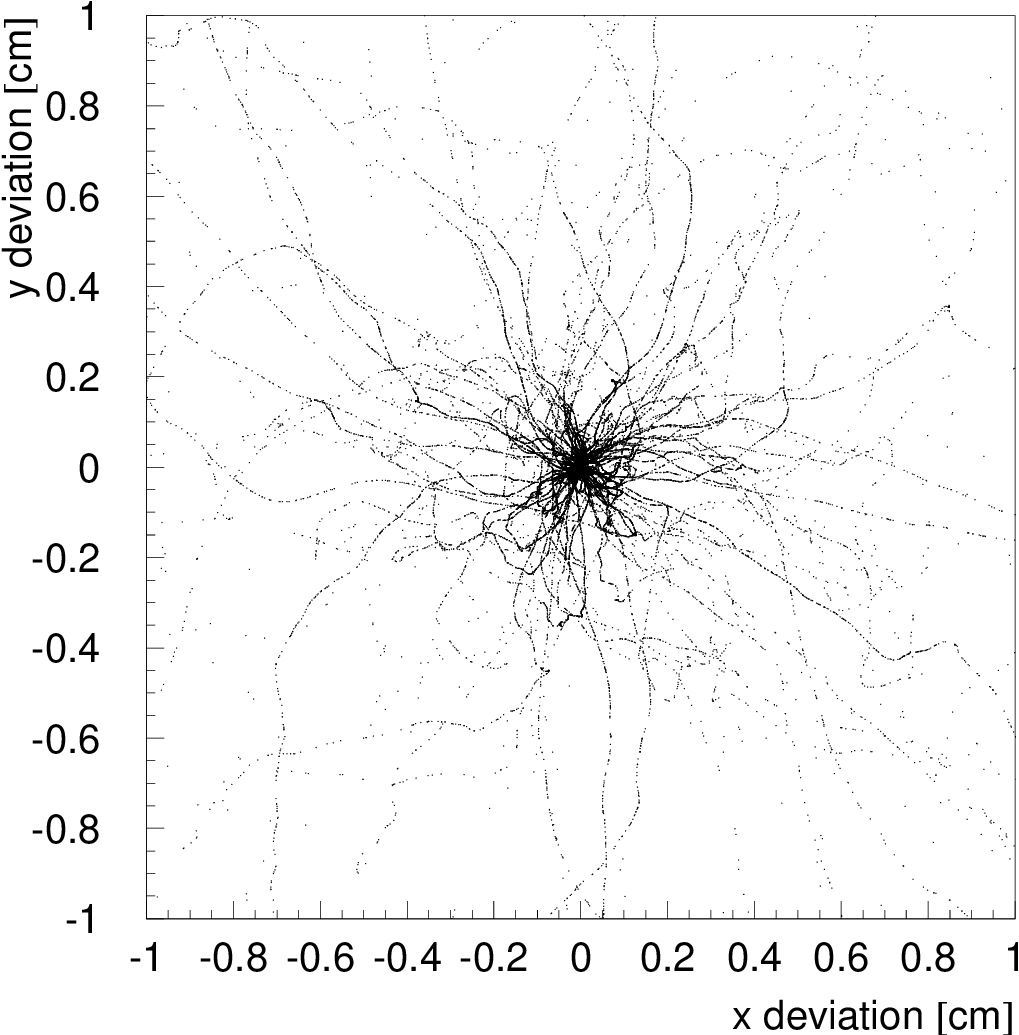,width=.45\textwidth}} & \ & \mbox{\epsfig{file=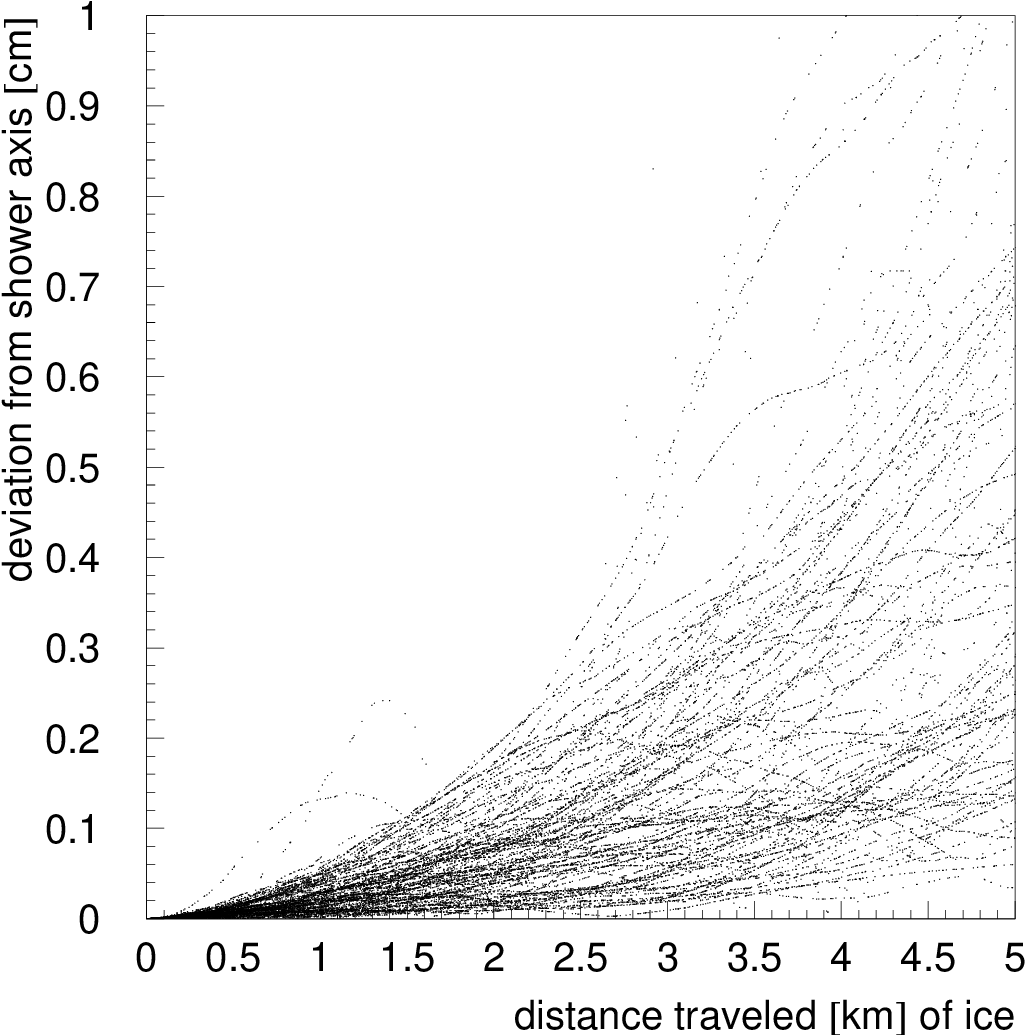,width=.45\textwidth}} \\
\end{tabular}
\parbox{.9\textwidth}{\caption[Moli\`{e}re scattering of one hundred 10 TeV muons going straight down through ice ]{\label{mmc_fig_22}Moli\`{e}re scattering of one hundred 10 TeV muons going straight down through ice }}
\end{center}\end{figure}

\subsection{Moli\`{e}re scattering}
\label{scat}
After passing through a distance x, the angular distribution is assumed Gaussian with a width $\sqrt{2} \theta_0$ \cite{pdb,lynch,mustafa}:
$$ \theta_0={13.6 MeV \over \beta c p} z \sqrt{x/X_0} \left[ 1+0.038 \ln(x/X_0) \right] $$
$$ X_0\quad \mbox{is evaluated as}\quad X_0=\left[ {\sigma_{brems}(E_{big}) \over E_{big}} \right]^{-1} \quad \mbox{for} \quad E_{big} \approx 10^{20} \mbox{eV} \mbox{.}$$
Deviations in two directions perpendicular to the muon track are independent, but for each direction the exit angle and lateral deviation are correlated:
$$ y_{plane}=z_1 x \theta_0/\sqrt{12} + z_2 x \theta_0/2 \quad \mbox{and} \quad \theta_{plane}=z_2 \theta_0 $$
\bce \mbox{\epsfig{file=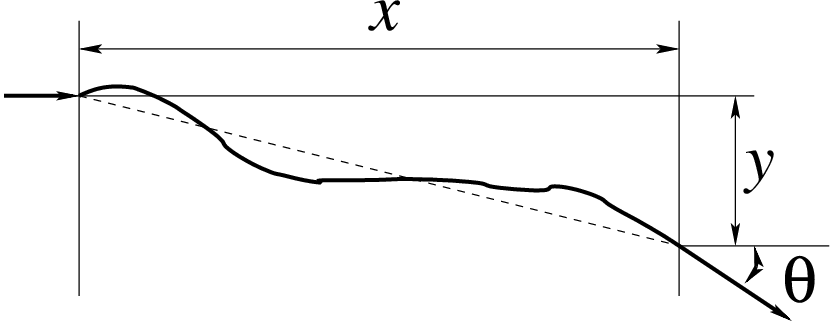,width=.5\textwidth}} \ece
for independent standard Gaussian random variables ($z_1$, $z_2$). A more precise treatment should take the finite size of the nucleus into account as described in \cite{scat}. See Figure \ref{mmc_fig_22} for an example of Moli\`{e}re scattering of a high energy muon.

\section{Conclusions}
A very versatile, clearly coded, and easy-to-use muon propagation Monte Carlo program (MMC) is presented. It is capable of propagating muon and tau leptons of energies from 105.7 MeV (muon rest mass, higher for tau) to $10^{11}$ GeV (or higher), which should be sufficient for the use as propagator in the simulations of the modern neutrino detectors. A very straightforward error control model is implemented, which results in computational errors being much smaller than uncertainties in the formulae used for evaluation of cross sections. It is very easy to ``plug in'' cross sections, modify them, or test their performance. The program was extended on many occasions to include new formulae or effects. MMC propagates particles in three dimensions and takes into account Moli\`{e}re scattering on the atomic centers, which could be considered as the zeroth order approximation to true muon scattering since bremsstrahlung and pair production are effects that appear on top of such scattering. A more advanced angular dependence of the cross sections can be implemented at a later date, if necessary.

Having been written in Java, MMC comes with the c/c++ interface package, which simplifies its integration into the simulation programs written in native languages. The distribution of MMC also includes a demonstration applet, which allows one to immediately visualize simulated events.

MMC was incorporated into the simulation of the AMANDA, IceCube, and Fr\'{e}jus experiments. It is distributed at \cite{mmc} with hope that the combination of precision, code clarity, speed, and stability will make it a useful tool in research where propagation of high energy particles through matter needs to be simulated.

A calculation of coefficients in the energy loss formula $dE/dx=a+bE$ and a similar formula for average range is presented for continuous (for energy loss) and stochastic (for average range calculation) energy loss treatments in ice and Fr\'{e}jus Rock. The calculated coefficients apply in the energy range from 20 GeV to $10^{11}$ GeV with an average deviation from the linear formula of 3.7\% and maximum of 6.6\%. Also, 99.9\% range of muons propagating in ice is estimated for energies from 1 GeV to $10^{11}$ GeV.

This work was supported by the German Academic Exchange Service (DAAD), U.S. NSF Grants OPP-020311 and OPP-0236449, and the U.S. DOE contract DE-AC-76SF00098.

\appendix

\section{Tables used by Muon Monte Carlo (MMC) }
\label{app_mmc}

\noindent All cross sections were translated to units $[1/\mbox{cm}]$ via multiplication by the number of molecules per unit volume. Many unit conversions (like eV $\rightarrow$ J) were achieved using values of $\alpha=e^2/\hbar c$ and $r_e=e^2/m_e c^2$.

\begin{center}
{\label{mmc_tab_4}Summary of physical constants employed by MMC (as in \cite{ephys})}
\vspace{3pt}
\nobreak
\begin{tabular}{|rl||rl|}
\hline
$\alpha$ & 1/137.03599976 & $r_e$ & $2.817940285\cdot 10^{-13}$ cm\\
\hline
\hline
$N_a$ & $6.02214199\cdot 10^{23}$ 1/mol & $K$ & 0.307075 $MeV\cdot cm^2/g$ \\
\hline
\hline
$c$ & $299792458\cdot 10^{10}$ cm/s & $R_y$ & 13.60569172 eV \\
\hline
\hline
$m_e$ & 0.510998902 MeV & $m_{\pi}$ & 139.57018 Mev \\
\hline
\hline
$m_p$ & 938.271998 MeV & $m_n$ & 939.56533 MeV \\
\hline
\hline
$m_{\mu}$ & 105.658389 MeV & $\tau_{\mu}$ & $2.19703\cdot 10^{-6}$ s \\
\hline
\hline
$m_{\tau}$ & 1777.03 MeV & $\tau_{\tau}$ & $290.6\cdot 10^{-15}$ s \\
\hline
\end{tabular}
\end{center}

\begin{sidewaysfigure}
\begin{center}
{\label{mmc_tab_5}Media constants (taken from \cite{cern85,tables})}
\vspace{3pt}
\nobreak
{\small
\begin{tabular}{|ccccccccccc|}
\hline
Material & $Z$ & $A$ & $I$, eV & $-C$ & $a$ & $m$ & $X_0$ & $X_1$ & $\rho$, $g/cm^2$ & $\delta_0$ \\
Water & 1 + & 1.00794 & 75.0 & 3.5017 & 0.09116 & 3.4773 & 0.2400 & 2.8004 & 1.000 & 0 \\
Ice & + 8 & 15.9994 & 75.0 & 3.5017 & 0.09116 & 3.4773 & 0.2400 & 2.8004 & 0.917 & 0 \\
Stand. Rock & 11 & 22 & 136.4 & 3.7738 & 0.08301 & 3.4120 & 0.0492 & 3.0549 & 2.650 & 0 \\
Fr\'{e}jus Rock & 10.12 & 20.34 & 149.0 & 5.053 & 0.078 & 3.645 & 0.288 & 3.196 & 2.740 & 0 \\
Iron & 26 & 55.845 & 286.0 & 4.2911 & 0.14680 & 2.9632 & -0.0012 & 3.1531 & 7.874 & 0.12 \\
Hydrogen & 1 & 1.00794 & 21.8 & 3.0977 & 0.13483 & 5.6249 & 0.4400 & 1.8856 & 0.07080 & 0 \\
Lead & 82 & 207.2 & 823.0 & 6.2018 & 0.09359 & 3.1608 & 0.3776 & 3.8073 & 11.350 & 0.14 \\
Uranium & 92 & 238.0289 & 890.0 & 5.8694 & 0.19677 & 2.8171 & 0.2260 & 3.3721 & 18.950 & 0.14 \\
\multicolumn{3}{|c}{\begin{tabular}{ccccc}
\multirow{3}{0.05in}{\begin{sideways}Air\end{sideways}} & N$_2$ & 78.1\% & 7 & 14.0067 \\
& O$_2$ & 21.0\% & 8 & 15.9994 \\
& Ar & 0.9\% & 18 & 39.948 \\
\end{tabular}
} & 85.7 & 10.5961 & 0.10914 & 3.3994 & 1.7418 & 4.2759 & \parbox{0.3in}{1.205 $\cdot 10^{-3}$} & 0 \\
\hline
\end{tabular}
}
\end{center}
\end{sidewaysfigure}

\begin{center}
{\label{mmc_tab_6}Radiation logarithm constant $B$ (taken from \cite{radlog})}
\vspace{3pt}
\nobreak
\begin{tabular}{|c||c||c||c||c|}
\hline
\begin{tabular}{cc}
$Z$ & $B$ \\
\hline
1 & 202.4 \\
2 & 151.9 \\
3 & 159.9 \\
4 & 172.3 \\
5 & 177.9 \\
6 & 178.3 \\
7 & 176.6 \\
\end{tabular} &
\begin{tabular}{cc}
$Z$ & $B$ \\
\hline
8 & 173.4 \\
9 & 170.0 \\
10 & 165.8 \\
11 & 165.8 \\
12 & 167.1 \\
13 & 169.1 \\
14 & 170.8 \\
\end{tabular} &
\begin{tabular}{cc}
$Z$ & $B$ \\
\hline
15 & 172.2 \\
16 & 173.4 \\
17 & 174.3 \\
18 & 174.8 \\
19 & 175.1 \\
20 & 175.6 \\
21 & 176.2 \\
\end{tabular} &
\begin{tabular}{cc}
$Z$ & $B$ \\
\hline
22 & 176.8 \\
26 & 175.8 \\
29 & 173.1 \\
32 & 173.0 \\
35 & 173.5 \\
42 & 175.9 \\
50 & 177.4 \\
\end{tabular} &
\begin{tabular}{cc}
$Z$ & $B$ \\
\hline
53 & 178.6 \\
74 & 177.6 \\
82 & 178.0 \\
92 & 179.8 \\
 & \\
other & 182.7 \\
 & \\
\end{tabular} \\
\hline
\end{tabular}
\end{center}

\begin{sidewaysfigure}
\begin{center}
{\label{mmc_tab_9}Parameterization coefficients of the hard component of the photonuclear cross section (as in \cite{bugaev_montaruli})}
\vspace{3pt}
\nobreak
{\small
\begin{tabular}{|c|c|c|c|c|c|c|c|}
\hline
$E$ & $10^3$ GeV & $10^4$ GeV & $10^5$ GeV & $10^6$ GeV & $10^7$ GeV & $10^8$ GeV & $10^9$ GeV \\
\hline
\multicolumn{8}{|c|}{muons} \\
\hline
$a_0$ & $7.174409 \cdot 10^{-4}$ & $1.7132 \cdot 10^{-3}$ & $4.082304 \cdot 10^{-3}$ & $8.628455 \cdot 10^{-3}$ & $0.01244159$ & $0.02204591$ & $0.03228755$ \\
$a_1$ & $-0.2436045$ & $-0.5756682$ & $-1.553973$ & $-3.251305$ & $-5.976818$ & $-9.495636$ & $-13.92918$ \\
$a_2$ & $-0.2942209$ & $-0.68615$ & $-2.004218$ & $-3.999623$ & $-6.855045$ & $-10.05705$ & $-14.37232$ \\
$a_3$ & $-0.1658391$ & $-0.3825223$ & $-1.207777$ & $-2.33175$ & $-3.88775$ & $-5.636636$ & $-8.418409$ \\
$a_4$ & $-0.05227727$ & $-0.1196482$ & $-0.4033373$ & $-0.7614046$ & $-1.270677$ & $-1.883845$ & $-2.948277$ \\
$a_5$ & $-9.328318 \cdot 10^{-3}$ & $-0.02124577$ & $-0.07555636$ & $-0.1402496$ & $-0.2370768$ & $-0.3614146$ & $-0.5819409$ \\
$a_6$ & $-8.751909 \cdot 10^{-4}$ & $-1.987841 \cdot 10^{-3}$ & $-7.399682 \cdot 10^{-3}$ & $-0.01354059$ & $-0.02325118$ & $-0.03629659$ & $-0.059275$ \\
$a_7$ & $-3.343145 \cdot 10^{-5}$ & $-7.584046 \cdot 10^{-5}$ & $-2.943396 \cdot 10^{-4}$ & $-5.3155 \cdot 10^{-4}$ & $-9.265136 \cdot 10^{-4}$ & $-1.473118 \cdot 10^{-3}$ & $-2.419946 \cdot 10^{-3}$ \\
\hline
\multicolumn{8}{|c|}{taus} \\
\hline
$a_0$ & $-1.269205 \cdot 10^{-4}$ & $-2.843877 \cdot 10^{-4}$ & $-5.761546 \cdot 10^{-4}$ & $-1.195445 \cdot 10^{-3}$ & $-1.317386 \cdot 10^{-3}$ & $-9.689228 \cdot 10^{-15}$ & $-6.4595 \cdot 10^{-15}$ \\
$a_1$ & $-0.01563032$ & $-0.03589573$ & $-0.07768545$ & $-0.157375$ & $-0.2720009$ & $-0.4186136$ & $-0.8045046$ \\
$a_2$ & $0.04693954$ & $0.1162945$ & $0.3064255$ & $0.7041273$ & $1.440518$ & $2.533355$ & $3.217832$ \\
$a_3$ & $0.05338546$ & $0.130975$ & $0.3410341$ & $0.7529364$ & $1.425927$ & $2.284968$ & $2.5487$ \\
$a_4$ & $0.02240132$ & $0.05496$ & $0.144945$ & $0.3119032$ & $0.5576727$ & $0.8360727$ & $0.8085682$ \\
$a_5$ & $4.658909 \cdot 10^{-3}$ & $0.01146659$ & $0.03090286$ & $0.06514455$ & $0.1109868$ & $0.1589677$ & $0.1344223$ \\
$a_6$ & $4.822364 \cdot 10^{-4}$ & $1.193018 \cdot 10^{-3}$ & $3.302773 \cdot 10^{-3}$ & $6.843364 \cdot 10^{-3}$ & $0.011191$ & $0.015614$ & $0.01173827$ \\
$a_7$ & $1.9837 \cdot 10^{-5}$ & $4.940182 \cdot 10^{-5}$ & $1.409573 \cdot 10^{-4}$ & $2.877909 \cdot 10^{-4}$ & $4.544877 \cdot 10^{-4}$ & $6.280818 \cdot 10^{-4}$ & $4.281932 \cdot 10^{-4}$ \\
\hline
\end{tabular}
}
\end{center}
\end{sidewaysfigure}

\begin{center}
{\label{mmc_tab_7}ALLM (`91) parameters (as in \cite{allm,allm3})}
\vspace{3pt}
\nobreak
\begin{tabular}{|cc||cc||cc|}
\hline
$a_{P1}$ & -0.04503 & $a_{P2}$ & -0.36407 & $a_{P3}$ & 8.17091 \\
$a_{R1}$ & 0.60408 & $a_{R2}$ & 0.17353 & $a_{R3}$ & 1.61812 \\
$b_{P1}$ & $0.49222^2$ & $b_{P2}$ & $0.52116^2$ & $b_{P3}$ & 3.55115 \\
$b_{R1}$ & $1.26066^2$ & $b_{R2}$ & $1.83624^2$ & $b_{R3}$ & 0.81141 \\
$c_{P1}$ & 0.26550 & $c_{P2}$ & 0.04856 & $c_{P3}$ & 1.04682 \\
$c_{R1}$ & 0.67639 & $c_{R2}$ & 0.49027 & $c_{R3}$ & 2.66275 \\
$m^2_P$ & $10.67564\cdot 10^6$ MeV$^2$ & $\Lambda^2$ & $0.06527\cdot 10^6$ MeV$^2$ & $m^2_0$ & $0.30508\cdot 10^6$ MeV$^2$ \\
$m^2_R$ & $0.20623\cdot 10^6$ MeV$^2$ & $Q^2_0-\Lambda^2$ & $0.27799\cdot 10^6$ MeV$^2$ & & \\
\hline
\end{tabular}
\end{center}

\begin{center}
{\label{mmc_tab_8}ALLM (`97) parameters (as in \cite{allm2,allm3})}
\vspace{3pt}
\nobreak
\begin{tabular}{|cc||cc||cc|}
\hline
$a_{P1}$ & -0.0808 & $a_{P2}$ & -0.44812 & $a_{P3}$ & 1.1709 \\
$a_{R1}$ & 0.58400 & $a_{R2}$ & 0.37888 & $a_{R3}$ & 2.6063 \\
$b_{P1}$ & $0.60243^2$ & $b_{P2}$ & $1.3754^2$ & $b_{P3}$ & 1.8439 \\
$b_{R1}$ & $0.10711^2$ & $b_{R2}$ & $1.9386^2$ & $b_{R3}$ & 0.49338 \\
$c_{P1}$ & 0.28067 & $c_{P2}$ & 0.22291 & $c_{P3}$ & 2.1979 \\
$c_{R1}$ & 0.80107 & $c_{R2}$ & 0.97307 & $c_{R3}$ & 3.4942 \\
$m^2_P$ & $49.457\cdot 10^6$ MeV$^2$ & $\Lambda^2$ & $0.06527\cdot 10^6$ MeV$^2$ & $m^2_0$ & $0.31985\cdot 10^6$ MeV$^2$ \\
$m^2_R$ & $0.15052\cdot 10^6$ MeV$^2$ & $Q^2_0-\Lambda^2$ & $0.46017\cdot 10^6$ MeV$^2$ & & \\
\hline
\end{tabular}
\end{center}

\begin{center}
{\label{mmc_tab_8}CKMT parameters of the Butkevich-Mikhailov parameterization (as in \cite{butmikh,butmikh2})}
\vspace{3pt}
\nobreak
\begin{tabular}{|c||c|}
\hline
\begin{tabular}{cc}
$a$ & $0.2513\cdot 10^6 \mbox{ MeV}^2$ \\
$b$ & $0.6186\cdot 10^6 \mbox{ MeV}^2$ \\
$c$ & $3.0292\cdot 10^6 \mbox{ MeV}^2$ \\
$d$ & $1.4817\cdot 10^6 \mbox{ MeV}^2$ \\
$A_s$ & $0.12$ \\
\end{tabular} &
\begin{tabular}{cc}
$\Delta_0$ & $0.0988$ \\
$\alpha_R$ & $0.4056$ \\
$\tau$ & $1.8152$ \\
$B_u$ & $1.2437$ \\
$B_d$ & $0.1853$ \\
\end{tabular} \\
\hline
\end{tabular}
\end{center}

\begin{center}
{\label{mmc_tab_10}Tsai's Radiation logarithms $L_{rad}$ and $L^{\prime}_{rad}$(as in \cite{pdb,tsai})}
\vspace{3pt}
\nobreak
\begin{tabular}{|ccc|}
\hline
Z & $L_{rad}$ & $L^{\prime}_{rad}$ \\
\hline
1 & 5.31 & 6.144 \\
2 & 4.79 & 5.621 \\
3 & 4.74 & 5.805 \\
4 & 4.71 & 5.924 \\
other & $\ln(184.15Z^{-1/3})$ & $\ln(1194Z^{-2/3})$ \\
\hline
\end{tabular}
\end{center}

\vspace{0.2cm}

\section[Comparison of Spectra of Secondaries Produced with MMC, MUM, LOH, and LIP]{Comparison of Spectra of Secondaries Produced with MMC,\\ MUM \cite{mum}, LOH \cite{cern85}, and LIP \cite{lip,propmu}}
\label{app_mmc2}

\begin{figure}[!h]\begin{center}
\caption[Spectra of the secondaries: the setup]{\label{mmc3_fig_1}Spectra of the secondaries: the setup}

\begin{tabular}{c}
\mbox{\epsfig{file=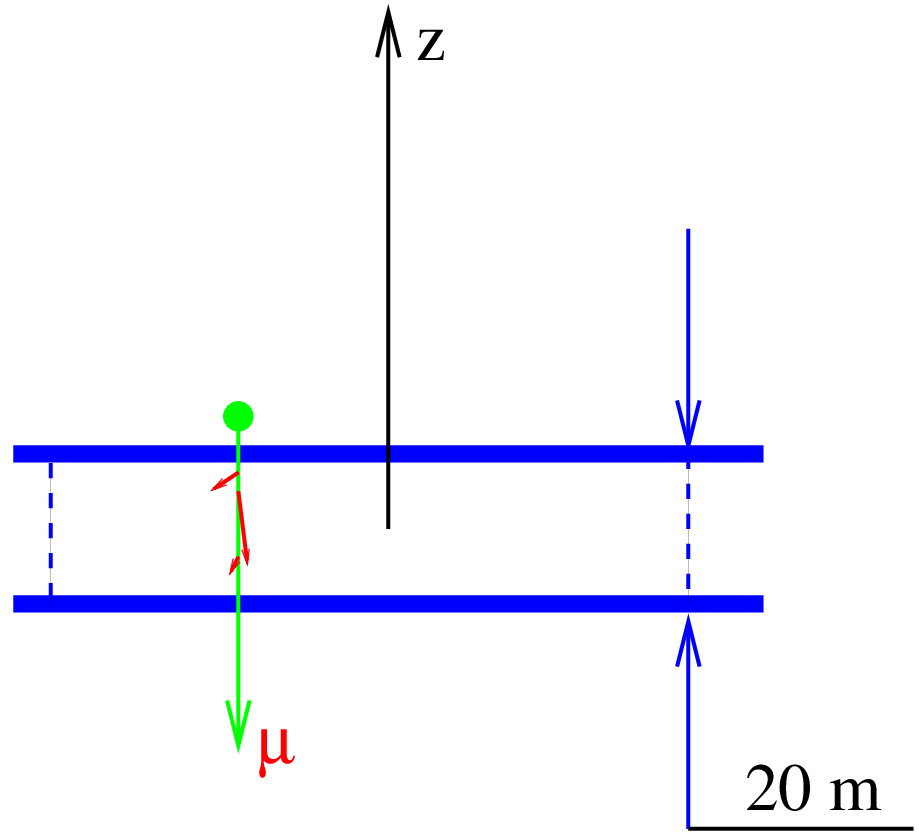,width=.45\textwidth}} \\
\end{tabular}
\end{center}\end{figure}

\subsection{Spectra of the secondaries}

In order to determine spectra of primaries consistently for all programs, the following setup was used. For each muon with fixed initial energy a first secondary created within the first 20 meters is recorded (Figure \ref{mmc3_fig_1}). This is somewhat different from what was done for Figure \ref{mmc_fig_7}, since the energy of the muon at the moment when the secondary is created is somewhat smaller than the initial energy due to continuous energy losses. These are smaller when $v_{cut}$ is smaller, and are generally negligible for all cases considered below.

\begin{figure}[!h]\begin{center}
\begin{tabular}{ccccc}
\mbox{\epsfig{file=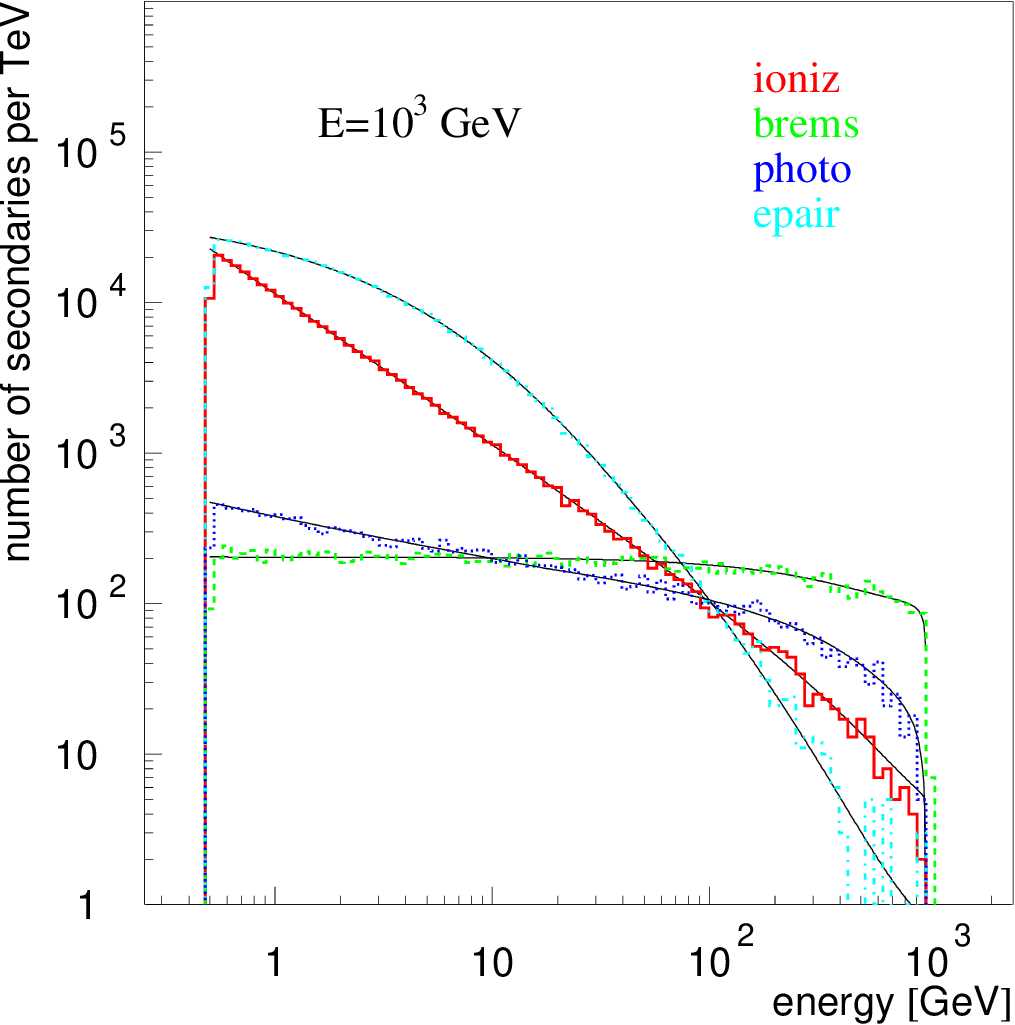,width=.30\textwidth}} & \ & \mbox{\epsfig{file=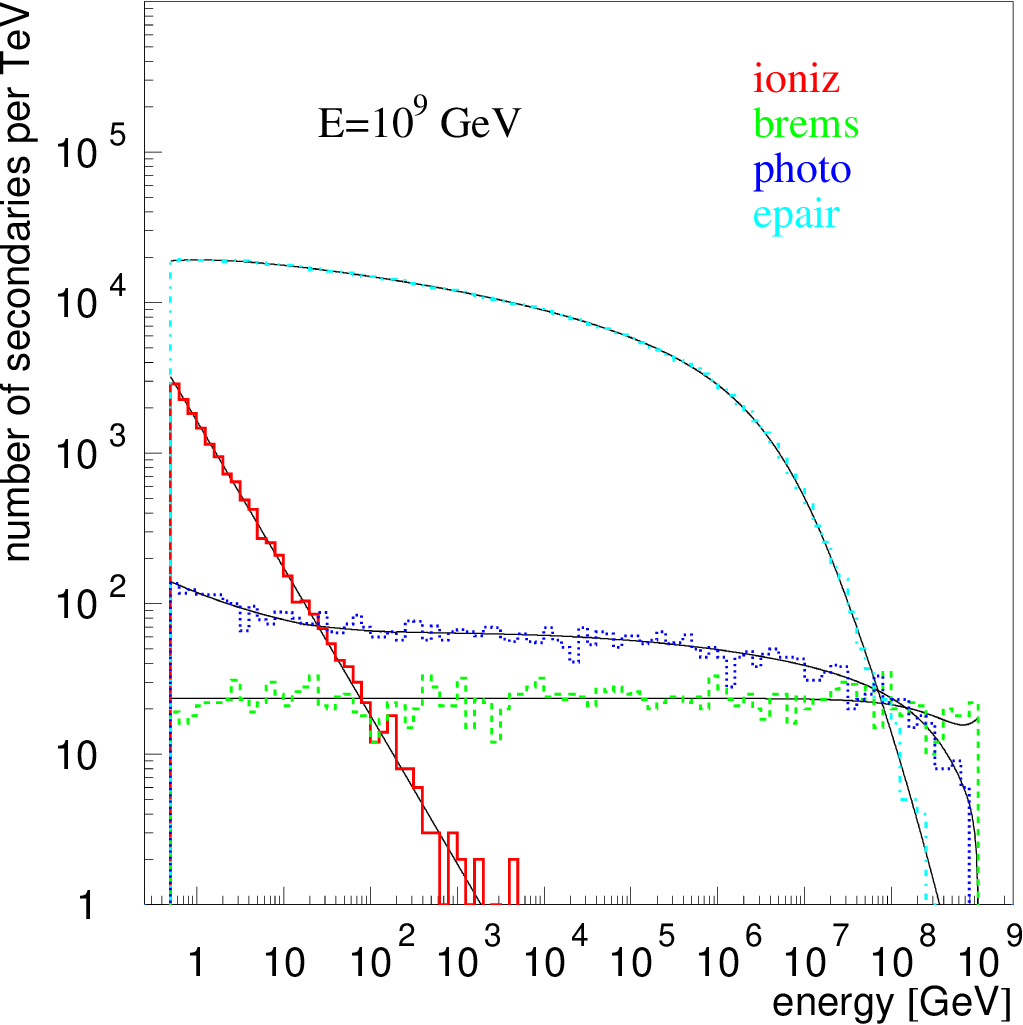,width=.30\textwidth}}  & \ & \mbox{\epsfig{file=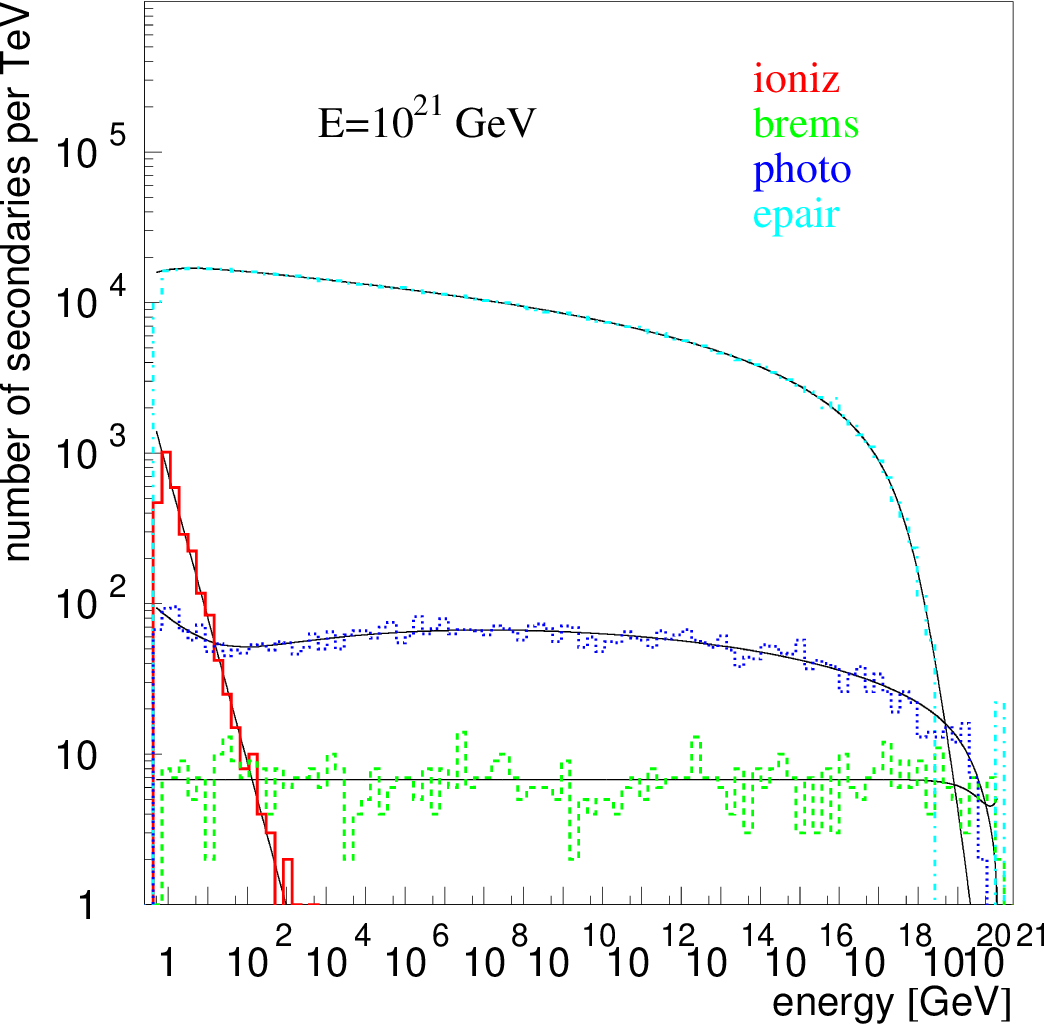,width=.30\textwidth}} \\
\end{tabular}
\parbox{.90\textwidth}{\caption[MMC: $E_{\mu}=10^3$ GeV, $E_{\mu}=10^9$ GeV, and  $E_{\mu}=10^{21}$ GeV with $E_{cut}=500$ MeV]{\label{mmc3_fig_2}MMC: $E_{\mu}=10^3$ GeV, $E_{\mu}=10^9$ GeV, and  $E_{\mu}=10^{21}$ GeV with $E_{cut}=500$ MeV}}
\end{center}\end{figure}

\begin{figure}[!h]\begin{center}
\begin{tabular}{ccccc}
\mbox{\epsfig{file=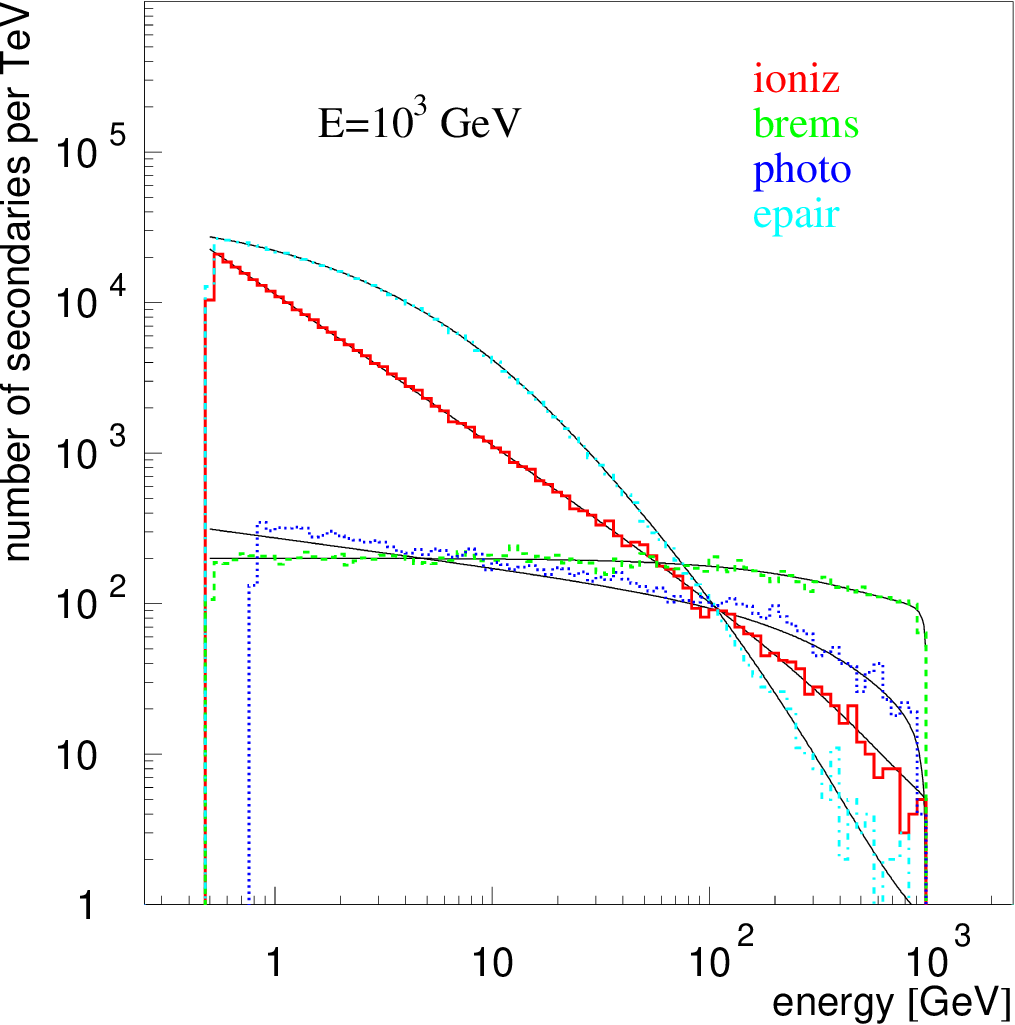,width=.30\textwidth}} & \ & \mbox{\epsfig{file=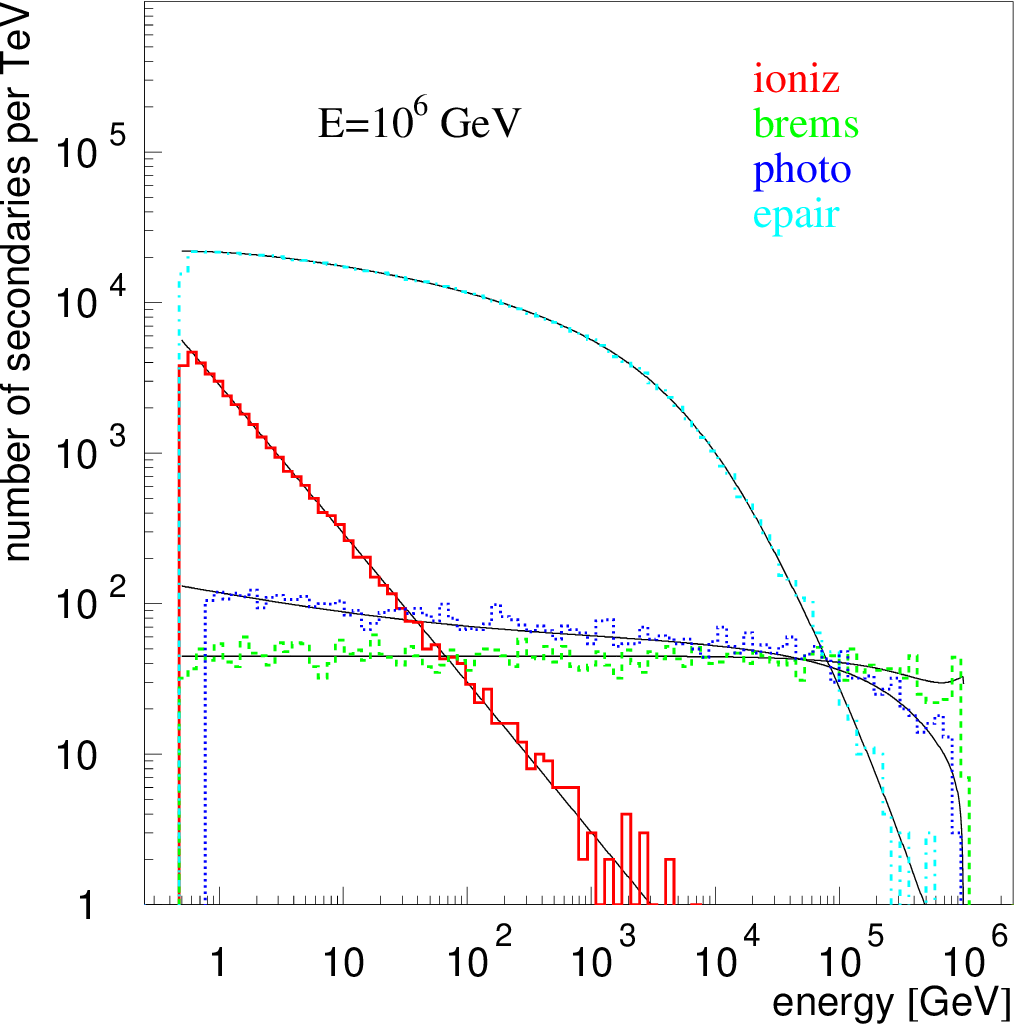,width=.30\textwidth}} & \ & \mbox{\epsfig{file=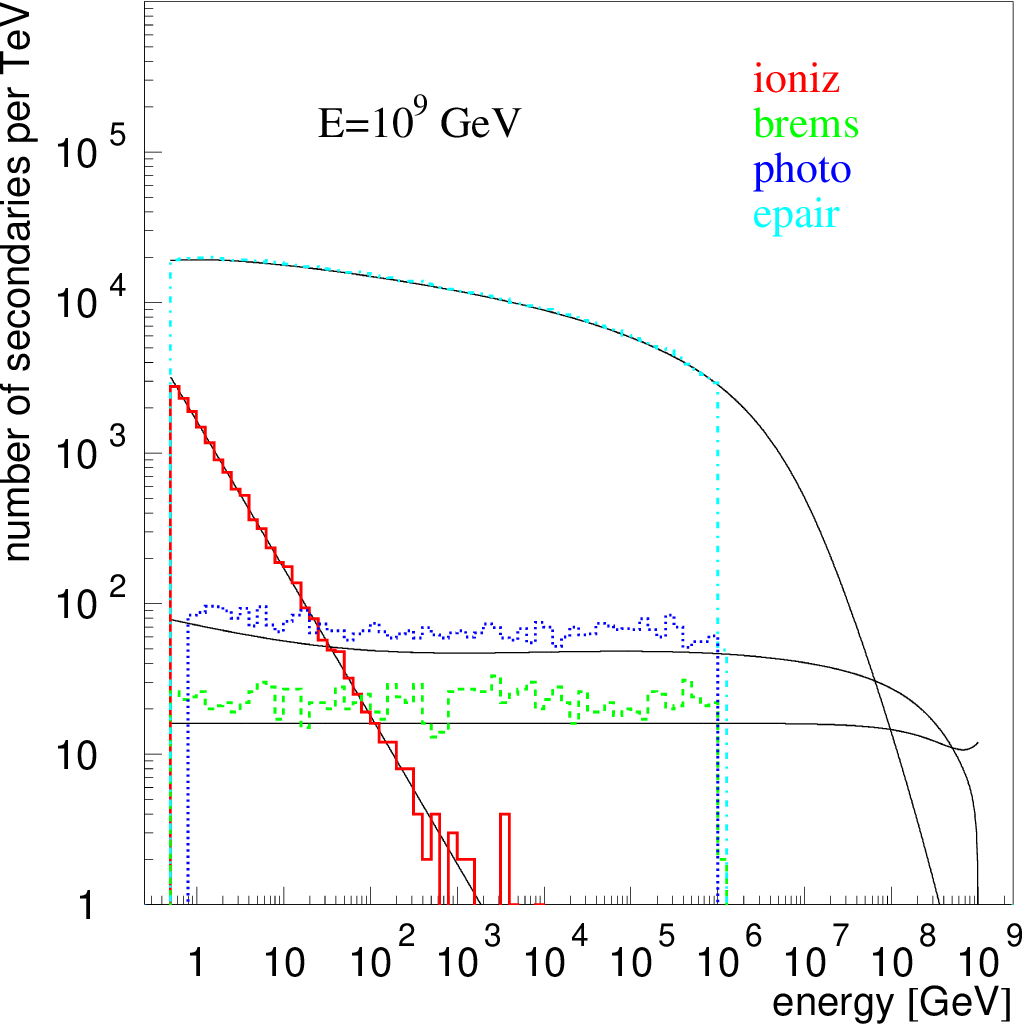,width=.30\textwidth}} \\
\end{tabular}
\parbox{.90\textwidth}{\caption[MUM: $E_{\mu}=10^3$ GeV, $E_{\mu}=10^6$ GeV, and $E_{\mu}=10^9$ GeV with $E_{cut}=500$ MeV]{\label{mmc3_fig_8}MUM: $E_{\mu}=10^3$ GeV, $E_{\mu}=10^6$ GeV, and $E_{\mu}=10^9$ GeV with $E_{cut}=500$ MeV}}
\end{center}\end{figure}

\begin{figure}[!h]\begin{center}
\begin{tabular}{ccccc}
\mbox{\epsfig{file=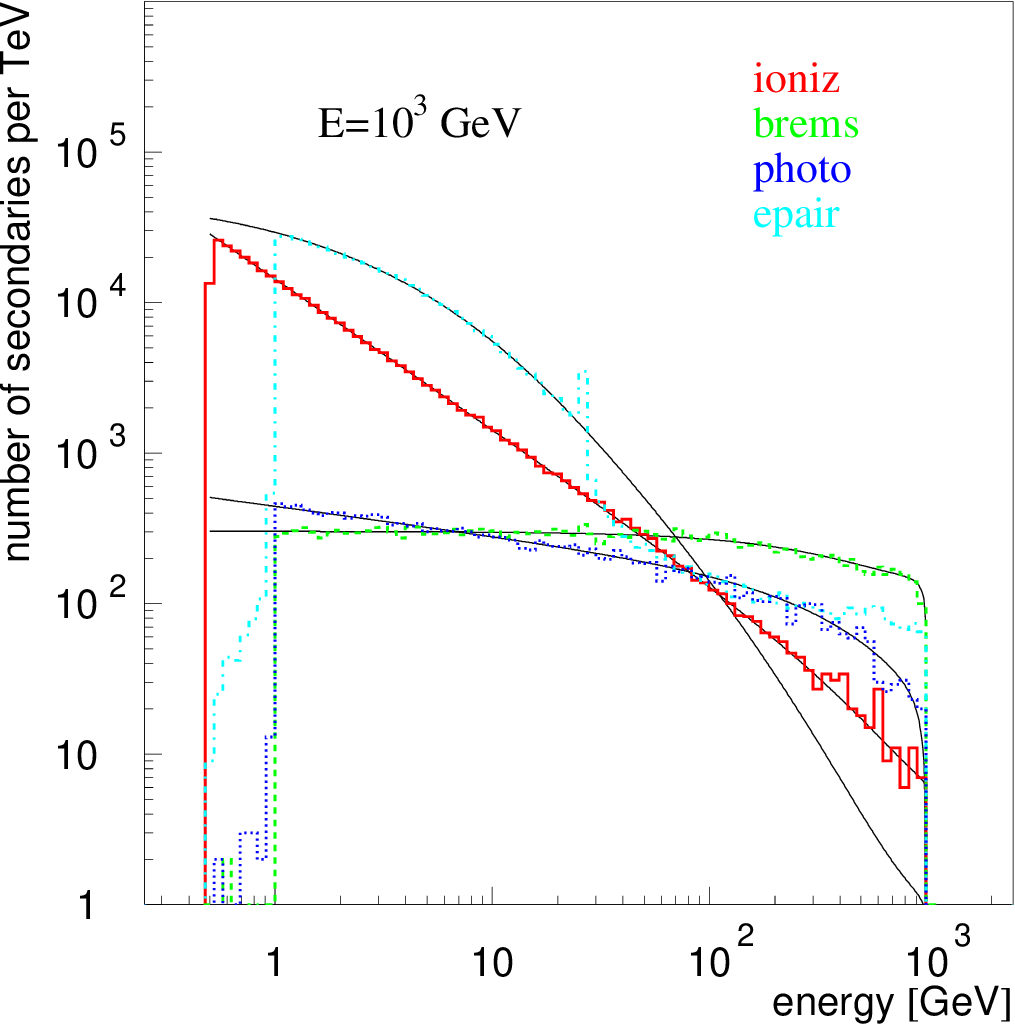,width=.30\textwidth}} & \ & \mbox{\epsfig{file=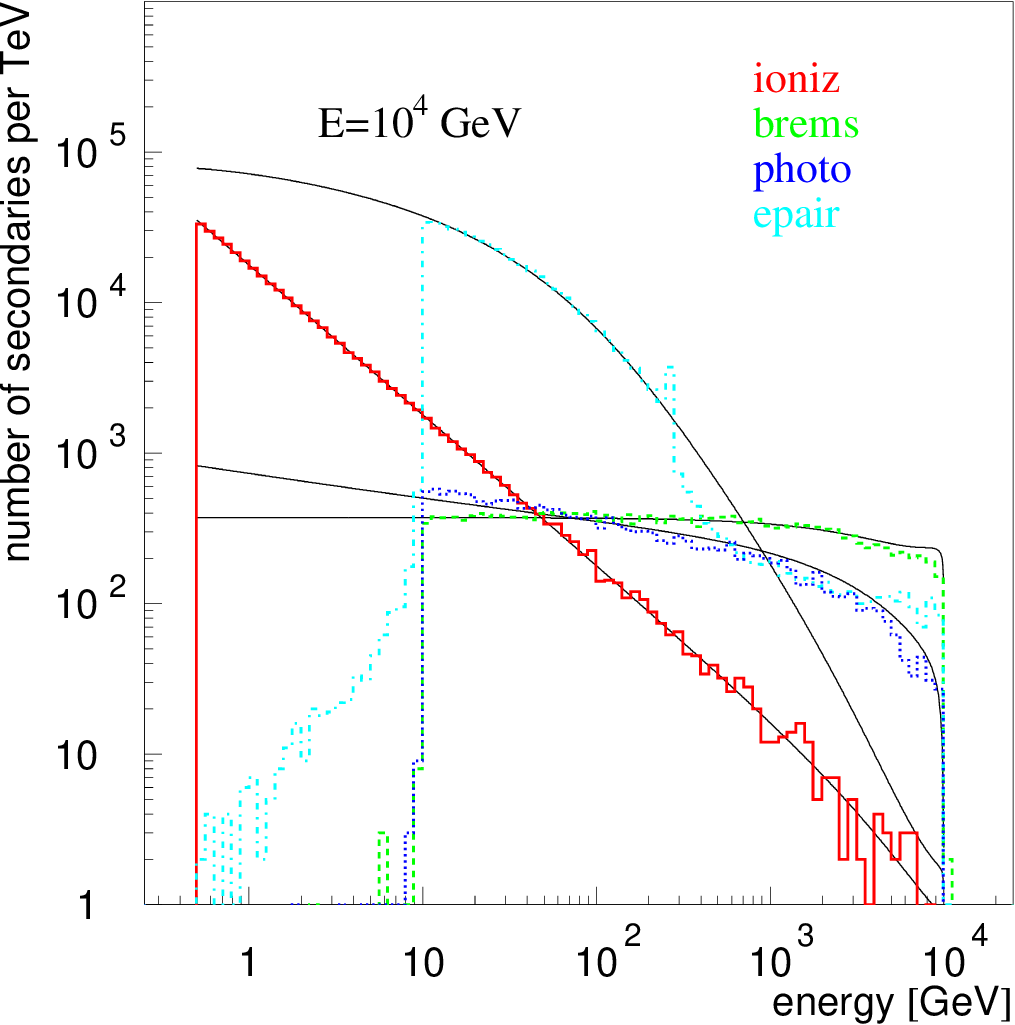,width=.30\textwidth}} & \ & \mbox{\epsfig{file=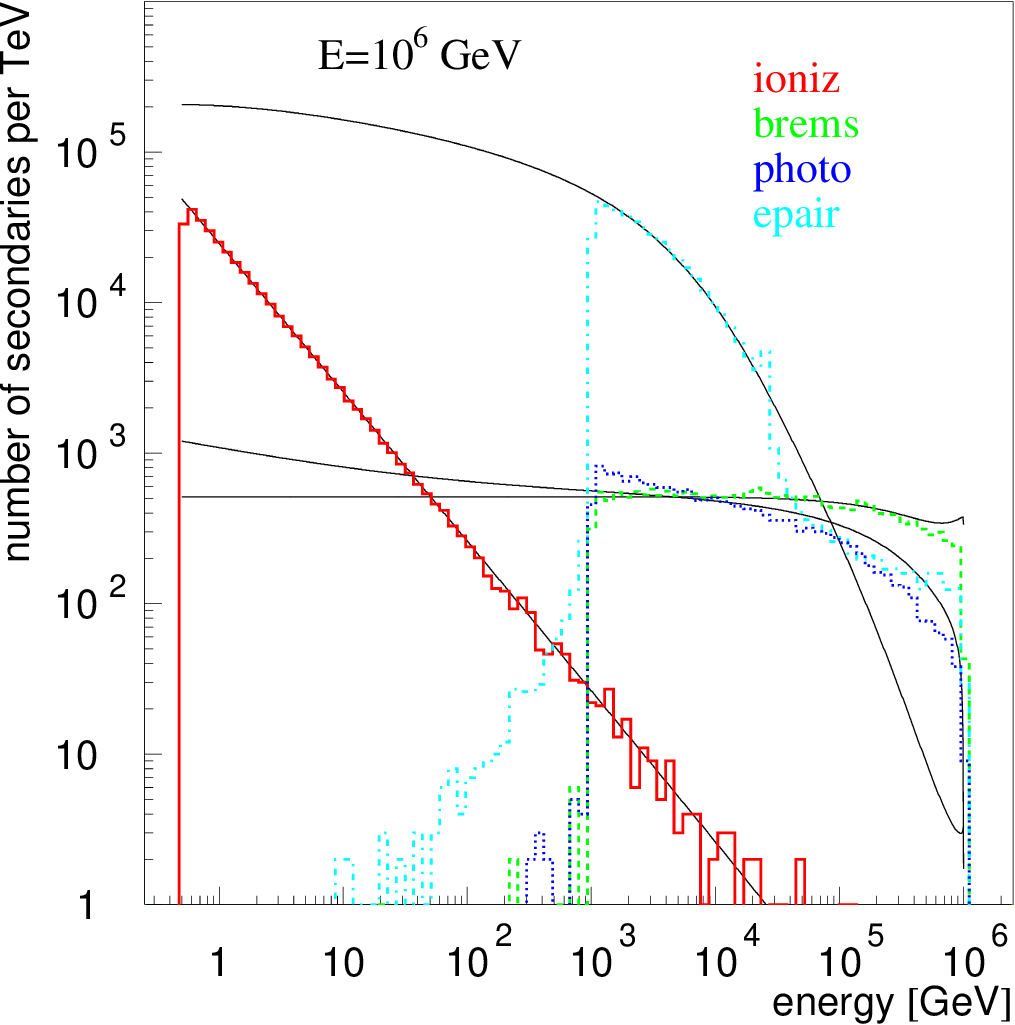,width=.30\textwidth}} \\
\end{tabular}
\parbox{.90\textwidth}{\caption[LOH: $E_{\mu}=10^3$ GeV, $E_{\mu}=10^4$ GeV, and $E_{\mu}=10^6$ GeV]{\label{mmc3_fig_12}LOH: $E_{\mu}=10^3$ GeV, $E_{\mu}=10^4$ GeV, and $E_{\mu}=10^6$ GeV}}
\end{center}\end{figure}

\begin{figure}[!h]\begin{center}
\begin{tabular}{ccccc}
\mbox{\epsfig{file=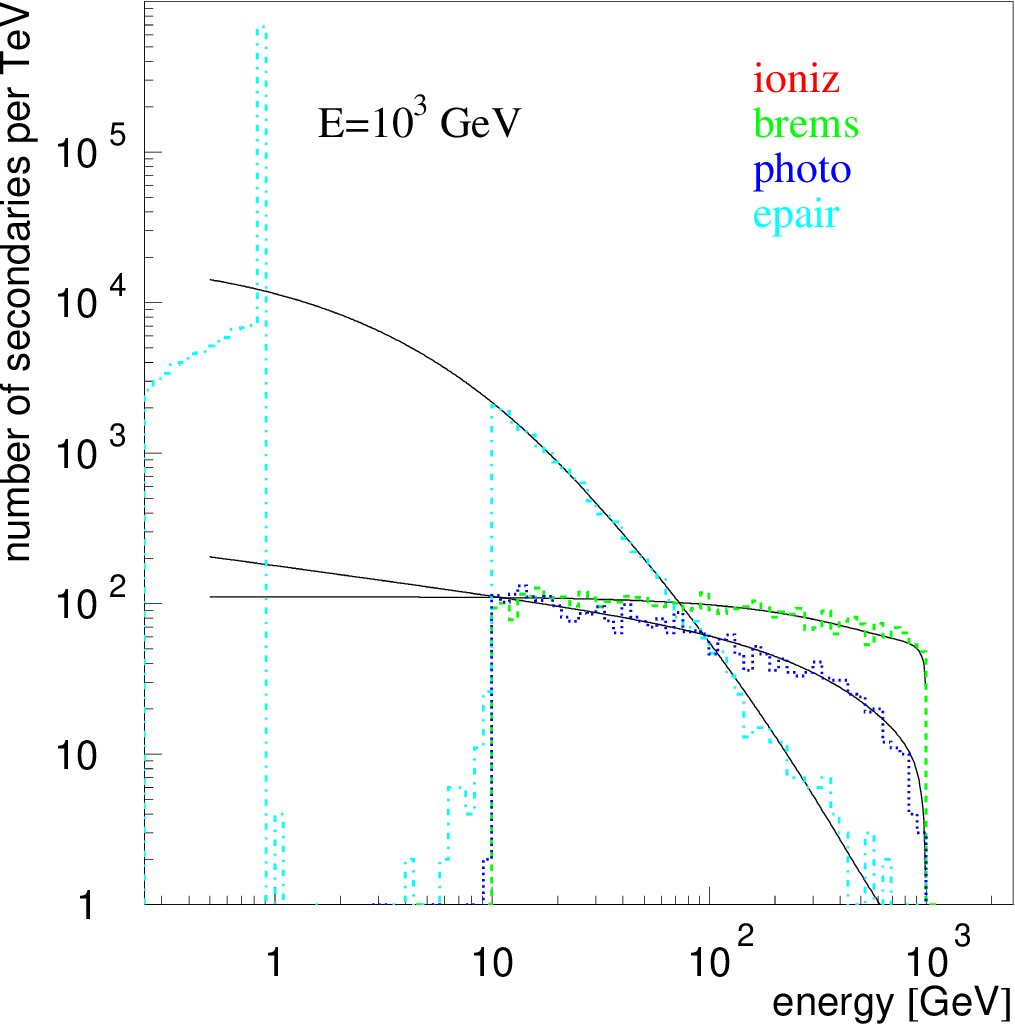,width=.30\textwidth}} & \ & \mbox{\epsfig{file=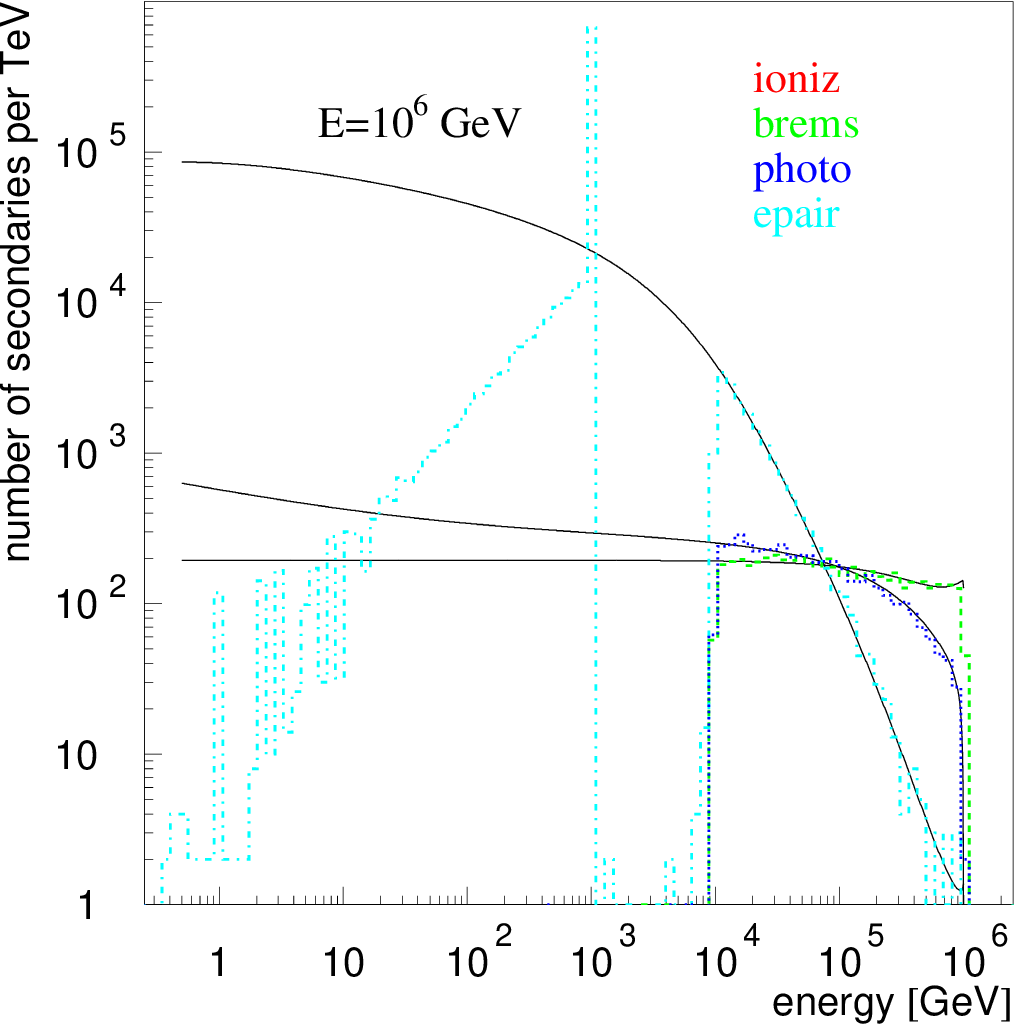,width=.30\textwidth}} & \ & \mbox{\epsfig{file=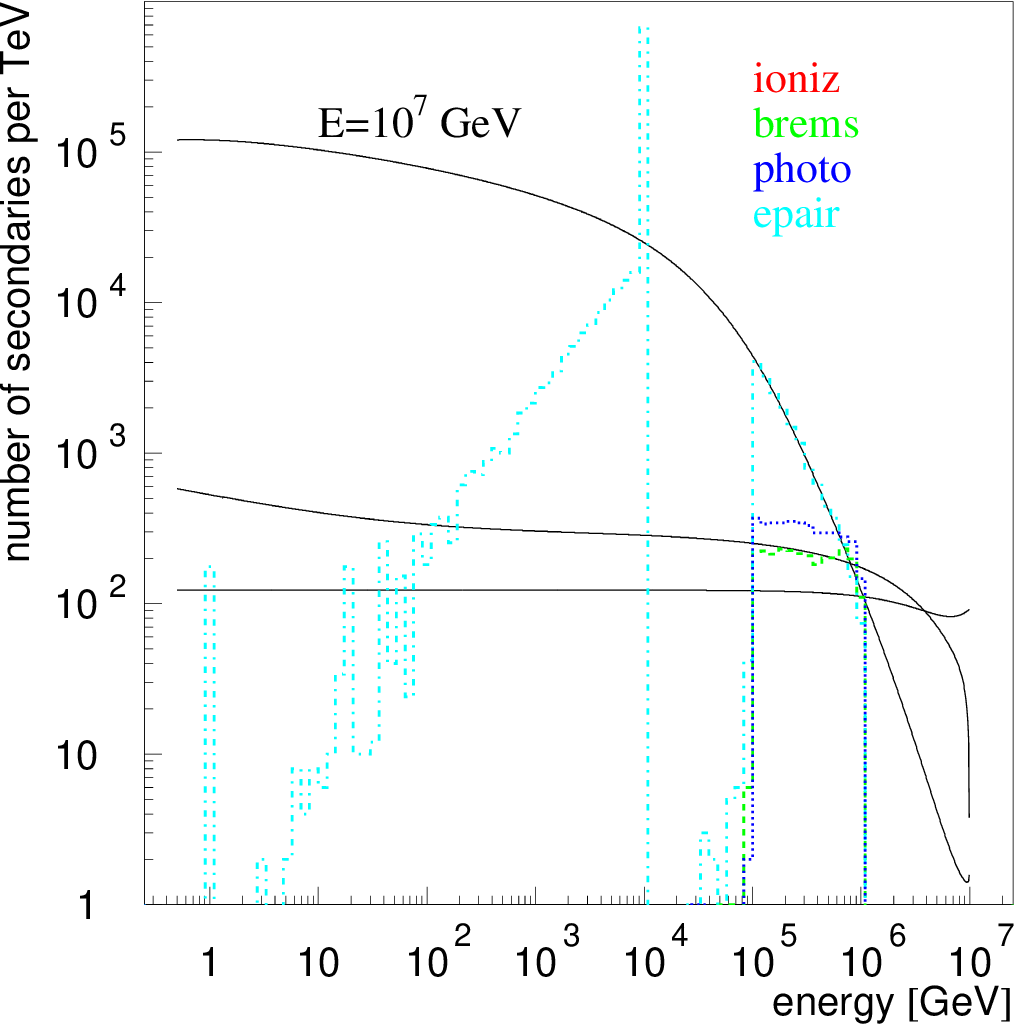,width=.30\textwidth}}\\
\end{tabular}
\parbox{.90\textwidth}{\caption[LIP: $E_{\mu}=10^3$ GeV, $E_{\mu}=10^6$ GeV, and $E_{\mu}=10^7$ GeV]{\label{mmc3_fig_16}LIP: $E_{\mu}=10^3$ GeV, $E_{\mu}=10^6$ GeV, and $E_{\mu}=10^7$ GeV}}
\end{center}\end{figure}

In Figure \ref{mmc3_fig_2} solid curves are probability functions normalized to the total number of secondaries above 500 MeV.
In Figure \ref{mmc3_fig_12} solid curves are probability functions normalized to the total number of secondaries above $10^{-3}\cdot E_{\mu}$.
In Figure \ref{mmc3_fig_16} solid curves are probability functions normalized to the total number of secondaries above $10^{-2}\cdot E_{\mu}$.
A setting of $E_{big}=10^{21}$ GeV is used for the third plot in Figure \ref{mmc3_fig_2} (default is $10^{11}$ GeV).

\subsection{Number and total energy of secondaries}

\begin{figure}[!h]
\begin{center}
\begin{tabular}{ccc}
\mbox{\epsfig{file=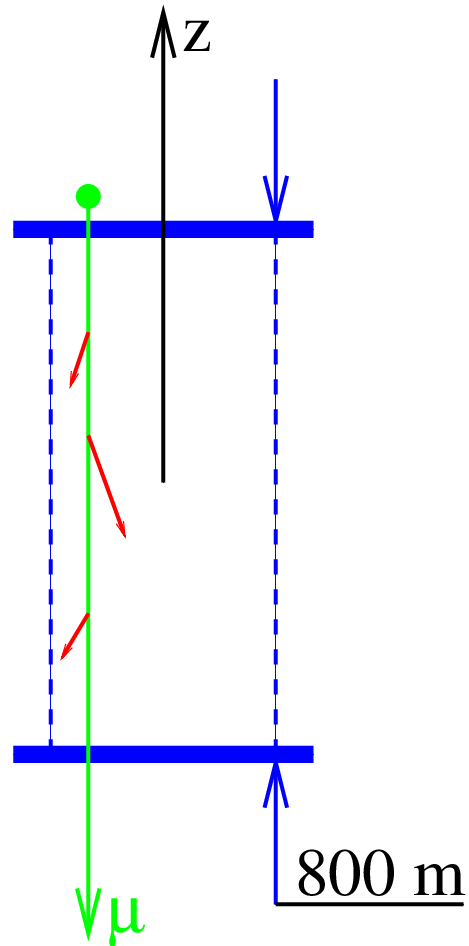,width=.22\textwidth}} & \ & \mbox{\epsfig{file=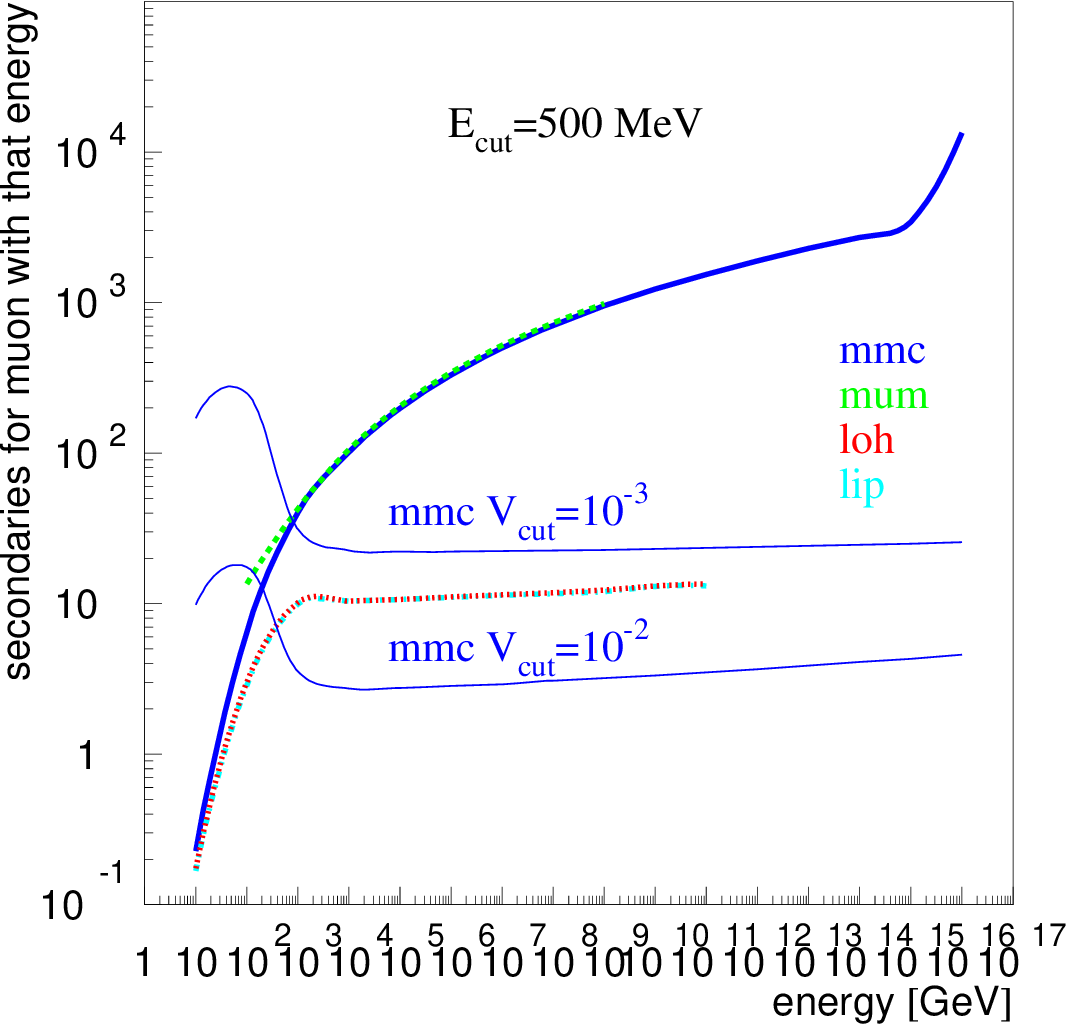,width=.45\textwidth}} \\
\parbox{.45\textwidth}{\caption[Number and total energy of secondaries: the setup]{\label{mmc3_fig_27}Number and total energy of secondaries: the setup}} & \ & \parbox{.45\textwidth}{\caption[Number of secondaries]{\label{mmc3_fig_28}Number of secondaries}} \\
\end{tabular}
\end{center}\end{figure}

\begin{figure}[!h]\begin{center}
\begin{tabular}{ccc}
\mbox{\epsfig{file=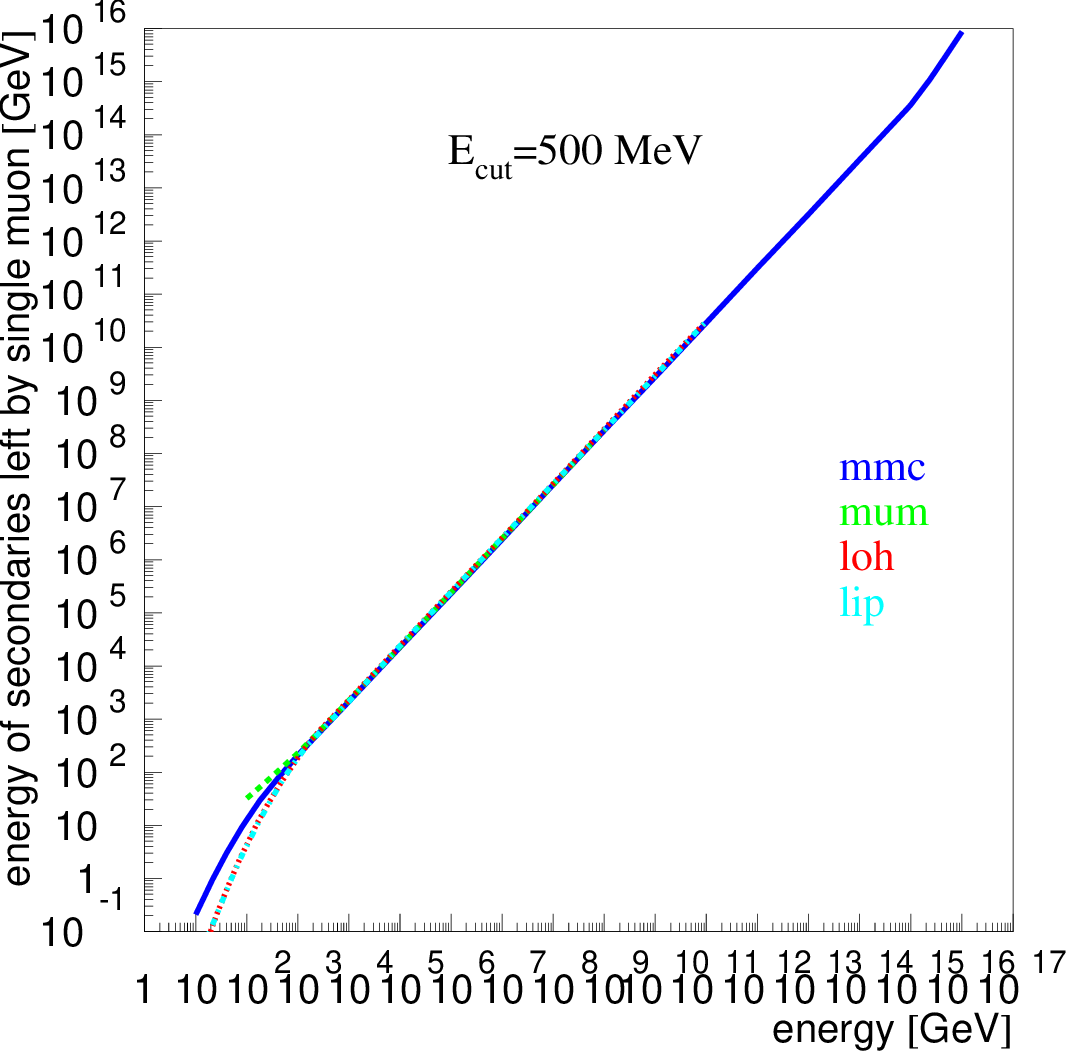,width=.45\textwidth}} & \ & \mbox{\epsfig{file=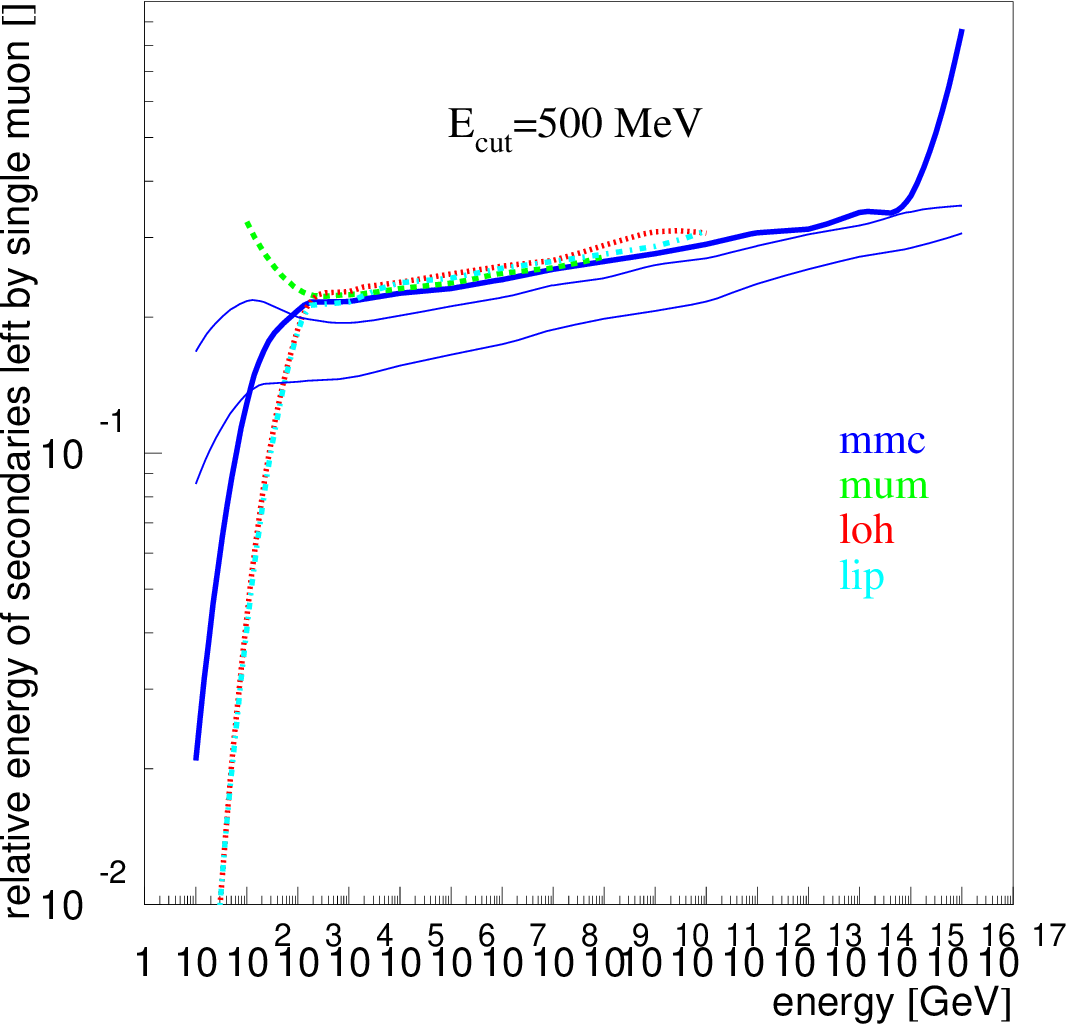,width=.45\textwidth}} \\
\parbox{.45\textwidth}{\caption[Total energy of secondaries]{\label{mmc3_fig_29}Total energy of secondaries}} & \ & \parbox{.45\textwidth}{\caption[Relative energy of secondaries]{\label{mmc3_fig_30}Relative energy of secondaries}} \\
\end{tabular}
\end{center}\end{figure}

In spite of the numerous problems with propagation codes other than MMC, shown in Figures \ref{mmc3_fig_8}$-$\ref{mmc3_fig_16}, it was possible to use these codes in the simulation of AMANDA-II. To understand why, the following setup is used. For each muon with fixed initial energy all secondaries created within the first 800 meters (equal to the height of the AMANDA-II detector) are recorded (Figure \ref{mmc3_fig_27}). Although the number of secondaries generated by propagators LOH and LIP is different from that generated by MMC or MUM (Figure \ref{mmc3_fig_28}), the total energy deposited in the volume of the detector is commensurable between all four propagators. The number of generated secondaries depends on the chosen value of $E_{cut}$ or $v_{cut}$. While MMC and MUM allow one to select this value, LOH and LIP have a built-in value which cannot be changed. From Figure \ref{mmc3_fig_28} it appears that these codes use a value of $v_{cut}$ which lies between $10^{-2}$ and $10^{-3}$ since their number of secondaries lies between that generated with MMC with $v_{cut}=10^{-2}$ and $v_{cut}=10^{-3}$. One would expect the total energy of secondaries generated with LOH or LIP to be somewhat lower than that generated with MMC or MUM with $E_{cut}=500$ MeV. This, however, is not true: the total energy of secondaries generated with LOH and LIP is somehow renormalized to match that of MMC and MUM (Figures \ref{mmc3_fig_29} and \ref{mmc3_fig_30}).

Figures \ref{mmc3_fig_28} and \ref{mmc3_fig_30} also demonstrate the span of energies over which MMC can be used with fixed $E_{cut}=0.5$ GeV. With such $E_{cut}$, MMC seems to work for energies up to $0.5\cdot 10^{15}$ GeV, which is determined by the computer precision with which double precision numbers can be added: $0.5/0.5\cdot 10^{15} \sim 10^{-15}$. When relative position increments fall below that, the muon ``gets stuck'' in one point until its energy becomes sufficiently low or it propagates without stochastic losses sufficiently far, so that it can advance again. A muon ``stuck'' in this fashion still looses the energy, which is why it appears that its losses go up. With fixed $v_{cut}=10^{-2}-10^{-3}$ (and apparently as low as $10^{-12}-10^{-15}$), MMC shows no signs of such deterioration.

\end{document}